\documentclass[aps,pre,twocolumn,epsfig,floats]{revtex4}
\usepackage{mathrsfs}

\usepackage{amsmath}
\usepackage{color}
\usepackage{graphicx}
\usepackage{epsfig}
\usepackage{amsfonts}
\usepackage{mathrsfs}
\usepackage{mathtools}
\usepackage{float}
\usepackage{physics}
\usepackage{hyperref}
\hypersetup{
colorlinks=true,
citecolor=blue,
linkcolor=blue,
urlcolor=blue}
\usepackage{upgreek}

\usepackage{fourier}
\usepackage{MnSymbol}
\usepackage{soul}
\usepackage{xcolor, soul}
\sethlcolor{red}
\hypersetup{
colorlinks=true,
citecolor=blue,
linkcolor=blue,
urlcolor=blue}

\begin{document}
\title{Fermi-Point Topology Determines Emergent Conformal Criticality in Extended Quantum Spin Chains 
}

\author{Mohammad Abbasi$^{1}$}
\author{Saeed Mahdavifar$^{1}$}
\email[]{smahdavifar@gmail.com}

\affiliation{$^{1}$Department of Physics, University of Guilan, 41335-1914, Rasht, Iran}

\begin{abstract}
	
Quantum criticality in one-dimensional quantum systems is characterized by emergent conformal field theories (CFTs), whose central charge characterizes the independent gapless degrees of freedom of the low-energy theory. Establishing a microscopic connection between this universal conformal structure and the momentum-space topology of the underlying quasiparticle spectrum remains an important challenge. Here, we uncover a direct correspondence between Fermi-point topology, conformal criticality, and quantum entanglement in an extended quantum spin chain with competing cluster interactions, exchange anisotropy, and a transverse magnetic field. We show that interaction- and field-driven Lifshitz transitions generate a hierarchy of conformal critical phases with effective central charges $c_{\rm eff}=1/2$, $1$, $3/2$, $2$, and $3$, including an unconventional multicritical point at which Ising and Luttinger-liquid sectors coexist. Importantly, we demonstrate that the central charge is not determined simply by the number of lattice gap closings or Fermi points. Instead, it is determined by the number and conformal content of the independent low-energy continuum sectors that emerge from the Fermi-point structure after accounting for lattice symmetries, reciprocal-lattice identifications, and mode equivalences. Consequently, Lifshitz transitions can either leave the conformal anomaly unchanged or modify the central charge, depending on whether the reconstruction of the gapless modes generates new independent continuum sectors. Real-space and momentum-space entanglement spectra provide complementary microscopic signatures of this correspondence, revealing both the conformal content and the momentum-space organization of the critical modes. Our results establish a microscopic framework linking quasiparticle Fermi-point topology to emergent conformal field theories and demonstrate how interaction-driven spectral reconstruction can generate higher-central-charge criticality and unconventional multicritical behavior in one-dimensional quantum systems.
	
\end{abstract}

\date{\today}
\maketitle

\section{Introsuction}

Understanding how microscopic interactions give rise to universal quantum critical behavior is a central problem in condensed matter physics \cite{sachdev2011quantum,vojta2003quantum,cardy1996scaling,liu2020quantum,kuwahara2020area}. In one-dimensional quantum systems, conformal field theory (CFT) provides the universal description of a broad class of gapless phases and quantum critical points, with the conformal central charge characterizing the independent gapless degrees of freedom of the emergent low-energy theory \cite{francesco2012conformal,cardy2010conformal,yang2023entanglement,calabrese2009entanglement,calabrese2009entanglement1,giamarchi2003quantum,tsvelik2007quantum,sukumar1985supersymmetric}. Beyond identifying universality classes, the central charge governs fundamental universal properties, including the logarithmic scaling of entanglement entropy and characteristic finite-size corrections to the ground-state energy \cite{calabrese2004entanglement,zhao2022scaling,vidal2002entanglement}. These universal signatures provide a powerful bridge between microscopic lattice Hamiltonians and their emergent continuum field theories, raising a fundamental question: how is the conformal structure of a critical phase encoded in the microscopic momentum-space structure of its gapless quasiparticle modes?

A broad class of conventional nearest-neighbor quantum spin chains is described at criticality by low-central-charge conformal theories, most notably the Ising universality class with $c_{\rm eff}=1/2$ and the Luttinger-liquid class with $c_{\rm eff}=1$ \cite{sachdev2011quantum,francesco2012conformal,affleck1988universal,blote1986conformal,giamarchi2003quantum}. Extending the interaction range can, under fermionization, generate longer-range hopping and pairing processes that substantially enrich the quasiparticle spectrum by producing additional Fermi points and multiple Lifshitz transitions \cite{vodola2014kitaev,alecce2017extended,viyuela2018chiral,kumar2026topologically,prembabu2025multicriticality,yang2026topological,verresen2018topology}. Such spectral reconstructions can provide a route to higher-central-charge critical phases and unconventional multicritical behavior \cite{prembabu2025multicriticality,yang2026topological,verresen2018topology,fendley2014free,huijse2010supersymmetric,bauer2012supersymmetric,huijse2012supersymmetric,roy2017quantum}. Yet the emergence of the conformal theory cannot, in general, be inferred from a simple count of gapless Fermi points. Additional Fermi points can introduce new low-energy modes without necessarily producing an equal number of independent continuum sectors, because lattice symmetries, reciprocal-lattice identifications, and mode equivalences can constrain how these excitations combine in the low-energy theory \cite{fendley2014free,huijse2010supersymmetric,bauer2012supersymmetric,huijse2012supersymmetric,roy2017quantum,volovik2003universe,yang2021topological}. This distinction raises a fundamental question: does the emergence of additional Fermi points necessarily imply additional independent conformal sectors, or is the effective central charge instead determined by the number and conformal content of the independent continuum sectors into which the critical modes reorganize?

Quantum entanglement provides a powerful set of diagnostics for quantum criticality \cite{calabrese2004entanglement,laflorencie2016quantum,amico2008entanglement,lepori2017long,laflorencie2022entanglement}. In one-dimensional critical systems described by a $(1+1)$-dimensional conformal field theory, the universal logarithmic scaling of the entanglement entropy provides a direct means of extracting the conformal central charge \cite{calabrese2004entanglement,calabrese2009entanglement,holzhey1994geometric,affleck1986exact,blote1986conformal}. Beyond this total central charge, the bulk entanglement spectrum  can provide information about the structure and degeneracies of the underlying conformal sectors, while the momentum-space entanglement spectrum  can retain signatures of the momentum-space organization of the critical modes \cite{li2008entanglement,turner2011topological,pollmann2009entanglement,lauchli2010disentangling,chandran2011bulk,mondragon2013characterizing,lundgren2014momentum,lundgren2016universal,wu2023floquet}. These complementary perspectives are particularly valuable when the number and arrangement of gapless modes change across Lifshitz transitions. Nevertheless, a unified characterization linking the reconstruction of Fermi points, the organization of independent conformal sectors, and the corresponding evolution of real- and momentum-space entanglement spectra across multiple critical phases and multicritical points has not yet been established \cite{prembabu2025multicriticality,yang2026topological,mondragon2013characterizing,
lundgren2014momentum,lundgren2016universal,wu2023floquet,zhou2024topological,
banerjee2026entanglement}.

In this work, we address these questions by studying an exactly solvable extended quantum spin chain with competing interactions that generate a rich hierarchy of conformal critical phases \cite{prosen2014exact,zhong2025quantum,wang2021universal}. We identify multiple critical lines connected by Lifshitz transitions \cite{wang2021universal,li2025emergent,efremov2019multicritical}, including an unconventional multicritical point characterized by an effective central charge $c_{\rm eff}=3/2$, as well as higher-central-charge critical phases with $c$ reaching 3. By combining exact diagonalization, finite-size scaling, low-energy continuum analysis, and bulk entanglement spectra in both real and momentum space, we establish a systematic connection between the microscopic Fermi-point structure, emergent conformal criticality, and quantum entanglement. Crucially, we show that the conformal central charge is not determined by a simple count of Fermi points, but by the conformal content of the independent low-energy continuum sectors that emerge after accounting for Brillouin-zone periodicity, lattice symmetries, and mode equivalences. In particular, Fermi points related through the Brillouin-zone boundary need not represent independent continuum sectors, but can correspond to the same low-energy degree of freedom. Taken together, our results provide a unified microscopic framework connecting interaction-driven spectral reconstruction, Fermi-point topology, Lifshitz transitions, emergent conformal field theories, and entanglement in an exactly solvable quantum spin chain.

The remainder of this paper is organized as follows. In Sec.~\ref{MH}, we introduce the microscopic model and present its exact fermionic mapping. Section~\ref{QCL} is devoted to the analysis of the quantum critical lines and the associated Fermi-point reconstructions. In Sec.~\ref{ECU}, we investigate the emergent critical behavior using continuum field theory, finite-size scaling, and real- and momentum-space bulk entanglement spectra, establishing the corresponding universality classes and conformal descriptions. Finally, our conclusions are presented in Sec.~\ref{conc}. Technical details of the analytical calculations and complementary numerical results are collected in the Appendices.

\section{Microscopic Hamiltonian and Exact Fermionic Mapping} \label{MH}

We consider a one-dimensional spin-$1/2$ XX chain subjected to a transverse magnetic field and supplemented by a four-spin cluster interaction. The nearest-neighbor contribution is described by
\begin{align}
	\mathcal H_{XX}
	=
	J\sum_{n=1}^{N}
	\left(
	S_n^xS_{n+1}^x+
	S_n^yS_{n+1}^y
	\right)
	-
	Jh\sum_{n=1}^{N}S_n^z,
\end{align}
where $S_n^\mu$ ($\mu=x,y,z$) denote spin-$1/2$ operators at site $n$, $J>0$ is the antiferromagnetic exchange coupling, and $h$ is the transverse magnetic field. Despite its apparent simplicity, the XX chain captures the essential physics of one-dimensional planar magnetism while remaining exactly solvable.

\begin{figure}[!ht]
	\centerline{\includegraphics[width=0.9\linewidth,height=0.4\linewidth]{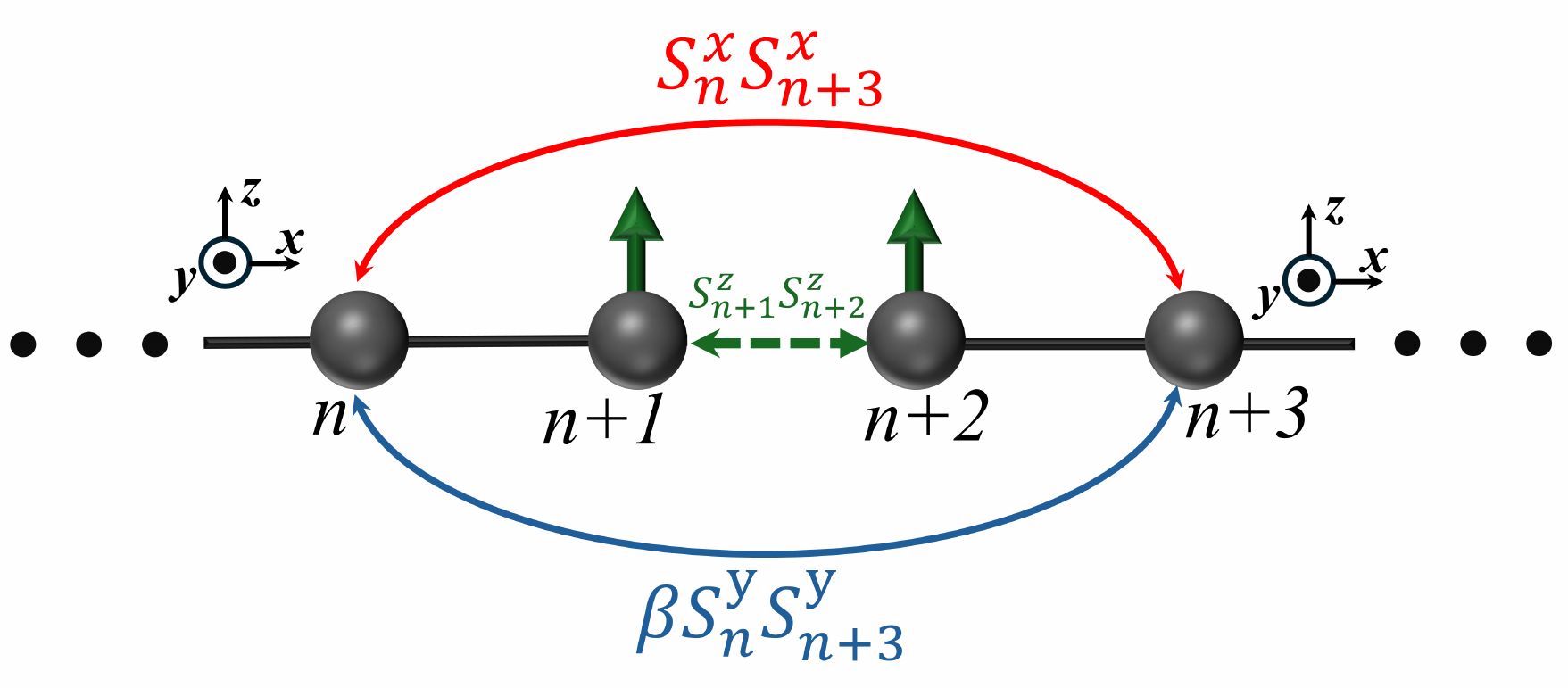}}
	\caption{Schematic representation of the four-spin cluster interaction Hamiltonian. }
	\label{Fig1}
\end{figure}

Beyond nearest-neighbor exchange, however, higher-order exchange processes naturally emerge in a wide variety of correlated quantum systems. In particular, four-spin cluster interactions have been shown to contribute to the renormalization of magnetic excitations and spectral properties in quasi-one-dimensional cuprates and ladder compounds, including $\mathrm{La_2CuO_4}$, $\mathrm{La_6Ca_{28}Cu_{24}O_{41}}$, and $\mathrm{La_4Sr_{10}Cu_{24}O_{41}}$ \cite{coldea2001spin,brehmer1999effects,matsuda2000magnetic,notbohm2007one}. Similar exchange mechanisms also arise in hydrogen-bonded ferroelectrics, squaric acid crystals, and certain polymeric materials \cite{chunlei1988green,wang1990critical,wang1989first,wang1989microscopic,silva1993pseudo}. More generally, cluster interactions provide effective descriptions for a broad class of many-body systems and can be engineered with a high degree of controllability in optical lattice platforms \cite{greiner2002quantum,zhou2015spin}. The diversity of these realizations highlights the important role played by multispin exchange processes in correlated quantum matter beyond simple pairwise interactions. Motivated by these observations, we supplement the conventional XX chain with a four-spin cluster interaction, schematically illustrated in Fig.~\ref{Fig1}, of the form

\begin{align}
	\mathcal H_{CI}
	=
	J'
	\sum_{n=1}^{N}
	\left[
	\left(
	S_n^xS_{n+3}^x
	+
	\beta S_n^yS_{n+3}^y
	\right)
	S_{n+1}^zS_{n+2}^z
	\right],
\end{align}
where $J'$ denotes the cluster-exchange strength and $\beta$ controls the anisotropy of the interaction.
\\
The complete Hamiltonian considered throughout this work is therefore given by
\begin{align}
	\mathcal H
	=
	\mathcal H_{XX}
	+
	\mathcal H_{CI}.
\end{align}
We introduce the dimensionless coupling $\alpha=J'/J$, which measures the relative strength of the four-spin interaction with respect to the nearest-neighbor exchange. The interplay among the magnetic field, the anisotropy parameter $\beta$, and the cluster coupling $\alpha$ gives rise to a rich variety of quantum critical phenomena. 

The model remains exactly solvable through a Jordan--Wigner transformation followed by a Bogoliubov diagonalization, which maps the spin Hamiltonian onto a quadratic fermionic Bogoliubov--de Gennes problem. The resulting quasiparticle excitation spectrum is

\begin{equation}
	\varepsilon(k)=\sqrt{\mathcal A_k^2+\mathcal B_k^2},
\end{equation}
where
\begin{align}
	\mathcal A_k &= \cos k-h+\frac{\alpha}{8}(1+\beta)\cos(3k), \nonumber \\
	\mathcal B_k &= \frac{\alpha}{8}(1-\beta)\sin(3k).
\end{align}
As shown throughout this work, the phase diagram, quantum critical behavior, and low-energy effective theory are completely determined by the momentum-space structure of this exact spectrum. The details of the Jordan--Wigner mapping and the Bogoliubov diagonalization are presented in Appendix~\ref{app:exact_solution}.

\section{Quantum Critical Lines and Fermi-Point Reconstruction} \label{QCL}

Having established the exact quasiparticle spectrum, we now investigate the critical properties of the model. Since all quantum phase transitions are associated with the closing of the excitation gap, the structure of the phase diagram can be understood directly from the evolution of the Fermi-point topology. We first consider the zero-field limit and subsequently discuss the effects induced by a transverse magnetic field.

\subsection{Zero-Field Lifshitz Criticality}
In the isotropic limit $\beta=1$, pairing correlations vanish identically and the quasiparticle spectrum reduces to
\begin{equation} 
	\varepsilon(k) = \cos k	+ \frac{\alpha}{4}\cos(3k).
	\label{eqbet1}
\end{equation}
The topology of the low-energy spectrum depends strongly on the cluster-interaction strength $\alpha$. For $\alpha<4/3$, the spectrum possesses two Fermi points located at $k_F=\pm\pi/2$, corresponding to a conventional gapless phase.

At the critical coupling $\alpha_c=4J/3$, additional zero-energy modes emerge. For $\alpha>4/3$, the spectrum contains six Fermi points, consisting of the original pair together with four additional nodes located at $k_F= \pm \pi/2$, $\pm k_0$, and  $\pm(\pi-k_0)$, where $k_0= \arccos \left( \sqrt{ (3/4)-(J/\alpha) } \right)$.

Therefore, the transition at $\alpha_c$ originates from a reconstruction of the Fermi-point topology rather than from symmetry breaking or gap opening, signaling a Lifshitz transition within the gapless phase.
We now turn to the more general anisotropic case $\beta\neq1$, where pairing correlations are present and the spectrum is determined by the simultaneous conditions $\mathcal A_k=0$ and $\mathcal B_k=0$. Solving these equations yields three critical lines in the $(\alpha,\beta)$ parameter space,

\begin{equation}
	\beta_{c_1}=1, \qquad \beta_{c_2}=- \left( 1+\frac{8}{\alpha} \right), \qquad \beta_{c_3}= \frac{4}{\alpha}-1.
\end{equation}
Each of these critical lines is associated with a distinct nodal structure of the quasiparticle spectrum. Along the isotropic line $\beta_{c_1}=1$, the system exhibits the Lifshitz transition discussed above, separating phases with two and six Fermi points. Along the second critical line, $\beta_{c_2}=- \left( 1+{8}/{\alpha} \right)$, the spectrum possesses three Fermi points located at $k_F=0,\ \pm\pi$. In contrast, along the third critical line, $\beta_{c_3}= {4}/{\alpha}-1$, the quasiparticle spectrum contains four Fermi points located at $k_F= \pm{\pi}/{3}$ and $k_F= \pm{2\pi}/{3}$.

These results demonstrate that each critical line is characterized by a different Fermi-point topology and therefore corresponds to a distinct universality class of gapless critical behavior. The number and arrangement of the Fermi points provide the key ingredients for understanding the low-energy theory, the associated central charge, and the nature of the Lifshitz criticalities that emerge in the model.

\subsection{Field-Driven Quantum Criticality and Momentum-Space Structure}
We next investigate the effect of a transverse magnetic field on the Lifshitz criticality established in the previous section. In the isotropic limit $\beta=1$ \cite{kurdadze2026critical}, the magnetic field simply shifts the exact single-particle dispersion,
\begin{equation}
	\varepsilon(k)
	=
	\cos k-h+\frac{\alpha}{4}\cos(3k),
\end{equation}
thereby continuously modifying the Fermi-point topology without introducing pairing correlations. Consequently, the zero-field Lifshitz transition broadens into a family of field-driven topological transitions governed entirely by the evolution of the Fermi surface. The low-energy properties are completely determined by the Fermi points satisfying $\varepsilon(k_F)=0$. Introducing $x=\cos k$, the Fermi-point condition can be recast as
\begin{equation}
	\alpha x^{3} + \left( 1-\frac{3\alpha}{4} \right)x -h =0.
\end{equation}
For fixed magnetic field, the number of real roots within the physical interval $x\in[-1,1]$ determines the number of Fermi points and hence the topology of the gapless phase. Changes in the root structure signal Lifshitz transitions associated with the emergence or annihilation of Fermi points.

The corresponding phase boundaries are obtained from the stationary points of the dispersion relation. One branch is located at
\begin{equation}
	h_{c}^{(\mathrm{sp})} = 1+\frac{\alpha}{4},
	\label{betc1}
\end{equation}
while for sufficiently strong cluster coupling, $\alpha\ge4/3$, an additional branch emerges,
\begin{equation}
	h_{c_L}^{(\mathrm{sp})}	= -\alpha c^3 - \left( 1 -\frac{3\alpha}{4} \right)c,
	\label{betc2}
\end{equation}
where
\begin{equation}
	c= \sqrt{\frac{1}{4}-\frac{1}{3\alpha}}.
\end{equation}
These critical lines separate gapless regions with different Fermi-point topologies and describe how the transverse field reshapes the Lifshitz structure of the isotropic model.

The anisotropic regime, $\beta\neq1$, exhibits qualitatively different behavior. In this case, the anomalous pairing term opens a quasiparticle gap, and quantum phase transitions are driven by the collapse of the Bogoliubov spectrum rather than by a reconstruction of the Fermi surface. The critical boundaries are therefore determined by the simultaneous conditions $\mathcal A_k=0$ and	$\mathcal B_k=0$, which define the phase diagram in the parameter space $(h,\alpha,\beta)$. Besides the anisotropy-driven critical line
\begin{equation}
	\beta_c=1,
	\qquad
	h\le 1+\frac{\alpha}{4},
\end{equation}
the field-induced transitions naturally organize into two pairs of branches,
\begin{align}
	h_c^{(L_1,R_1)}
	&=
	\mp 1
	\mp
	\frac{\alpha}{8}(\beta+1),
	\\
	h_c^{(L_2,R_2)}
	&=
	\mp\frac{1}{2}
	\pm
	\frac{\alpha}{8}(\beta+1),
\end{align}
leading to four distinct critical fields,
\begin{align}
	h_{c_1}
	&=
	-1
	-
	\frac{\alpha}{8}(\beta+1),
	\qquad
	\beta\le
	-\left(
	1+\frac{8}{\alpha}
	\right),
	\nonumber\\
	h_{c_2}
	&=
	1
	+
	\frac{\alpha}{8}(\beta+1),
	\qquad
	\beta>
	-\left(
	1+\frac{8}{\alpha}
	\right),
	\nonumber\\
	h_{c_3}
	&=
	\frac{1}{2}
	-
	\frac{\alpha}{8}(\beta+1),
	\qquad
	\beta\le
	\left(
	\frac{4}{\alpha}-1
	\right),
	\nonumber\\
	h_{c_4}
	&=
	-\frac{1}{2}
	+
	\frac{\alpha}{8}(\beta+1),
	\qquad
	\beta>
	\left(
	\frac{4}{\alpha}-1
	\right).
\end{align}
The physical origin of these critical branches becomes transparent from the momentum-space structure of the BdG spectrum. Since $\mathcal B_k\propto\sin(3k)$, gap closing can occur only at the discrete momenta $k_F= n\pi/3$, with $n=0,\pm1,\pm2,\pm3$.
Each critical branch is therefore associated with a distinct critical mode. The line $h_{c_1}$ corresponds to gap closing at the Brillouin-zone boundary, $k_F=\pm\pi$, whereas $h_{c_2}$ originates from a zero-momentum instability at $k_F=0$. The remaining branches are associated with finite-momentum critical modes: $h_{c_3}$ is characterized by gap closing at $k_F=\pm\pi/3$, while $h_{c_4}$ corresponds to $k_F=\pm2\pi/3$.

From this perspective, the field-driven phase diagram can be viewed as arising from competing critical modes located at high-symmetry points of the Brillouin zone. The four-spin cluster interaction not only generates pairing correlations but also selects a discrete set of momenta at which the quasiparticle gap may collapse, thereby determining the topology of the quantum critical boundaries and the structure of the low-energy excitations.

\section{Emergent Criticality and Universality} \label{ECU}
With the quantum critical boundaries and the corresponding Fermi-point topology now identified, we now focus on the emergent low-energy theories associated with these transitions. While all critical points correspond to gap closings of the quasiparticle spectrum, the resulting critical modes need not belong to the same universality class. Their scaling properties can be naturally understood within the framework of conformal field theory. In the following, we investigate both the zero-field and field-induced regimes, revealing a rich hierarchy of Lifshitz and quantum critical behaviors characterized by distinct central charges and low-energy excitation structures.

\subsection{Zero-Field Critical Regime}

\begin{figure}[h]
	\centering
	\includegraphics[width=0.8\linewidth,height=0.6\linewidth]{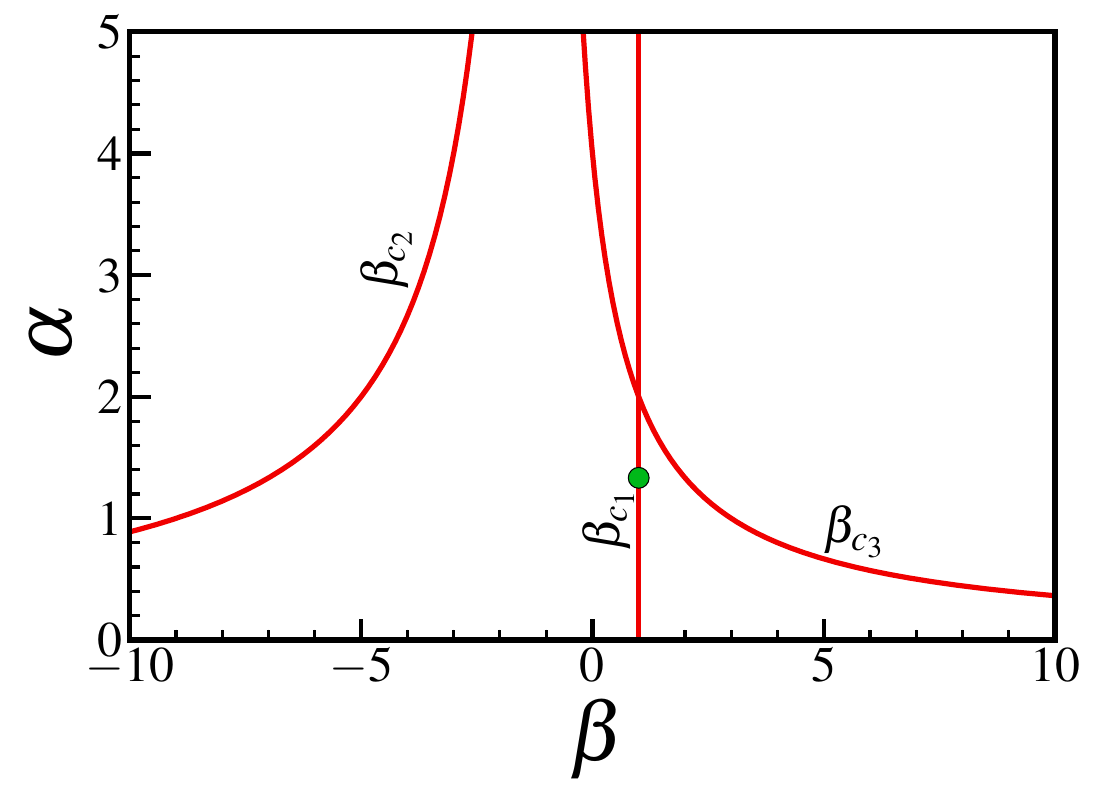}
	\caption{Ground-state phase diagram of the generalized spin-$1/2$ XX chain with four-spin cluster interactions in the absence of a transverse magnetic field ($h=0$).}
	\label{fig2}
\end{figure}

We begin by considering the zero-field limit, where the interplay between the cluster interaction and the anisotropy gives rise to a sequence of quantum critical lines with distinct nodal structures. As discussed in the previous section, the exact solution predicts three critical boundaries in the $(\alpha,\beta)$ plane, corresponding to different configurations of gapless modes and Fermi-point topologies. These analytical results are summarized in Fig.~\ref{fig2}, which was established previously in Ref.~\cite{mahdavifar2026topological,katibzadah2026quantum,mahdavifar2026chiral,kheiri2026dynamical,valeh2026spin} and is reproduced here for completeness.

The isotropic line $\beta=1$ plays a special role, as it preserves the $U(1)$ symmetry and hosts a Lifshitz transition at $\alpha_c=4/3$, where the Fermi surface is reconstructed from two to six gapless Fermi points. This reconstruction provides an ideal setting for exploring how microscopic changes in the Fermi-point topology are encoded in the emergent conformal field theory. Away from the isotropic limit, the pairing term generated by the cluster interaction lifts the $U(1)$ symmetry and qualitatively reconstructs the nodal structure of the quasiparticle spectrum, giving rise to additional critical boundaries characterized by three and four gap-closing momenta. Although all these boundaries originate from the same microscopic gap-closing condition, they correspond to qualitatively distinct low-energy theories because the gapless modes organize into different effective continuum sectors. In the following, we analyze these critical lines through the singular behavior of the ground-state energy, the evolution of the quasiparticle spectrum, and their corresponding conformal-field-theory signatures.

We next investigate the momentum-resolved quasiparticle spectrum, which directly reveals the microscopic origin of the low-energy excitations governing each critical phase. Rather than the number of gap-closing momenta alone, it is the organization of these modes into independent continuum sectors that determines the emergent conformal theory. Consequently, the evolution of the quasiparticle dispersion across the Lifshitz transitions provides a microscopic picture of how the cluster interaction reshapes the Fermi-point topology and controls the resulting conformal criticality.

\begin{figure}[h]
	\centerline{\includegraphics[width=0.5\linewidth,height=0.42\linewidth]{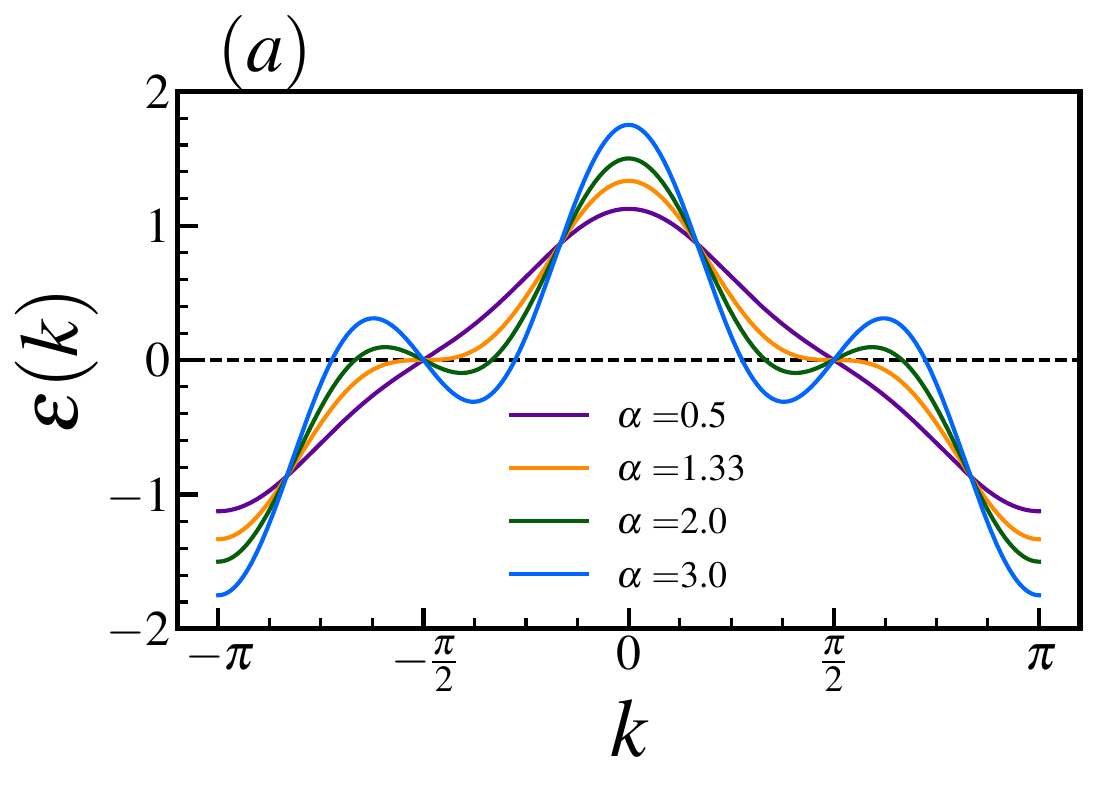} \includegraphics[width=0.5\linewidth,height=0.42\linewidth]{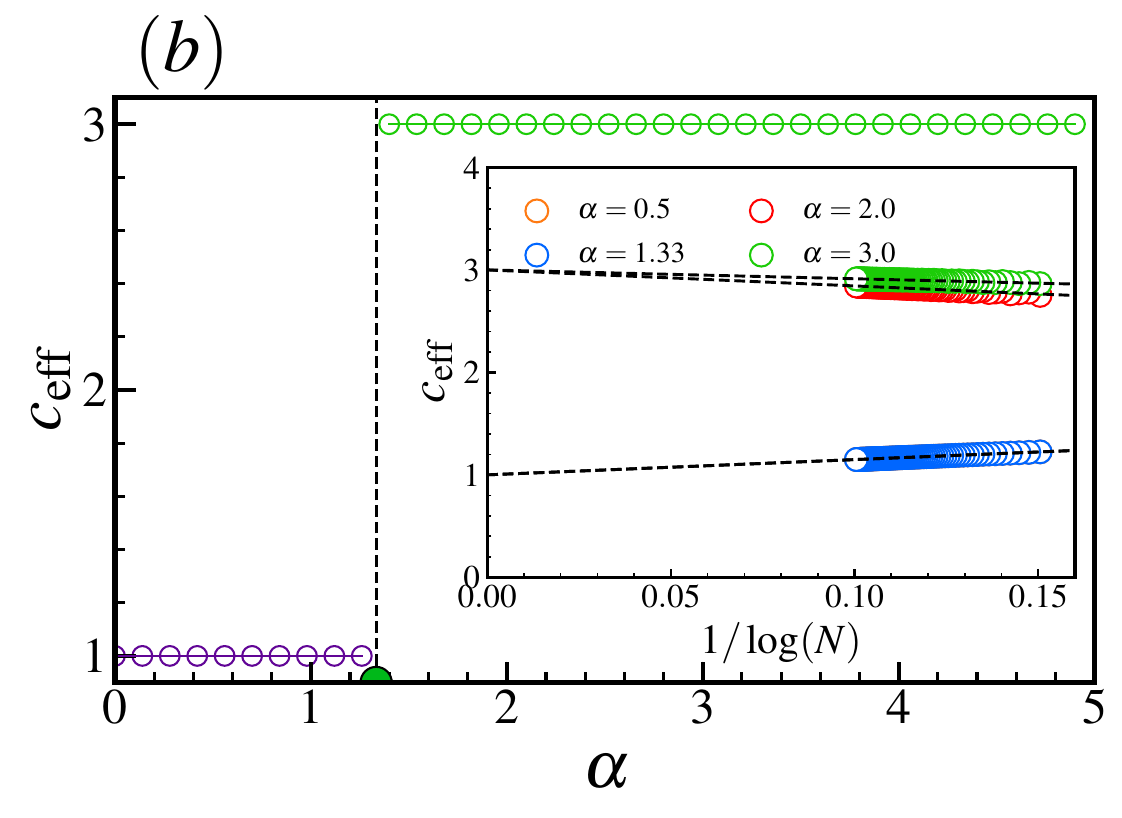} 
		
	}
	\centerline{\includegraphics[width=0.5\linewidth,height=0.42\linewidth]{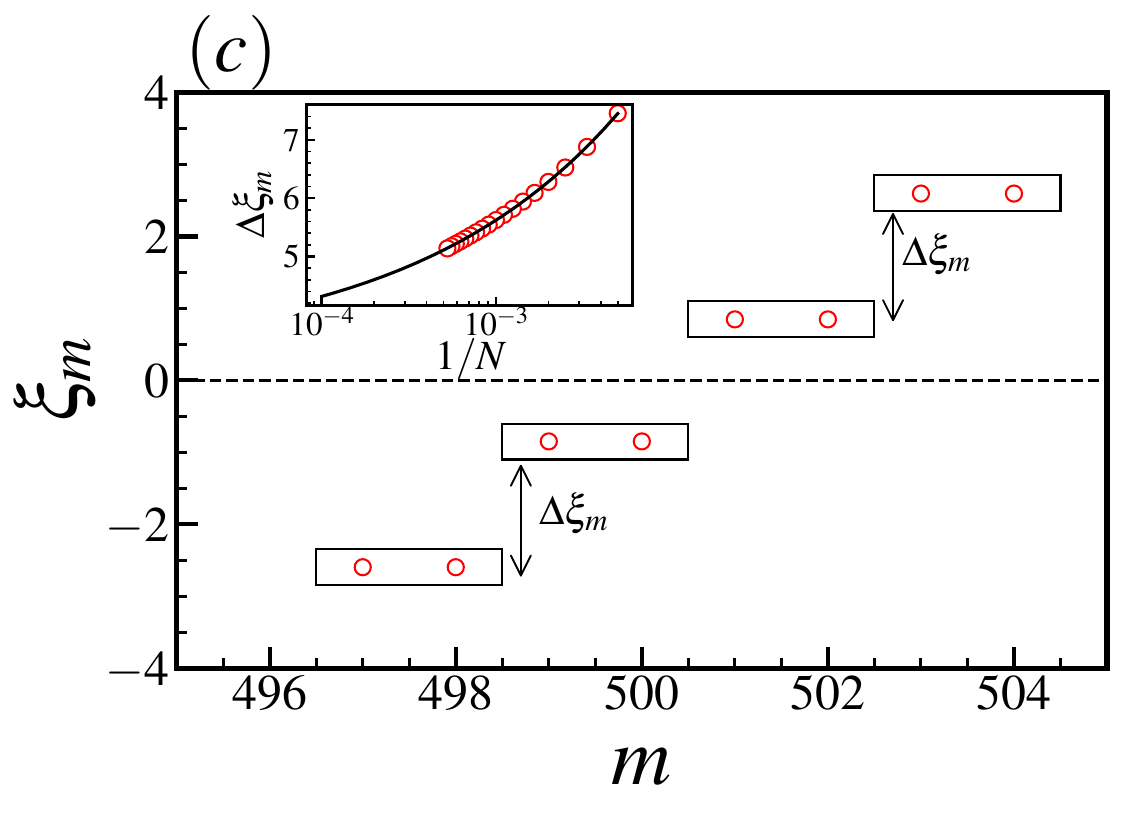} \includegraphics[width=0.5\linewidth,height=0.42\linewidth]{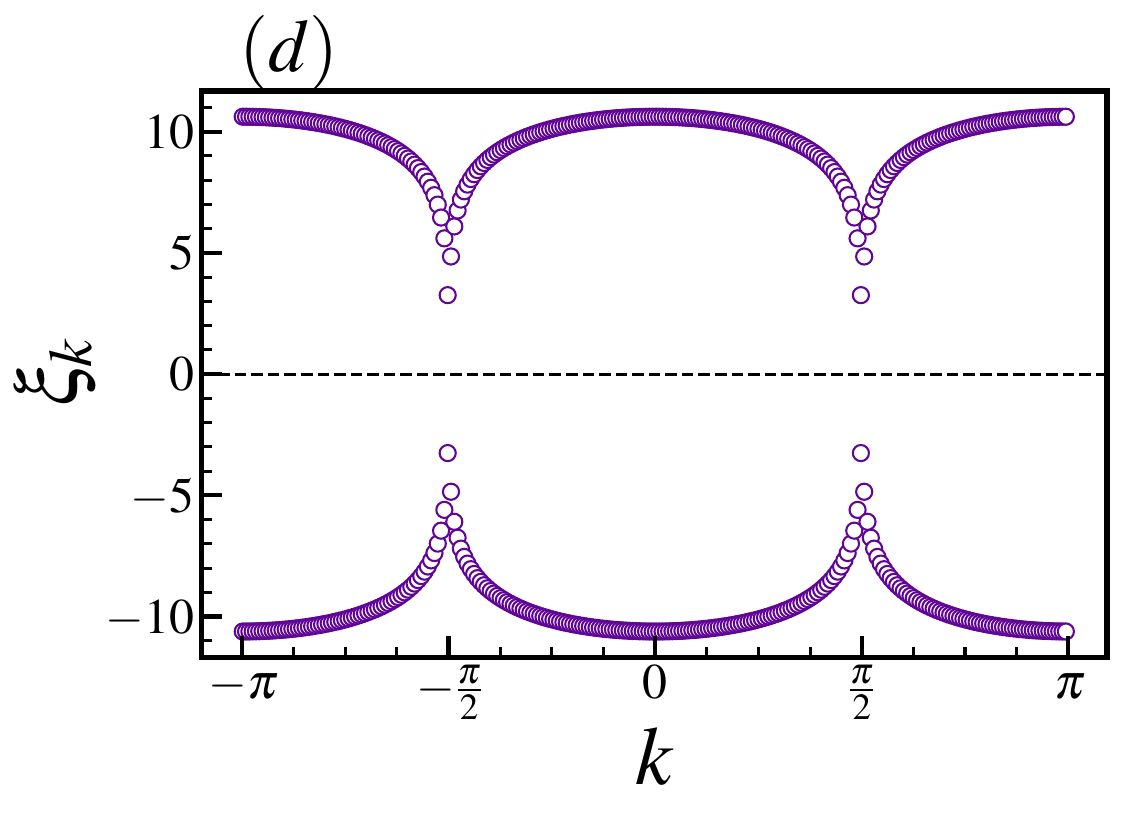} 
		
	}
	\centerline{\includegraphics[width=0.5\linewidth,height=0.42\linewidth]{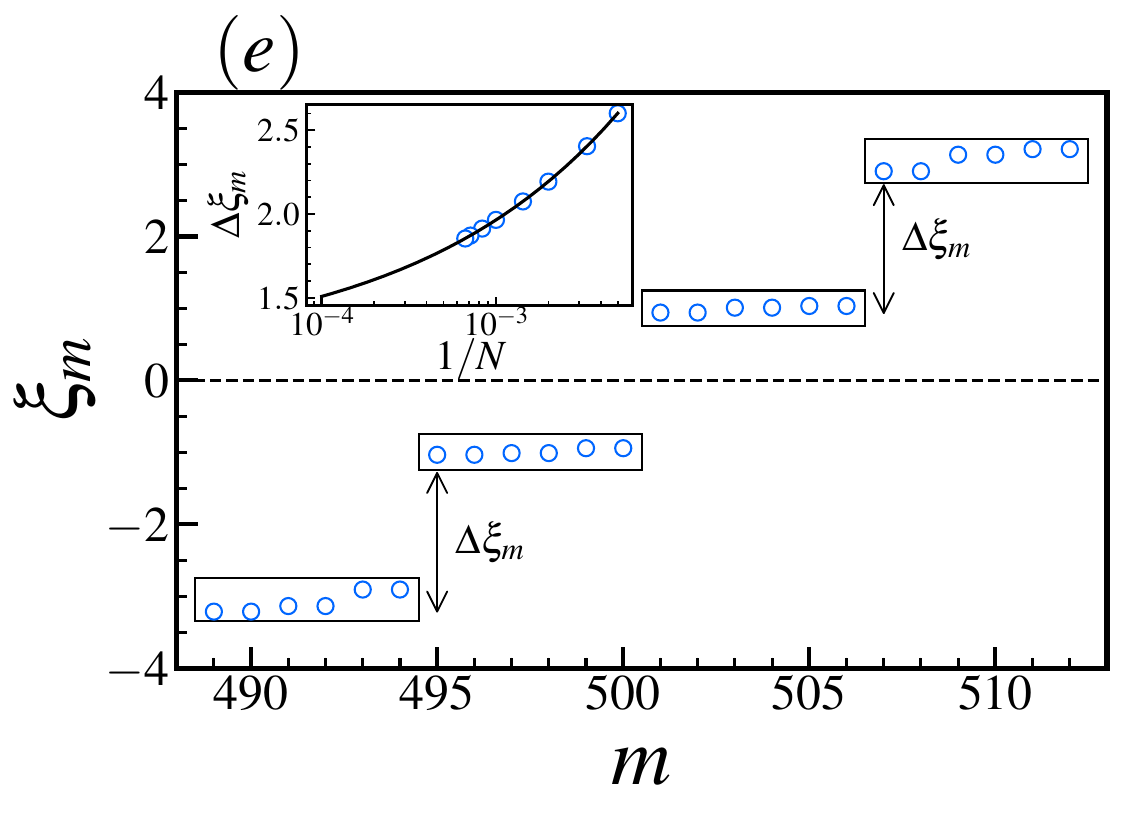} \includegraphics[width=0.5\linewidth,height=0.42\linewidth]{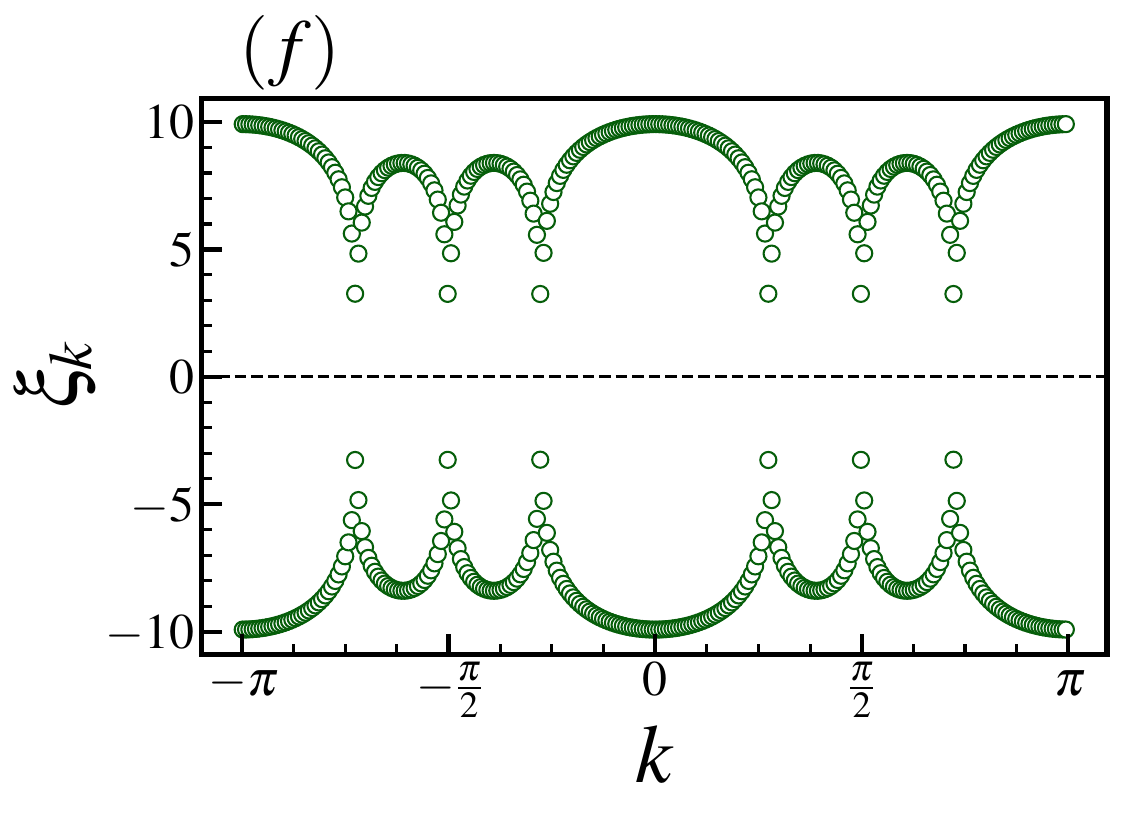} 
		
	}

	\caption{(a) Quasiparticle spectrum across the Lifshitz transition on the isotropic line ($\beta=1$). (b) Effective central charge extracted from entanglement-entropy scaling. (c,d) Real-space (RES) and momentum-space (MES) bulk entanglement spectra in the $c_{\rm eff}=1$ phase ($\alpha<\alpha_L$). (e,f) Real-space (RES) and momentum-space (MES) bulk entanglement spectra in the $c_{\rm eff}=3$ phase ($\alpha>\alpha_L$).
	}
	\label{Fig3}
	
\end{figure}

We first focus on the isotropic line $\beta=1$, where the $U(1)$ symmetry is preserved and the quasiparticle spectrum reduces to a purely dispersive band. Figure~\ref{Fig3}~(a) shows the evolution of the single-particle spectrum as the cluster coupling $\alpha$ is tuned across the Lifshitz point $\alpha_c=4/3$. For $\alpha<\alpha_c$, exemplified by $\alpha=0.5$, the spectrum intersects the Fermi level only at $k_F=\pm\pi/2$, giving rise to two gapless modes with linear dispersion. As the critical coupling is approached, the low-energy band progressively flattens around the Fermi points, and at $\alpha=\alpha_c$ the spectrum develops a higher-order band touching. This multicritical point marks the onset of a topological reconstruction of the Fermi surface.

Beyond the transition, the original pair of Fermi points persists while two additional pairs of gapless nodes emerge, yielding a total of six zero-energy crossings for $\alpha>\alpha_c$. The transition therefore corresponds to a Lifshitz reconstruction of the Fermi surface from two to six Fermi points rather than to symmetry breaking or gap opening. This profound change in the momentum-space structure constitutes the microscopic origin of the unconventional critical behavior discussed below.

The critical behavior is determined by the local quasiparticle dispersion near the gapless Fermi points. Away from the Lifshitz point, the quasiparticle spectrum in the vicinity of each Fermi point exhibits the linear scaling $\varepsilon(k)\propto |k-k_F|$, corresponding to the dynamical exponent $z=1$. Thus, both the two-node phase ($\alpha<\alpha_c$) and the six-node phase ($\alpha>\alpha_c$) are governed by massless Dirac excitations. In contrast, exactly at the Lifshitz point $\alpha_c=4/3$, the linear Dirac velocity vanishes and the low-energy spectrum around $k_F=\pm\pi/2$ acquires the higher-order form $\varepsilon(k)\propto |k-k_F|^{3}$, yielding the unconventional dynamical exponent $z=3$ characteristic of the Lifshitz critical mode. The corresponding momentum-resolved spectra and the analytical derivation of these scaling forms are presented in Appendix~\ref{app:bet1}. The emergence of the cubic band touching at $\alpha_c$ drives the nucleation of the additional Dirac nodes and, as we demonstrate below, has profound consequences for the conformal structure of the critical theory.

The reconstruction of the low-energy mode structure revealed above  is directly reflected in the conformal properties of the critical theory. To characterize the resulting universality classes, we employ three complementary entanglement diagnostics: the effective central charge, the real-space entanglement spectrum, and the momentum-space entanglement spectrum. The effective central charge $c_{\rm eff}$ is extracted from the finite-size scaling of the bipartite von Neumann entanglement entropy. According to CFT, the bipartite entanglement entropy of a one-dimensional critical system with periodic boundary conditions obeys the Calabrese–Cardy scaling form:
\begin{equation}
	S(l) = \frac{c_{\rm eff}}{3} \log_2 \left[ \frac{N}{\pi} \sin \left( \frac{\pi l}{N} \right) \right] +s_0 ,
	\label{eq:CCm}
\end{equation}
where $c_{\rm eff}$ is the central charge and $s_0$ is a nonuniversal constant. Throughout this work, the fitted value is denoted by $c_{\rm eff}$ and provides a quantitative characterization of the low-energy degrees of freedom of the underlying CFT that governs the critical behavior. The real-space entanglement spectrum provides complementary information about the organization of the low-energy excitations and the underlying conformal structure, while the momentum-space entanglement spectrum probes correlations between quasiparticle modes associated with different Fermi points and is particularly sensitive to changes in Fermi-surface topology. Owing to the quadratic fermionic representation of the model, all these quantities can be computed directly from the correlation matrix. Details of the correlation-matrix formulation and the corresponding expressions for the entropy and entanglement spectra are summarized in Appendix~\ref{app:entanglement}.
%%%%%%%%%%%%%%%%%%%%%%%%%%%%%%
Figure~\ref{Fig3}~(b) presents the effective central charge extracted from the finite-size scaling of the entanglement entropy along the isotropic line ($\beta=1$) at zero magnetic field. As the cluster coupling $\alpha$ is varied, two distinct conformal regimes are observed, separated by the Lifshitz transition at $\alpha_c=4/3$. Throughout the region $\alpha<\alpha_c$, the numerical data consistently extrapolate to the universal value $c_{\rm eff}=1$. Remarkably, no enhancement of the central charge occurs exactly at the Lifshitz point, where the extracted value remains compatible with $c_{\rm eff}=1$. Crossing the transition, however, produces an abrupt reconstruction of the low-energy critical theory, with the effective central charge rapidly increasing to the universal value $c_{\rm eff}=3$ for $\alpha>\alpha_c$. The robustness of these results is demonstrated in the inset of Fig.~\ref{Fig3}~(b), where finite-size estimates for representative couplings $\alpha=0.5$, $4/3$, $2$, and $3$ are extrapolated as a function of $1/\log N$. The extrapolated thermodynamic values clearly separate into two universality classes: all points below and at the Lifshitz transition converge to $c_{\rm eff}=1$, whereas those beyond the transition converge to $c_{\rm eff}=3$. These values therefore reflect intrinsic properties of the thermodynamic critical theory rather than finite-size effects.

The origin of this reconstruction can be understood directly from the evolution of the Fermi-point topology discussed above. For $\alpha<\alpha_c$, the low-energy spectrum contains a single pair of independent Fermi points, corresponding to one gapless low-energy Dirac sector and consequently $c_{\rm eff}=1$. Beyond the Lifshitz transition, two additional symmetry-related pairs of gapless Fermi points emerge, producing three independent low-energy Dirac sectors. In the present model, each independent gapless sector contributes one unit to the conformal central charge, yielding the universal value $c_{\rm eff}=3$. The Lifshitz transition therefore reconstructs not only the topology of the Fermi surface but also the conformal structure of the low-energy theory by changing the number of independent gapless sectors. This establishes a direct microscopic correspondence between the Fermi-point topology and the emergent conformal field theory, demonstrating that the effective central charge is governed by the number of independent low-energy conformal sectors generated by the underlying Fermi-point structure.

To identify corresponding conformal field theories and the effective central charge of the underlying CFT, we derive the continuum low-energy field theory associated with the microscopic Bogoliubov--de Gennes Hamiltonian. The construction proceeds by expanding the lattice fermion operator around all symmetry-related gapless momenta and retaining only the slowly varying long-wavelength modes. If the spectrum contains $N_f$ pairs of gapless Fermi points $\pm k_i$, the lattice operator may be written as
\begin{equation}
	c_n = \sum_{i=1}^{N_f} \left[ \psi_{R,i}(x)e^{ik_i n} + \psi_{L,i}(x)e^{-ik_i n} \right],	\qquad 	x=na ,
\end{equation}
where $\psi_{R,i}$ and $\psi_{L,i}$ denote slowly varying right- and left-moving continuum fields.
\\
Substituting this expansion into the lattice Hamiltonian and retaining the leading gradient terms yields the effective continuum BdG Hamiltonian
\begin{align}
	H_{\rm eff} = \sum_{i=1}^{N_f} \int dx\, \Big[
	&
	M_i \left( \psi^\dagger_{R,i}\psi_{R,i} + \psi^\dagger_{L,i}\psi_{L,i} \right) \nonumber\\
	& +i v_{F,i} \left( \psi^\dagger_{R,i}\partial_x\psi_{R,i} - \psi^\dagger_{L,i}\partial_x\psi_{L,i} \right) \nonumber\\
	&
	+i \Delta_i \left( \psi^\dagger_{R,i}\psi^\dagger_{L,i} - \psi_{L,i}\psi_{R,i} \right) \nonumber\\
	& +\tilde v_i	\left( \psi^\dagger_{R,i}\partial_x\psi^\dagger_{L,i} + \psi_{L,i}\partial_x\psi_{R,i} \right) \Big].
\end{align}

The coefficients $M_i$, $\Delta_i$, $v_{F,i}$, and $\tilde v_i$ are determined directly by the microscopic BdG Hamiltonian evaluated at the corresponding gapless momentum $k_i$, representing the effective mass, pairing amplitude, normal velocity, and anomalous velocity of each low-energy sector, respectively. Their explicit dependence on the microscopic couplings follows from the continuum expansion of the lattice Hamiltonian and is presented in Appendix~\ref{app:general_continuum}. Throughout this work, we will show that the effective central charge is controlled by the number of independent low-energy continuum sectors. In the present model, this corresponds to the number of independent Fermi-point pairs after identifying symmetry-related points connected through the Brillouin-zone boundary.
%%%%%%%%%%%%%%%%%%%%%%%%%%%%%%%%%%%%%%5

Specializing now to the line $\beta=1$ and $h=0$, the pairing amplitude vanishes identically, $\Delta_i=\tilde v_i=0$, and the continuum theory reduces to a sum of independent gapless fermionic sectors. The low-energy physics is therefore determined by the structure of the Fermi surface and, in particular, by the number of symmetry-related pairs of gapless momenta satisfying Eq.~(\ref{eqbet1}).
For $\alpha<\alpha_c=4/3$, Eq.~(\ref{eqbet1}) admits only a single pair of Fermi points, $ k_F=\pm{\pi}/{2}$, with a finite Fermi velocity
\begin{equation}
	v_F = -\left.\frac{d\mathcal A_k}{dk}\right|_{k_F} = \pm 1 \mp \frac{3\alpha}{4}.
\end{equation}
Expanding around the two Fermi points yields the standard massless Dirac Hamiltonian
\begin{equation}
	H_{\rm eff} = i v_F \int dx\, \left( \psi_R^\dagger \partial_x \psi_R - \psi_L^\dagger \partial_x \psi_L \right),
\end{equation}
which describes a single gapless Dirac fermion. The corresponding low-energy critical theory is therefore the  $c_{\rm eff}=1$ Luttinger-liquid universality class, consistent with the entanglement-entropy results shown in Fig.~\ref{Fig3}~(b). At the Lifshitz transition $\alpha=\alpha_c$, the Fermi points remain located at $\pm \pi/2$, but the linear coefficient vanishes,$v_F=0$. The leading low-energy dispersion is then cubic, $E(q)\sim q^3$, while the number of independent gapless sectors remains unchanged,  signaling a Lifshitz critical point at which the topology of the Fermi surface changes. Importantly, no additional gapless branches emerge exactly at the transition, explaining why the effective central charge remains compatible with $c_{\rm eff}=1$. For $\alpha>\alpha_c$, the Lifshitz transition reconstructs the Fermi surface and generates three distinct pairs of gapless Fermi points,$\pm{\pi}/{2}$, $\pm k_0$, and $\pm(\pi-k_0)$. Each pair possesses a finite linear velocity and therefore gives rise to an independent massless Dirac sector. The continuum Hamiltonian becomes
\begin{equation}
	H_{\rm eff}	=	i	\sum_{a=1}^{3}	v_a	\int dx	\left( \psi_{R,a}^{\dagger}\partial_x\psi_{R,a}	- \psi_{L,a}^{\dagger}\partial_x\psi_{L,a} \right),
	\label{eqheffbet1}
\end{equation}
corresponding to three decoupled gapless Dirac fermions. The resulting CFT is therefore a direct product of three independent $c_{\rm eff}=1$ conformal field theories.

Because the three Dirac sectors are completely decoupled, the corresponding conformal field theory is the direct product of three independent $ c_{\rm eff}=1$ theories. The conformal central charge is therefore additive, giving
\begin{equation}
	c_{\rm eff}	= \sum_{a=1}^{3} c_a = 3.
\end{equation}
\\
Equivalently, the strong-coupling phase may be viewed as three decoupled copies of the $c_{\rm eff}=1$ Luttinger-liquid fixed point. The abrupt change in the number of linearly dispersing Dirac modes across the Lifshitz transition therefore provides the microscopic field-theoretic origin of the observed jump of the effective central charge from $c_{\rm eff}=1$ to $c_{\rm eff}=3$. This result illustrates a central conclusion of the present work: the effective conformal central charge is controlled by the number of independent low-energy continuum sectors generated by the Fermi-point topology, rather than by the mere occurrence of additional band crossings. The detailed derivation of the continuum theory and the explicit construction of the low-energy Hamiltonians for the $\beta=1$, $h=0$ line are presented in Appendix~\ref{app:bet1_continuum}.

%%%%%%%%%%%%%%%%%%%%

The bulk entanglement spectrum provides an independent probe of the low-energy degrees of freedom and complements the effective central charge by revealing the organization of the underlying conformal sectors. It therefore offers a direct fingerprint of the Fermi-surface reconstruction across the Lifshitz transition. The relation between the bulk entanglement spectrum and the correlation matrix is summarized in Appendix~\ref{app:entanglement}.

In the weak-coupling regime $\alpha<\alpha_c$, where the low-energy theory contains a single massless Dirac fermion and $c_{\rm eff}=1$, the real-space bulk entanglement spectrum shown in Fig.~\ref{Fig3}~(c) exhibits a characteristic sequence of doubly degenerate low-lying levels, indicating that the low-energy entanglement excitations organize into a single conformal tower. The finite-size splitting $\Delta\xi_m$ between successive entanglement multiplets decreases systematically with increasing system size, as illustrated in the inset. This behavior indicates that the low-energy entanglement sector becomes asymptotically gapless in the thermodynamic limit. The observed twofold degeneracy reflects the presence of a single independent Dirac sector formed by one pair of left- and right-moving gapless modes of the underlying lattice model.

Additional insight is obtained from the momentum-space bulk entanglement spectrum shown in Fig.~\ref{Fig3}~(d). The spectrum develops pronounced minima precisely at the gapless Fermi momenta $k=\pm k_F$, demonstrating that the momentum-space entanglement spectrum faithfully resolves the microscopic locations of the low-energy critical modes. These minima identify the momentum-space locations carrying the strongest quantum entanglement and therefore directly track the gapless degrees of freedom of the system. The spectrum is symmetric around $\xi_k=0$, reflecting the particle-hole structure caused by the Bogoliubov-de Gennes description. Most importantly, only a single pair of entanglement minima is present, consistent with the existence of a single Dirac sector and the effective central charge $c_{\rm eff}=1$.

Across the Lifshitz transition, the bulk entanglement spectrum undergoes a qualitative reorganization, reflecting the reconstruction of the underlying low-energy conformal sectors. As shown in Fig.~\ref{Fig3}~(e), the doubly degenerate pattern is replaced by a degenerate pattern of sixfold multiplets in the strong-coupling regime $\alpha>\alpha_c$. The residual splitting between the members of each multiplet decreases continuously with increasing system size, indicating that the exact sixfold degeneracy is recovered in the thermodynamic limit. The sixfold degeneracy reflects the coexistence of three independent Dirac sectors, each contributing an identical pair of low-energy entanglement excitations. This enhanced multiplicity originates from the Lifshitz reconstruction of the Fermi surface, which produces three independent gapless Dirac sectors and, consequently, three distinct low-energy entanglement channels.

The momentum-space bulk entanglement spectrum in Fig.~\ref{Fig3}~(f) reveals the microscopic origin of this degeneracy enhancement. In contrast to the $c_{\rm eff}=1$ phase, the spectrum now exhibits three distinct pairs of minima located at the reconstructed Fermi points $\pm\pi/2$, $\pm k_0$, and $\pm(\pi-k_0)$. Each pair of entanglement minima identifies an independent gapless Dirac sector in the continuum theory. Consequently, the low-energy theory contains three decoupled massless Dirac fermions, yielding $c_{\rm eff}=3$. The sixfold multiplet structure observed in the real-space entanglement spectrum therefore constitutes a direct entanglement signature of the Fermi-surface reconstruction and provides an independent entanglement confirmation of the continuum field theory derived above in Appendix~\ref{app:general_continuum}. 
Collectively, Fig.~\ref{Fig3} reveals that the cluster interaction fundamentally enriches the low-energy critical structure of the system. By introducing longer-range correlations beyond nearest-neighbor couplings, it drives a Lifshitz reconstruction of the Fermi surface that increases the number of gapless Dirac sectors from one to three.

%%%%%%%%%%%%%%%%%%%%%%%%%%%%%

\begin{figure*}
	\centerline{\includegraphics[width=0.25\linewidth,height=0.2\linewidth]{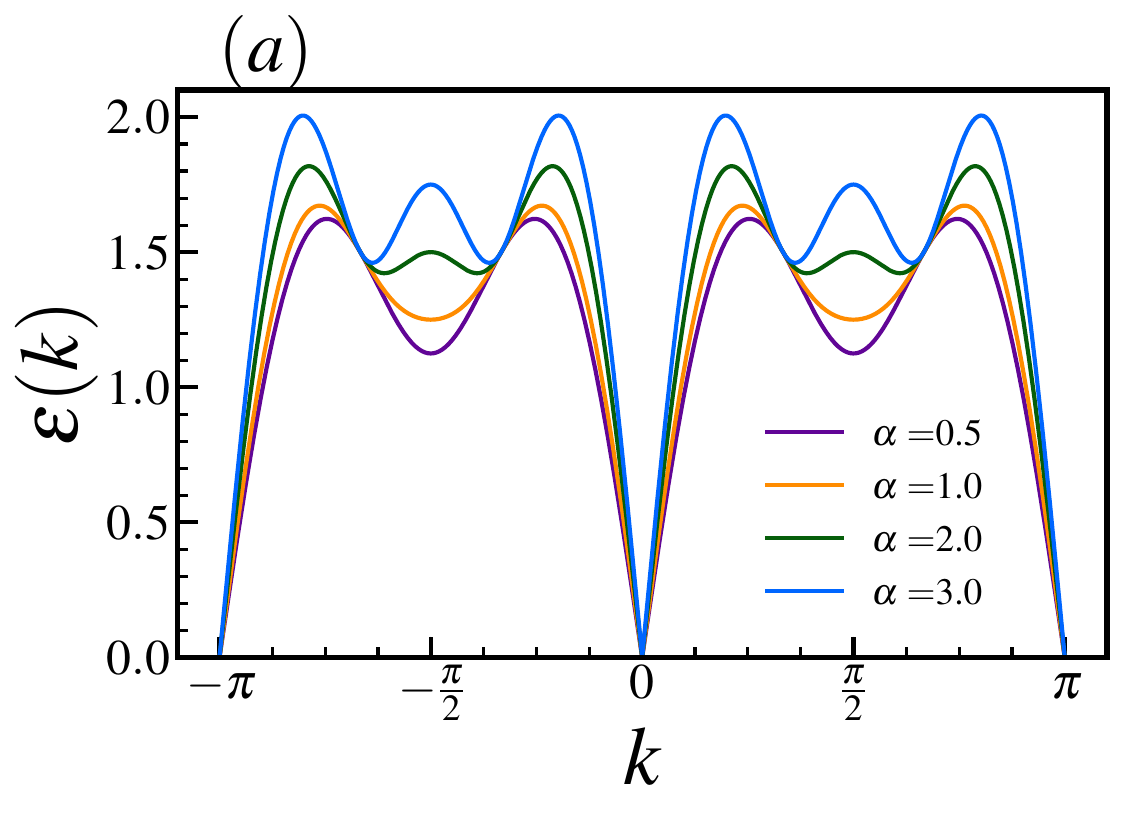} \includegraphics[width=0.25\linewidth,height=0.2\linewidth]{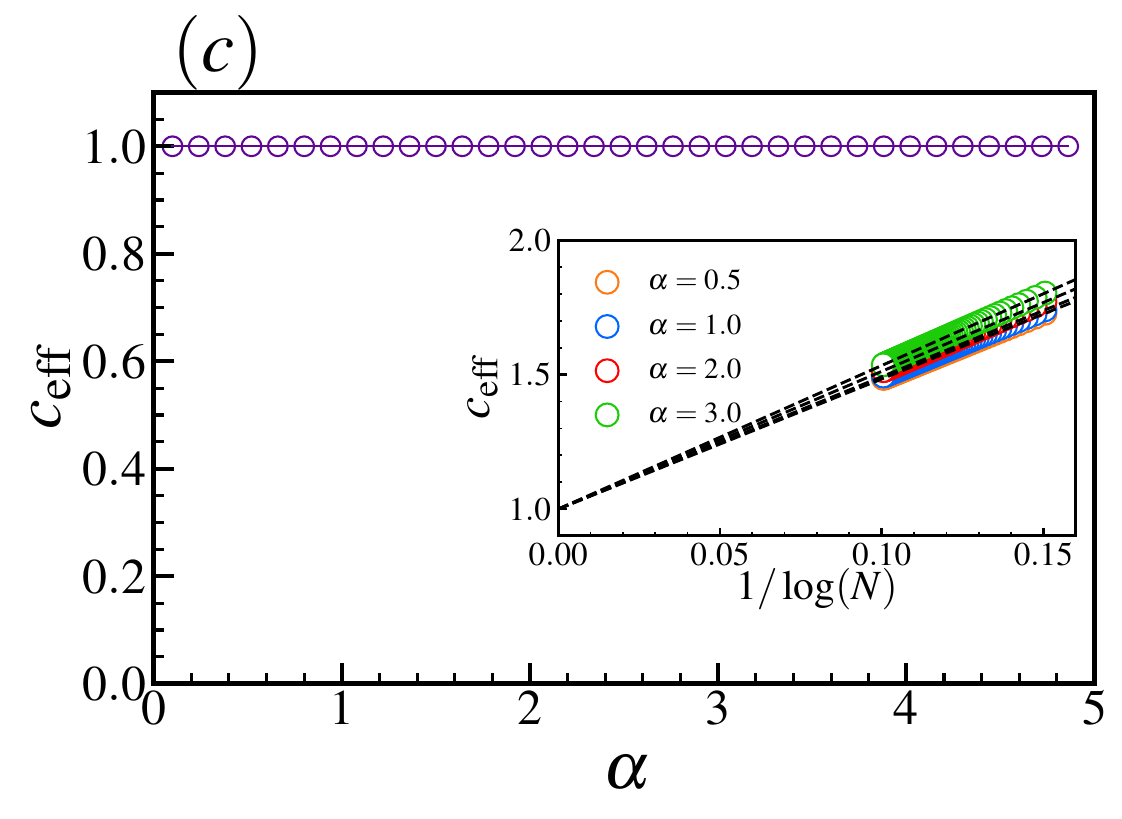} 
		\includegraphics[width=0.25\linewidth,height=0.2\linewidth]{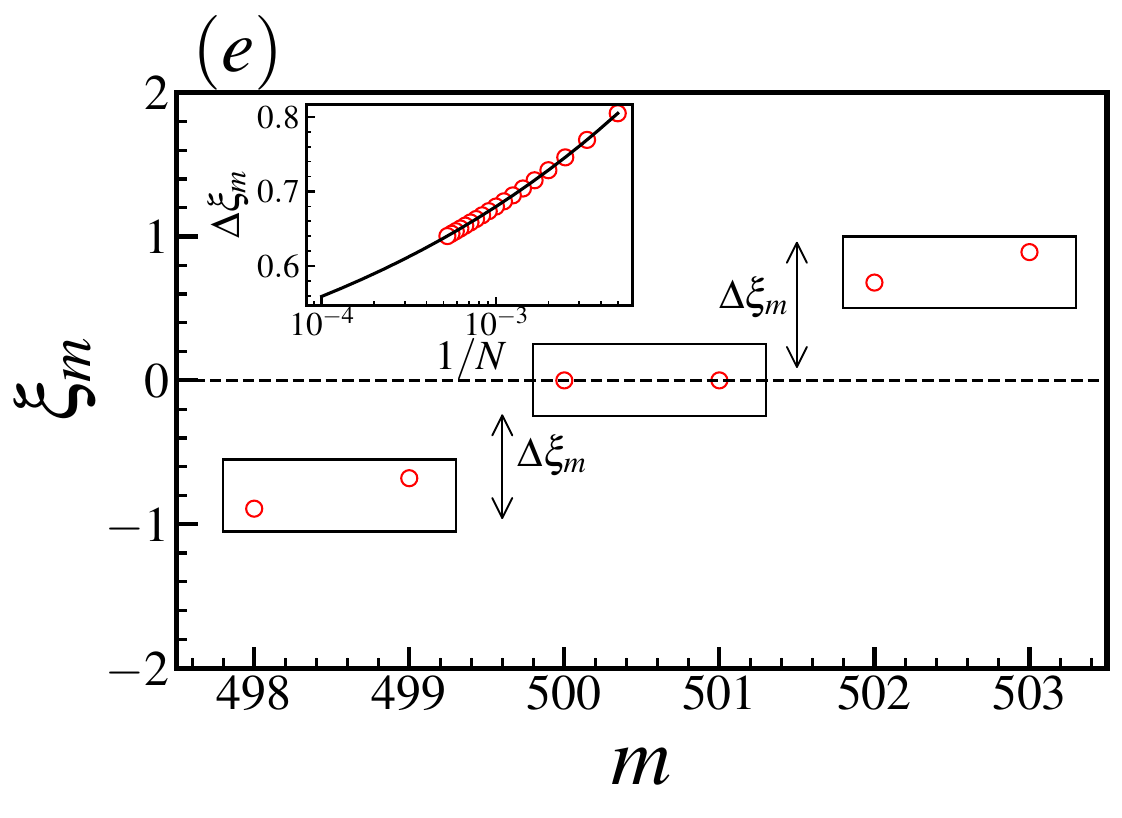}
		\includegraphics[width=0.25\linewidth,height=0.2\linewidth]{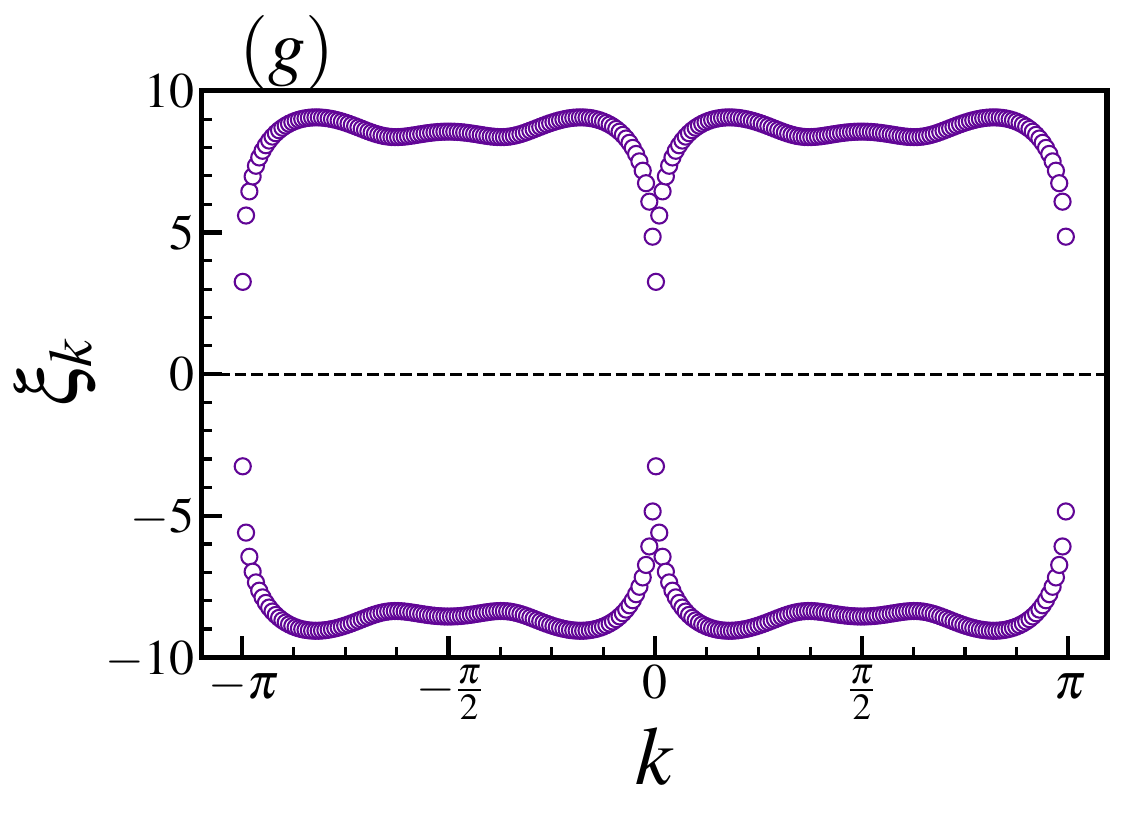}
	}
	
	\centerline{\includegraphics[width=0.25\linewidth,height=0.2\linewidth]{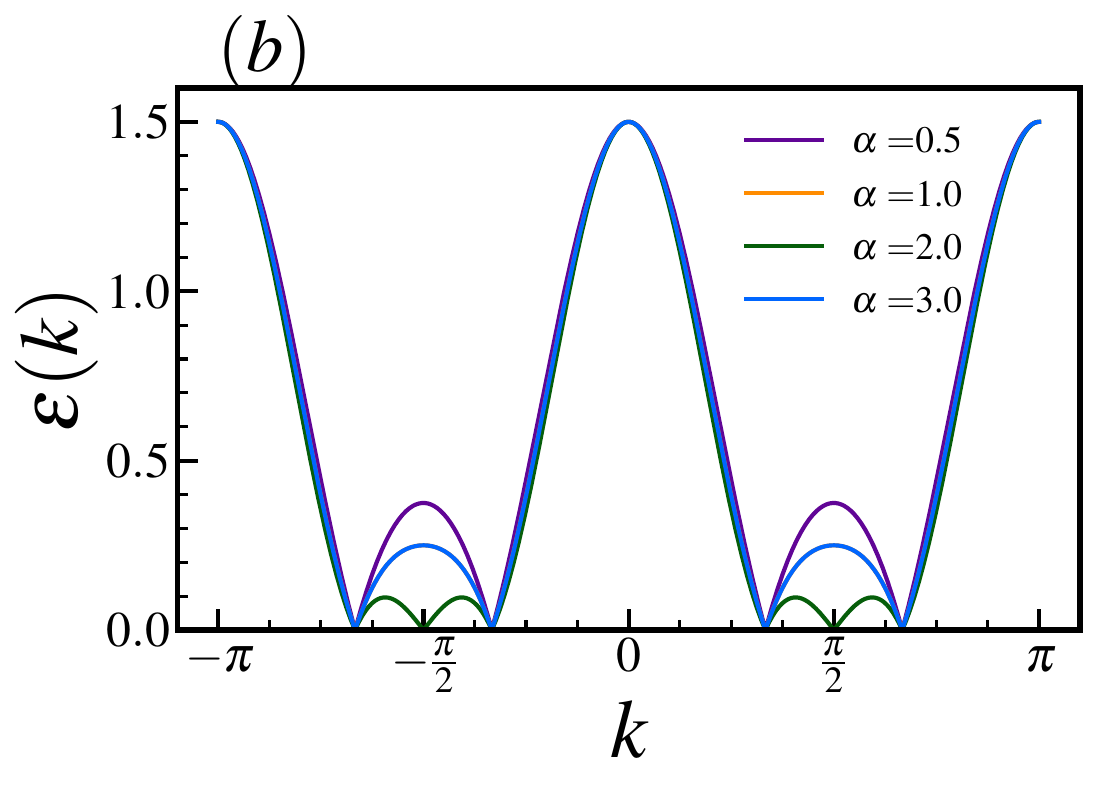} \includegraphics[width=0.25\linewidth,height=0.2\linewidth]{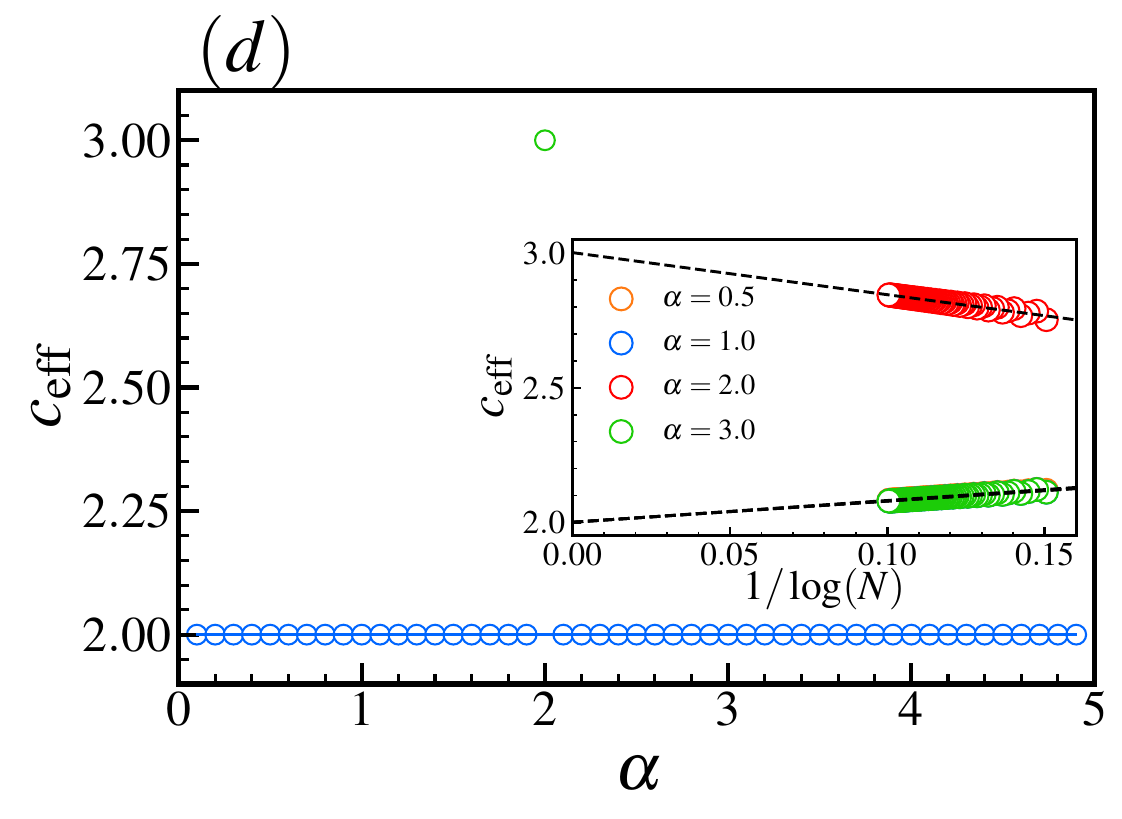} 
		\includegraphics[width=0.25\linewidth,height=0.2\linewidth]{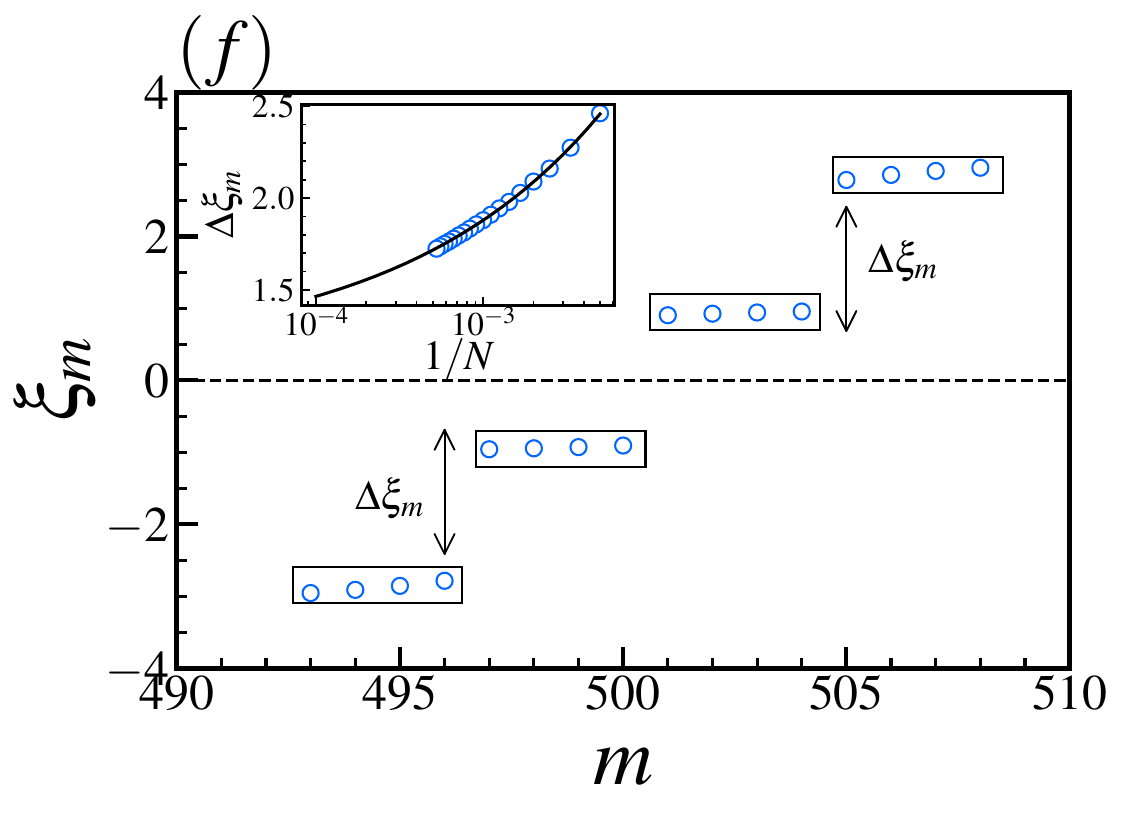}
		\includegraphics[width=0.25\linewidth,height=0.2\linewidth]{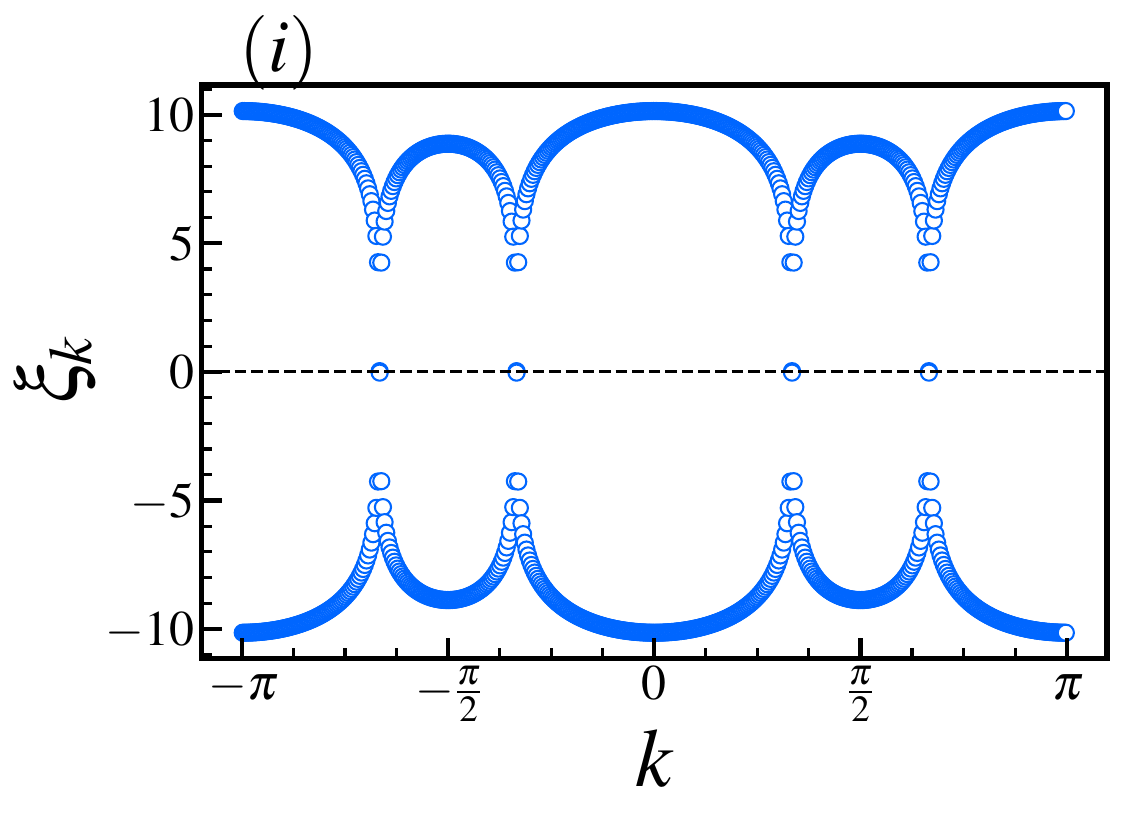}
	}
	
	\caption{(a,b) Quasiparticle spectrum along the critical lines $\beta_{c_2}$ and $\beta_{c_3}$, respectively. (c,d) Effective central charges extracted from entanglement-entropy scaling. (e,f) Real-space (RES) bulk entanglement spectra, and (g,i) momentum-space (MES) bulk entanglement spectra for $\beta_{c_2}$ and $\beta_{c_3}$, respectively.
	}
	\label{Fig4}
	
\end{figure*}

By identifying the microscopic origin of the $c_{\rm eff}=1\rightarrow3$ transition along the exactly solvable line $\beta=1$, we now demonstrate that the interplay between cluster interactions and Fermi-surface topology gives rise to a considerably richer hierarchy of conformal critical phases throughout the full phase diagram. Figure~\ref{Fig4} summarizes the entanglement and low-energy properties along the remaining critical lines, denoted $\beta_{c_2}$ and $\beta_{c_3}$, where the number of gapless Dirac sectors undergoes a sequence of distinct topological reconstructions.  As we demonstrate below, the anisotropic regime reveals that the relation between Fermi-point topology and conformal criticality is more subtle than a simple counting of gapless momenta. Instead, the effective central charge is determined by the organization of the gapless modes into independent low-energy conformal sectors, leading to distinct universality classes despite similar Fermi-point topologies.

Figure~\ref{Fig4}~(a) and Fig.~\ref{Fig4}~(b) present the quasiparticle spectra along the critical lines $\beta_{c_2}$ and $\beta_{c_3}$, respectively. The spectrum exhibits three gapless points located at $k_F=0$ and $k_F=\pm\pi$ on the $\beta_{c_2}$ line, whereas four gapless points emerge at $k_F=\pm\pi/3$ and $k_F=\pm2\pi/3$ along the $\beta_{c_3}$ line. In both cases, a low-energy expansion about each gapless momentum yields $\varepsilon(k)\propto |k-k_F|$, demonstrating that the quasiparticle excitations remain linearly dispersing with dynamical exponent $z=1$. A qualitatively different behavior emerges along the $\beta_{c_3}$ line upon approaching the multicritical point $(\alpha,\beta)=(2,1)$. At this point, the Fermi-surface topology is reconstructed, with the number of gapless Fermi points increasing from four to six before returning to four beyond the multicritical point.  Unlike the Lifshitz transition on the $\beta=1$ line, where the critical point is characterized by a cubic band touching $z=3$, the present reconstruction preserves linear Dirac dispersions throughout $z=1$. The multicritical point therefore represents an unconventional Lifshitz reconstruction driven purely by changes in the topology of the Fermi surface rather than by a modification of the dynamical scaling. The corresponding signatures in the quasiparticle spectrum and in the second derivative of the ground-state energy, together with the analytical continuum expansion for both critical lines, are presented in Appendix~\ref{app:betc2} and \ref{app:betc3}.

Figure~\ref{Fig4}~(c) presents the effective central charge extracted from the finite-size scaling of the entanglement entropy along the critical line $\beta_{c_2}$. As shown in the inset, the numerical extrapolation converges to the universal value $c_{\rm eff}=1$ over the entire critical line, demonstrating that it belongs to a single $c_{\rm eff}=1$ conformal universality class. This result is particularly noteworthy because the corresponding quasiparticle spectrum contains three gapless momenta, located at $k_F=0$ and $k_F=\pm\pi$. At first sight, one might therefore expect multiple independent critical sectors. However, the continuum theory demonstrates that the modes at $k_F=\pm\pi$ are not independent low-energy degrees of freedom. This follows from the lattice identity $ e^{i\pi n}=e^{-i\pi n}=(-1)^n$, which implies that the two Brillouin-zone boundaries are represented by the same slowly varying continuum field. Consequently, they describe a single continuum sector rather than two independent critical modes.  The corresponding low-energy continuum theory is therefore described by a single massless Dirac sector,
\begin{equation}
	H_{\rm eff}
	=
	2i
	\sum_{a=0,\pi}
	\tilde v_a
	\int dx
	\left(
	\psi^\dagger_{R,a}
	\partial_x
	\psi^\dagger_{L,a}
	-
	\psi_{L,a}
	\partial_x
	\psi_{R,a}
	\right),
\end{equation}
where the Brillouin-zone boundary contributes only one independent continuum flavour. Consequently, $c_{\rm eff}=1$, where the Brillouin-zone boundary contributes only a single continuum flavor.  Equivalently, in the continuum description, this Dirac fermion may be viewed as two Majorana fields, each contributing $c_{\rm eff}=1/2$, giving the total conformal central charge $c_{\rm eff}=1$. Consequently, despite the apparent presence of three gapless crossings in the lattice spectrum, the low-energy limit theory contains only one independent conformal degree of freedom, fully consistent with the numerical value $c_{\rm eff}=1$. This example demonstrates that the conformal central charge is determined not by the total number of gapless lattice crossings, but by the number of independent continuum degrees of freedom sectors.  This distinction becomes essential for understanding the more complex critical structures discussed below. The derivation of the corresponding low-energy effective Hamiltonian and the reduction from the lattice description to a single Dirac mode are presented in Appendix~\ref{app:betc2_cft}.

Figure~\ref{Fig4}~(e) shows the real-space bulk entanglement spectrum along the critical line $\beta_{c_2}$. The low-lying entanglement levels organize into doubly degenerate multiplets, reflecting the universal structure of the underlying critical theory rather than a microscopic lattice symmetry. As established by the continuum theory, the low-energy sector is governed by a single massless Dirac fermion, or equivalently two gapless Majorana modes, whose combined conformal contribution yields $c_{\rm eff}=1$. Equivalently, the Dirac field is decomposed into two gapless Majorana fermions, whose combined conformal contribution yields the observed $c_{\rm eff}=1$ criticality. The observed twofold degeneracy therefore constitutes a direct entanglement signature of this universal field-theoretical structure rather than the presence of additional independent gapless channels. 
Complementary information is provided by the momentum-space bulk entanglement spectrum shown in Fig.~\ref{Fig4}~(g). The spectrum develops three pronounced minima located precisely at the gapless momenta $k_F=0$ and $k_F=\pm\pi$, demonstrating that the momentum-space entanglement faithfully resolves the Fermi-point structure of the critical state. Although the lattice spectrum contains three gapless crossings, the continuum theory identifies the two zone-boundary points, $k_F=\pm\pi$, as a single physical sector because they are related by a reciprocal lattice vector. Consequently, the three entanglement minima encode the microscopic distribution of gapless excitations, while only two independent real fermionic degrees of freedom survive in the continuum limit. These combine into a single massless Dirac fermion, so that the universal long-wavelength description remains that of a $c_{\rm eff}=1$ conformal field theory. The combined real- and momentum-space entanglement spectra therefore provide complementary evidence that the critical line $\beta_{c_2}$ belongs to the $c_{\rm eff}=1$ Luttinger-liquid universality class.
%%%%%%%%%%%%%%%%%%%%%%%%%%%%%%%%%%%%%%

Figure~\ref{Fig4}~(d) shows the effective central charge extracted from finite-size scaling along the critical line $\beta_{c_3}$. Away from the multicritical point, the extrapolated values converge to $c_{\rm eff}=2$, demonstrating that the low-energy theory along the entire critical line consists of two independent massless Dirac sectors and is therefore described by a $c_{\rm eff}=2$ conformal field theory. This is fully consistent with the continuum theory derived in Appendix~\ref{app:betc3_cft}, where the four lattice Fermi points, $k_F=\pm\pi/3$ and $\pm2\pi/3$, form two independent valleys.
Unlike the $\beta_{c_2}$ line, all Fermi points lie inside the first Brillouin zone, so no reciprocal-lattice identification occurs. Each valley therefore generates an independent massless Dirac fermion, yielding a total central charge $c_{\rm eff}=2$. The corresponding low-energy continuum Hamiltonian takes the form
\begin{align}
	\mathcal{H}_{\rm eff}
	&=
	\sum_{a=1}^{2}
	\int dx
	\Big[
	i v_{F,a}
	\left(
	\psi^\dagger_{R,a}\partial_x\psi_{R,a}
	-
	\psi^\dagger_{L,a}\partial_x\psi_{L,a}
	\right)
	\nonumber\\
	&
	\qquad\qquad
	+
	\tilde v_a
	\left(
	\psi^\dagger_{R,a}\partial_x\psi^\dagger_{L,a}
	+
	\psi_{L,a}\partial_x\psi_{R,a}
	\right)
	\Big],
\end{align}
where both flavor sectors remain independent because no Brillouin-zone identification relates the corresponding Fermi points. Consequently,
\begin{equation}
	c_{\rm eff}	= \sum_{a=1}^{2} c_a = 2.
\end{equation}
A striking exception occurs at $\alpha=2$, where the effective central charge exhibits a sharp enhancement to $c_{\rm eff}=3$. This isolated point coincides precisely with the unconventional Lifshitz multicritical point, where the topology of the quasiparticle spectrum changes through the reconstruction of the Fermi surface. At this singular point, the Lifshitz reconstruction nucleates an additional independent Dirac sector, increasing the number of continuum conformal sectors from two to three. The enhancement of the central charge therefore originates from the appearance of an additional gapless continuum sector induced by the Lifshitz reconstruction, rather than from a change of the universality class along the critical line itself.  Away from the multicritical point, this additional channel becomes gapped and the continuum theory immediately reverts to two independent massless Dirac fermions with $c_{\rm eff}=2$. The finite-size scaling therefore provides direct numerical evidence that the Lifshitz multicritical point hosts an emergent conformal degree of freedom generated by the Fermi-surface reconstruction, establishing a direct correspondence between microscopic Fermi-point topology and the effective continuum theory.
%%%%%%%%%%%%%%%%

Figure~\ref{Fig4}~(f) presents the real-space bulk entanglement spectrum along the critical line $\beta_{c_3}$. The low-lying entanglement levels organize into characteristic fourfold multiplets, in clear contrast to the twofold structure observed for the $c_{\rm eff}=1$ phase. As shown in the inset, the residual splitting within each multiplet decreases with increasing system size, indicating that the exact fourfold degeneracy is recovered in the thermodynamic limit. This enhanced multiplicity directly reflects the underlying continuum theory. As demonstrated in Appendix~\ref{app:betc3_cft}, the low-energy spectrum consists of two independent massless Dirac fermions associated with the two symmetry-related valleys centered at $k_F=\pm\pi/3$ and $k_F=\pm2\pi/3$. Each Dirac sector contributes an identical pair of low-energy entanglement excitations, so that their coexistence naturally gives rise to the observed fourfold multiplet structure. The real-space entanglement spectrum therefore provides a direct fingerprint of the two independent conformal sectors and independently confirms the $c_{\rm eff}=2$ criticality obtained from the finite-size scaling analysis.

The momentum-space bulk entanglement spectrum is shown in Fig.~\ref{Fig4}~(i). Four pronounced entanglement minima appear precisely at the gapless Fermi momenta $k_F=\pm\pi/3$ and $k_F=\pm2\pi/3$, demonstrating that the momentum-space entanglement faithfully resolves the complete Fermi-point topology of the critical phase. These minima naturally group into two symmetry-related valleys, each corresponding to an independent low-energy Dirac sector in the continuum limit. Consequently, although the lattice spectrum contains four gapless Fermi points, they organize into only two independent Dirac sectors, each contributing one unit to the conformal central charge. The momentum-space entanglement spectrum therefore provides a direct microscopic fingerprint of the underlying Fermi-surface topology and independently confirms the $c_{\rm eff}=2$ criticality established from both the finite-size scaling analysis and the continuum field-theoretical description.

\subsection{Field-Induced Critical Regime}

\begin{figure}[h]
	\centering
	\includegraphics[width=0.8\linewidth,height=0.6\linewidth]{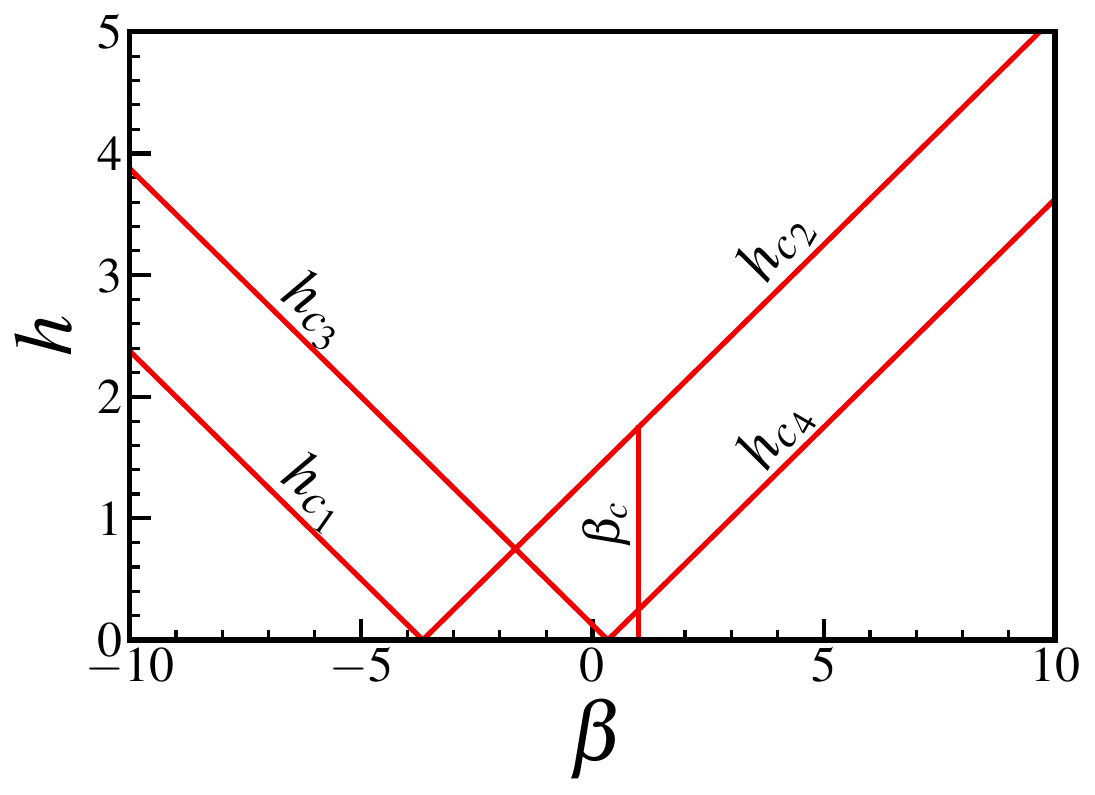}
	\caption{ Ground-state phase diagram of the generalized spin-$1/2$ XX chain with four-spin cluster interactions in the presence of a transverse magnetic field ($h \neq 0$).}
	\label{fig5}
\end{figure}

We now investigate the effects of a transverse magnetic field, which dramatically enlarges the landscape of conformal criticality. The resulting schematic phase diagram is shown in Fig.~\ref{fig5}, where the interplay between the magnetic field, the exchange anisotropy, and the four-spin cluster interaction gives rise to several distinct critical boundaries separating gapped and gapless quantum phases. Unlike the zero-field phase diagram, where the critical behavior is controlled solely by the competition between anisotropy and cluster interactions, the magnetic field continuously reshapes the Fermi-point topology, producing distinct gap-closing structures and a richer hierarchy of effective conformal field theories. As we demonstrate below, these topological reconstructions give rise to several universality classes with different conformal central charges.

We begin with the isotropic critical line, $\beta=1$, whose critical points are determined analytically by Eqs.~(\ref{betc1}) and (\ref{betc2}). Along this line, the transverse magnetic field induces a sequence of Lifshitz transitions that successively reconstruct the Fermi-point topology and alter the low-energy excitation structure. We first analyze the evolution of the quasiparticle spectrum before examining the corresponding conformal and entanglement properties.

\begin{figure*}
	\centerline{\includegraphics[width=0.3\linewidth,height=0.24\linewidth]{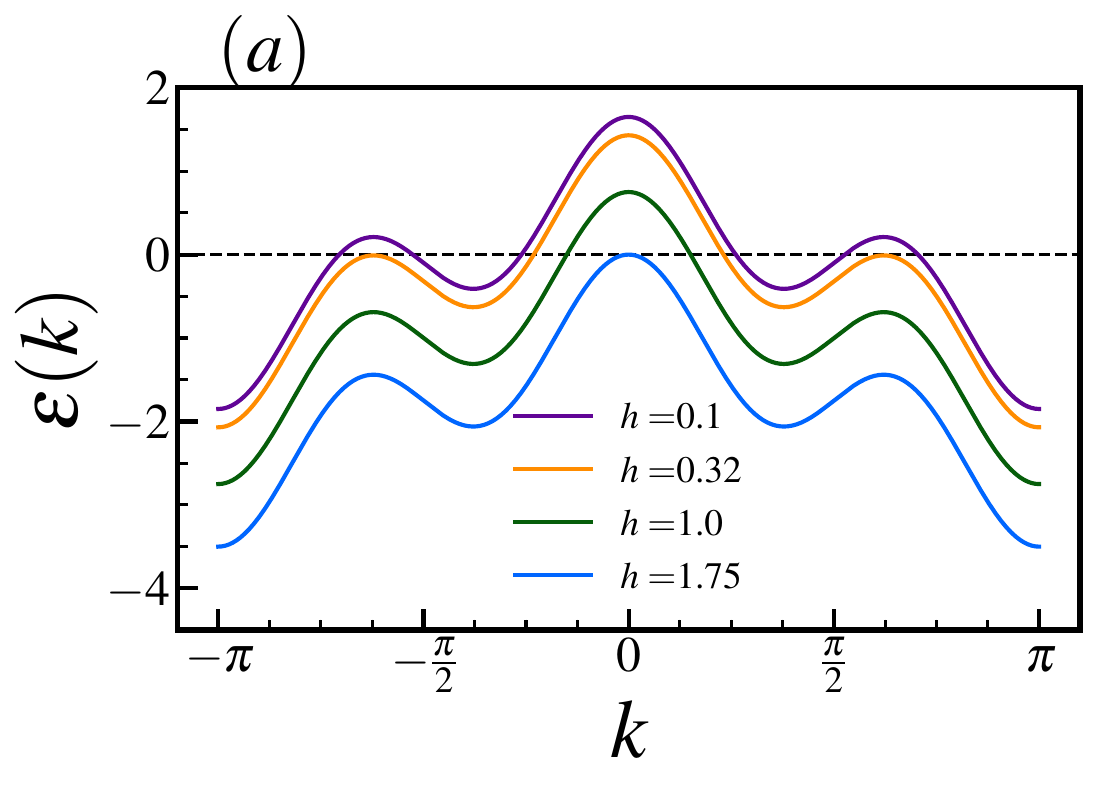} \includegraphics[width=0.3\linewidth,height=0.24\linewidth]{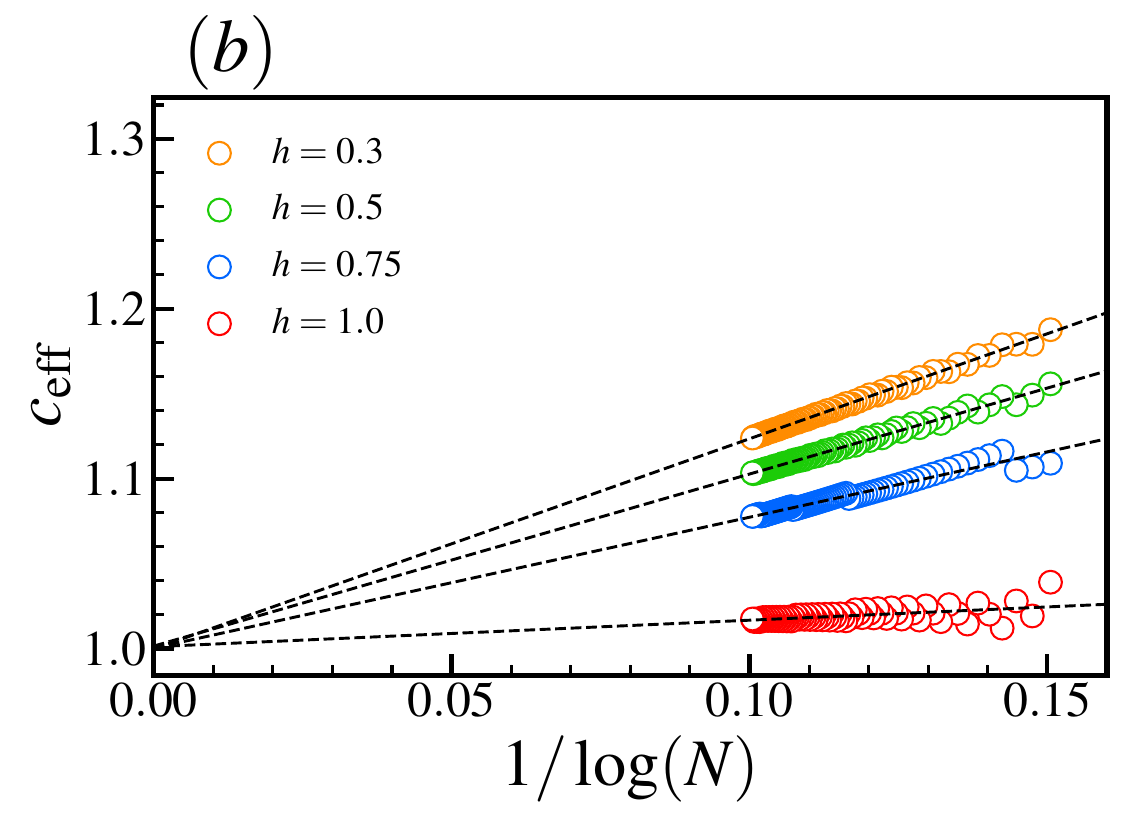} 
		\includegraphics[width=0.3\linewidth,height=0.24\linewidth]{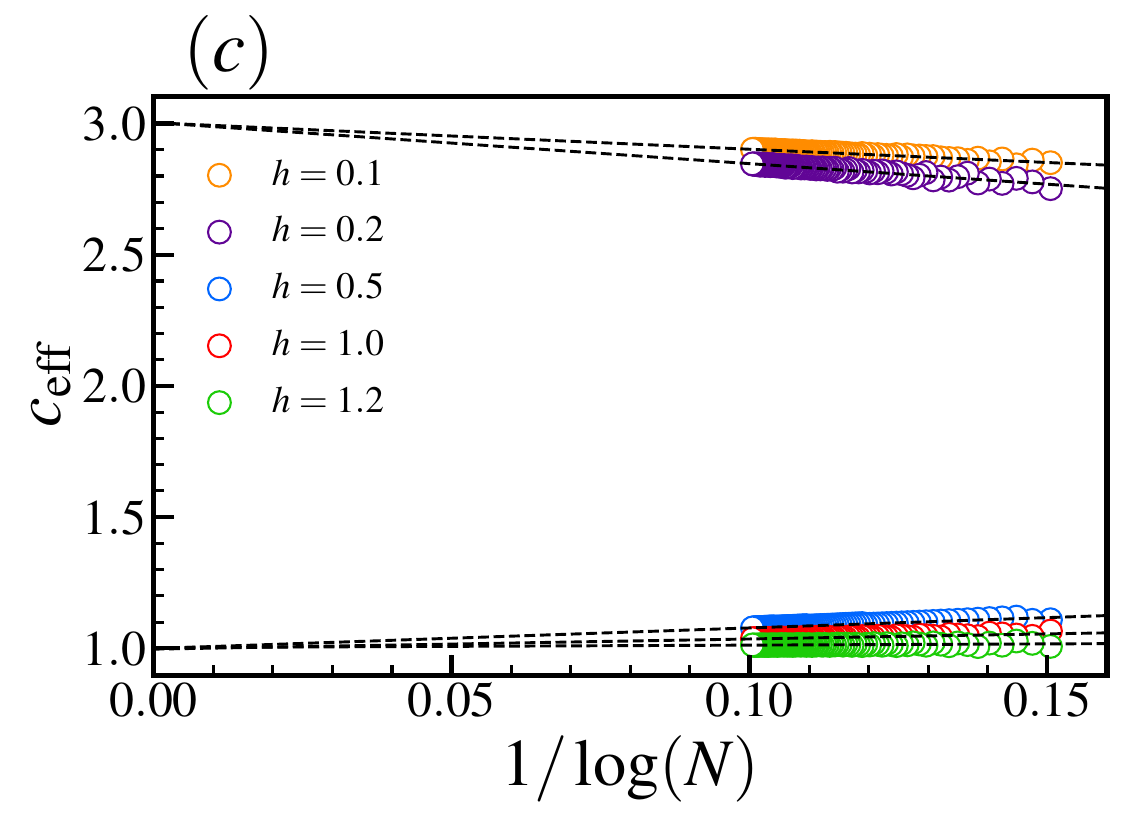}
	}
	\caption{(a) Quasiparticle spectrum across the Lifshitz transition on the isotropic line ($\beta=1$). (b) Effective central charge extracted from entanglement-entropy scaling in the phase ($\alpha<4/3$). (b) Effective central charge extracted from entanglement-entropy scaling in the phase ($\alpha>4/3$).}
	\label{fig6}
	
\end{figure*}

Figure~\ref{fig6}~(a) illustrates the evolution of the quasiparticle spectrum along the isotropic critical line, $\beta=1$, as the transverse magnetic field is increased. For weak fields, the system remains in the Lifshitz phase generated by the four-spin cluster interaction and exhibits six gapless Fermi points. Increasing the magnetic field continuously deforms the dispersion until the field-driven Lifshitz transition at $h=h_c$, where the Fermi surface is reconstructed. At this multicritical point, four gapless modes remain: two retain linear dispersion with dynamical exponent $z=1$, while the other two exhibit quadratic band touching characterized by $z=2$, as derived analytically in Appendix~\ref{app:betc1h}. Beyond this transition, the quadratic nodes disappear and only a single pair of linearly dispersing Fermi points survives. Finally, at $h=h_{c}^{(\mathrm{sp})}$, this remaining pair annihilates, driving the system into the fully gapped phase.

The corresponding evolution of the effective central charge is shown in Figs.~\ref{fig6}~(b) and \ref{fig6}~(c). For $\alpha=1$, i.e., below the zero-field Lifshitz threshold, the magnetic field continuously shifts the Fermi points without changing their topology. The number of independent gapless sectors therefore remains unchanged, and the entire critical line is described by a single massless Dirac fermion with $c_{\rm eff}=1$.

A qualitatively different behavior emerges for $\alpha=3$, where the zero-field system initially hosts six Fermi points and three independent Dirac sectors, yielding $c_{\rm eff}=3$. The magnetic field preserves this topology up to the Lifshitz transition, where two pairs of Fermi points annihilate simultaneously. As a consequence, the continuum theory is reduced from three independent Dirac sectors to one, leading to the abrupt transition $c_{\rm eff}=3 \rightarrow 1$. The finite-size extrapolation shown in Fig.~\ref{fig6}~(c) confirms these quantized values with excellent accuracy.

These field-driven transitions further reinforce the central conclusion of this work: the effective conformal central charge is controlled by the number of independent low-energy continuum sectors generated by the Fermi-point topology. External parameters such as the magnetic field modify $c_{\rm eff}$ only through changes in this topology, providing a direct microscopic mechanism for engineering different conformal critical theories. The derivation of the corresponding low-energy effective Hamiltonians, together with the analysis of the real- and momentum-space bulk entanglement spectra, is deferred to Appendix~\ref{app:bet1_field_cft}.

\begin{figure}[h]
	\centerline{\includegraphics[width=0.5\linewidth,height=0.42\linewidth]{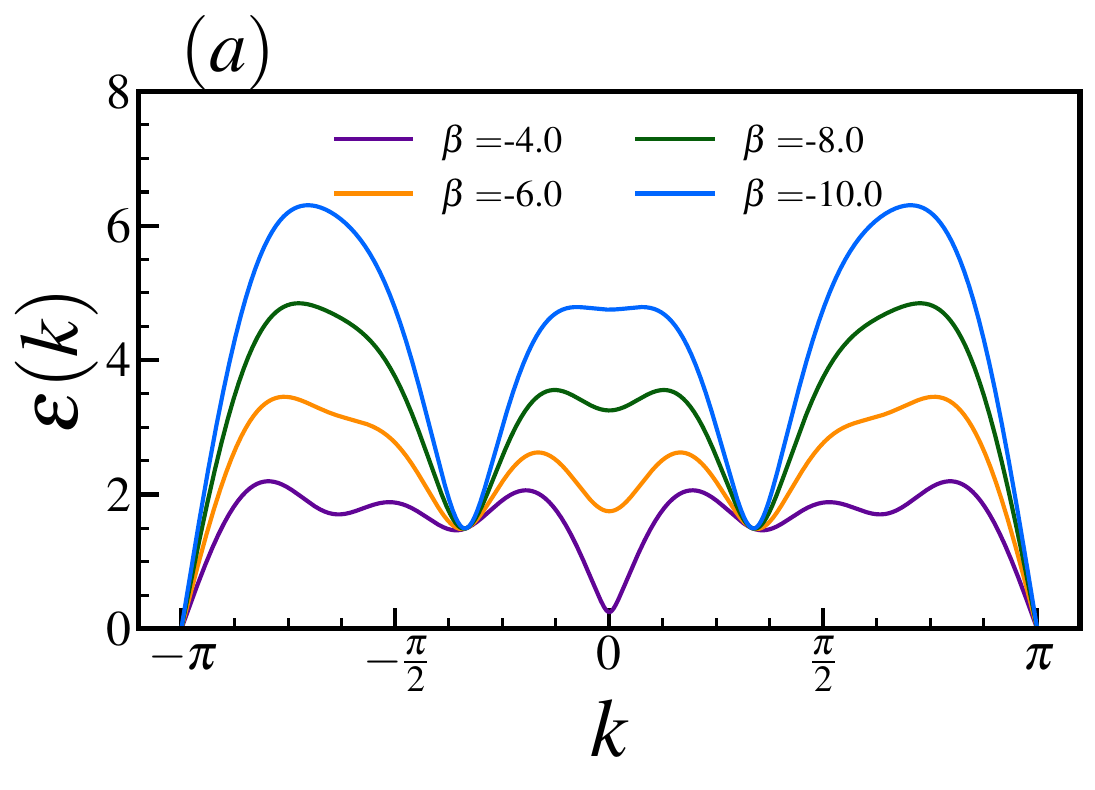} \includegraphics[width=0.5\linewidth,height=0.42\linewidth]{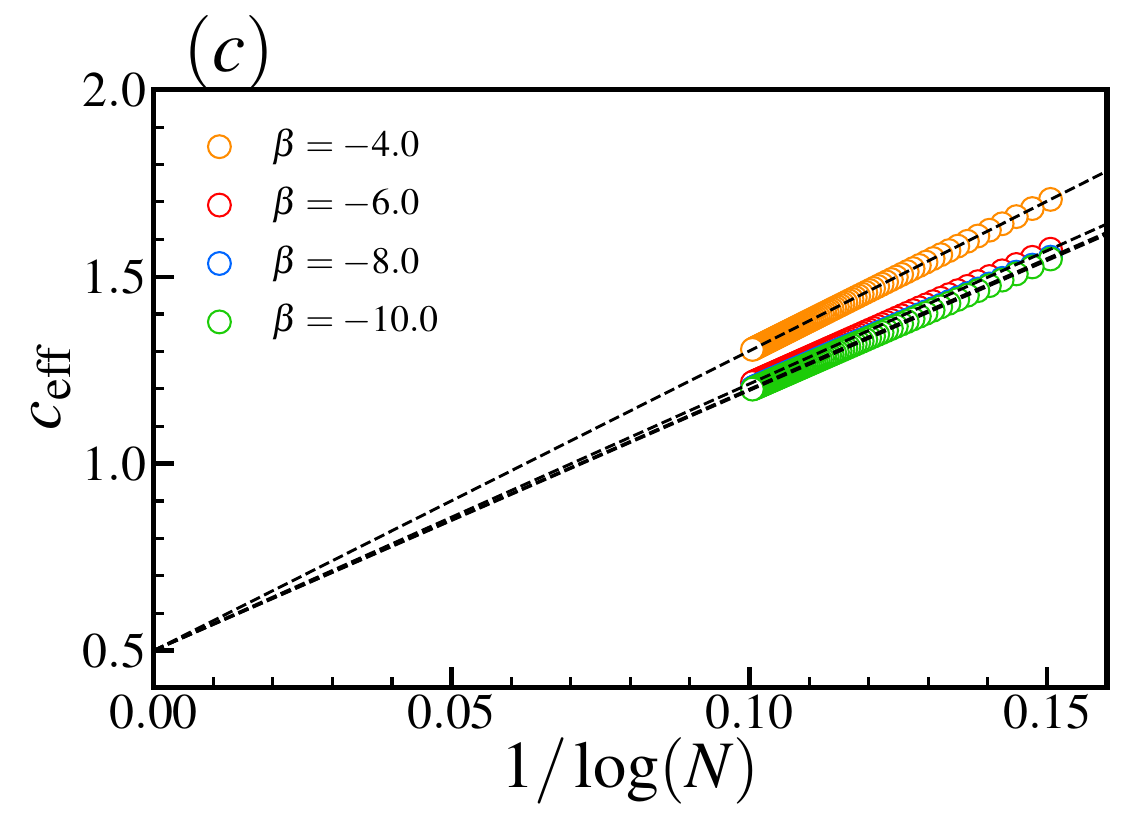} 
		
	}
	\centerline{\includegraphics[width=0.5\linewidth,height=0.42\linewidth]{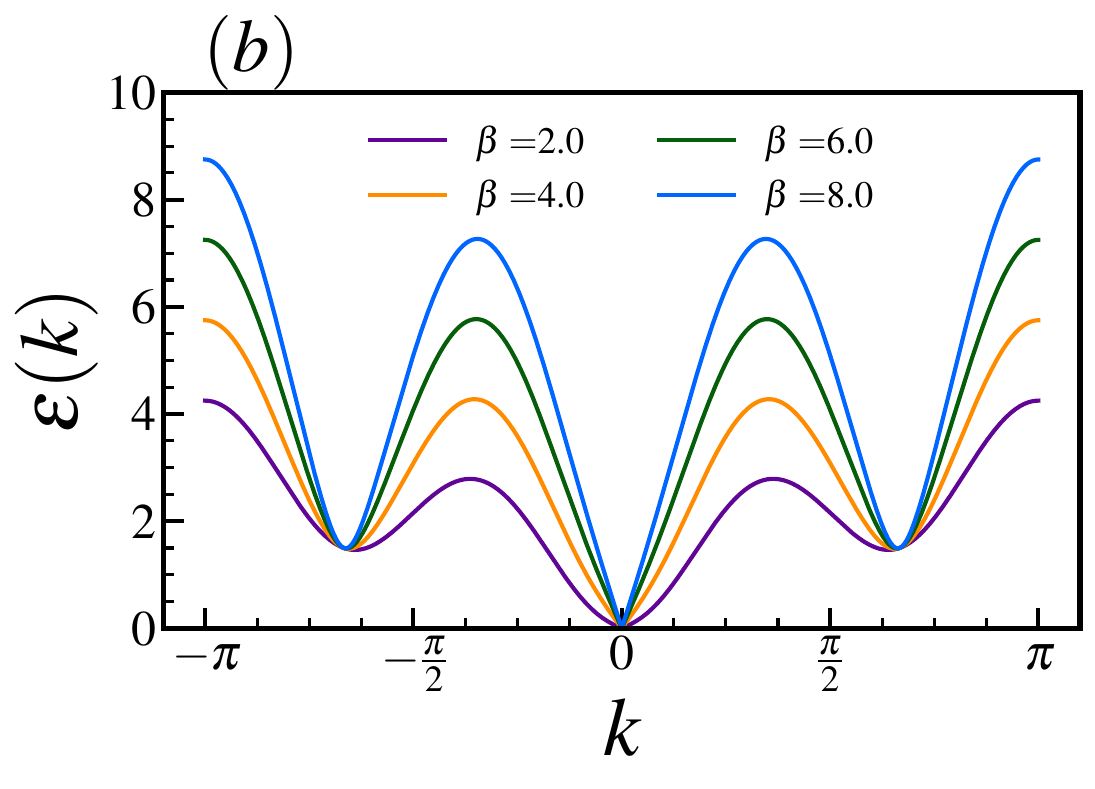} \includegraphics[width=0.5\linewidth,height=0.42\linewidth]{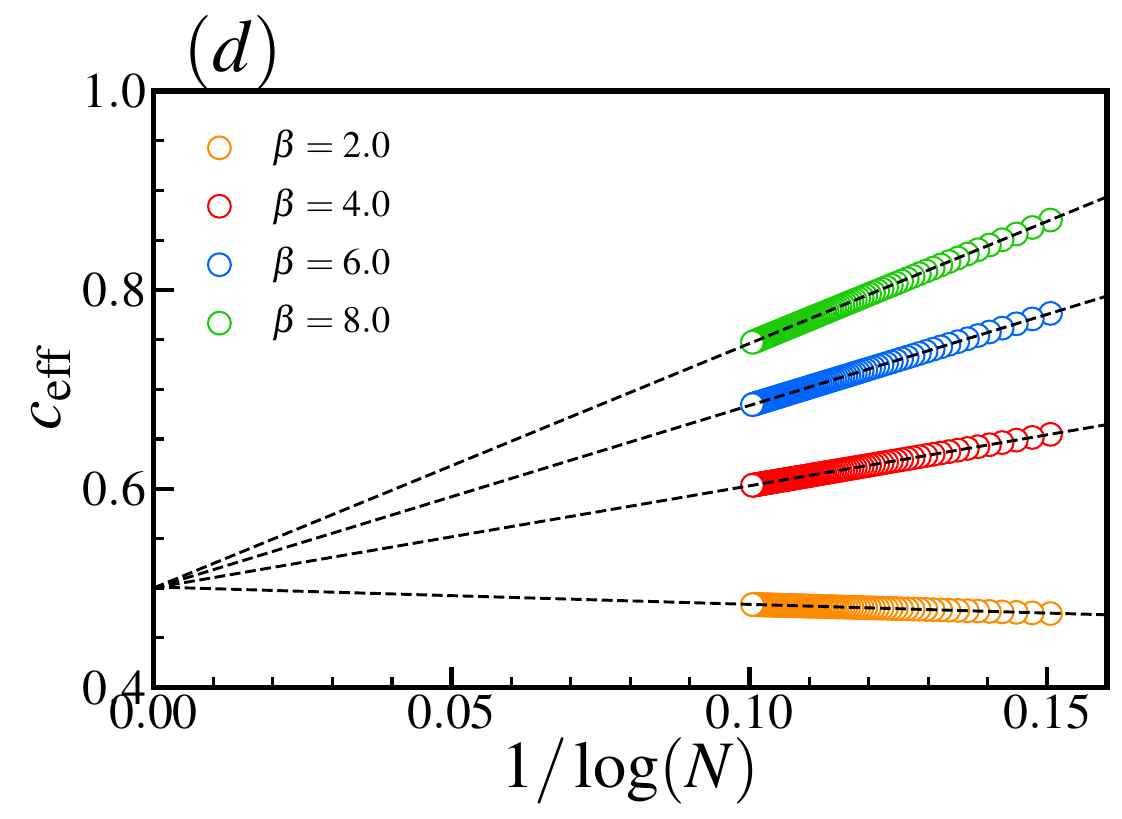} 
		
	}

	\caption{ (a,b) Quasiparticle spectrum and (c,d) Effective central charges extracted from entanglement-entropy scaling, along the critical lines $h_{c_1}$ and $h{c_2}$, respectively.}
	\label{fig7}
	
\end{figure}

We next investigate the remaining field-induced critical branches, $h_{c_1}$ and $h_{c_2}$, whose critical behavior differs qualitatively from the Lifshitz transitions discussed above. Figure~\ref{fig7}~(a) shows the quasiparticle spectrum along the $h_{c_1}$ critical branch. Throughout the entire branch, the quasiparticle gap closes exclusively at the Brillouin-zone boundary, $k_F=\pm\pi$, while the remainder of the spectrum remains gapped. Varying the exchange anisotropy continuously renormalizes the quasiparticle velocity without modifying either the location or the topology of the Fermi points. Consequently, the low-energy excitation remains a single linearly dispersing Dirac mode obeying $\varepsilon(k)\propto|k-k_F|$, corresponding to the dynamical exponent $z=1$, as derived analytically in Appendix~\ref{app:hc12}. An analogous behavior is observed along the $h_{c_2}$ critical branch, whose quasiparticle spectrum is presented in Fig.~\ref{fig7}~(b). Here the gap closes exclusively at the Brillouin-zone center, $k_F=0$, while all other momenta remain gapped. Although the anisotropy continuously deforms the dispersion, the Fermi-point topology is preserved throughout the critical line. The low-energy theory is therefore again governed by a single linearly dispersing Dirac mode with dynamical exponent $z=1$, consistent with the continuum expansion presented in Appendix~\ref{app:hc12}.

The corresponding effective central charges are shown in Figs.~\ref{fig7}~(c) and \ref{fig7}~(d). In both cases, the finite-size extrapolation converges to the universal value $c_{\rm eff}=1/2$, demonstrating that the two field-induced transitions belong to the Ising universality class. The origin of this result becomes transparent within the continuum description. Although the gap closes at the Brillouin-zone boundary $k_F=\pm\pi$ along $h_{c_1}$ and at the zone center $k_F=0$ along $h_{c_2}$, the low-energy theory in each case contains only a single independent gapless Majorana mode. Consequently, both critical lines are governed by the same $c_{\rm eff}=1/2$ conformal field theory despite their distinct microscopic band structures. This demonstrates that the momentum-space location of the gap closing is not relevant to the conformal universality class; rather, the effective central charge is determined solely by the number of independent gapless continuum sectors. The explicit continuum derivation and the reduction of the lattice Hamiltonian to the corresponding single-Majorana field theory are presented in Appendix~\ref{app:hc12_cft}.

The bulk entanglement spectra in both real and momentum space provide an independent confirmation of this interpretation. Along both critical branches, the entanglement structure is fully consistent with a single gapless Majorana degree of freedom and therefore with the Ising universality class characterized by $c_{\rm eff}=1/2$. The detailed analysis of the real-space conformal towers and the momentum-space bulk entanglement spectra is presented in Appendix~\ref{app:hc12_cft}, where these features are shown to faithfully track the underlying low-energy critical modes.

\begin{figure}[h]
	\centerline{\includegraphics[width=0.5\linewidth,height=0.42\linewidth]{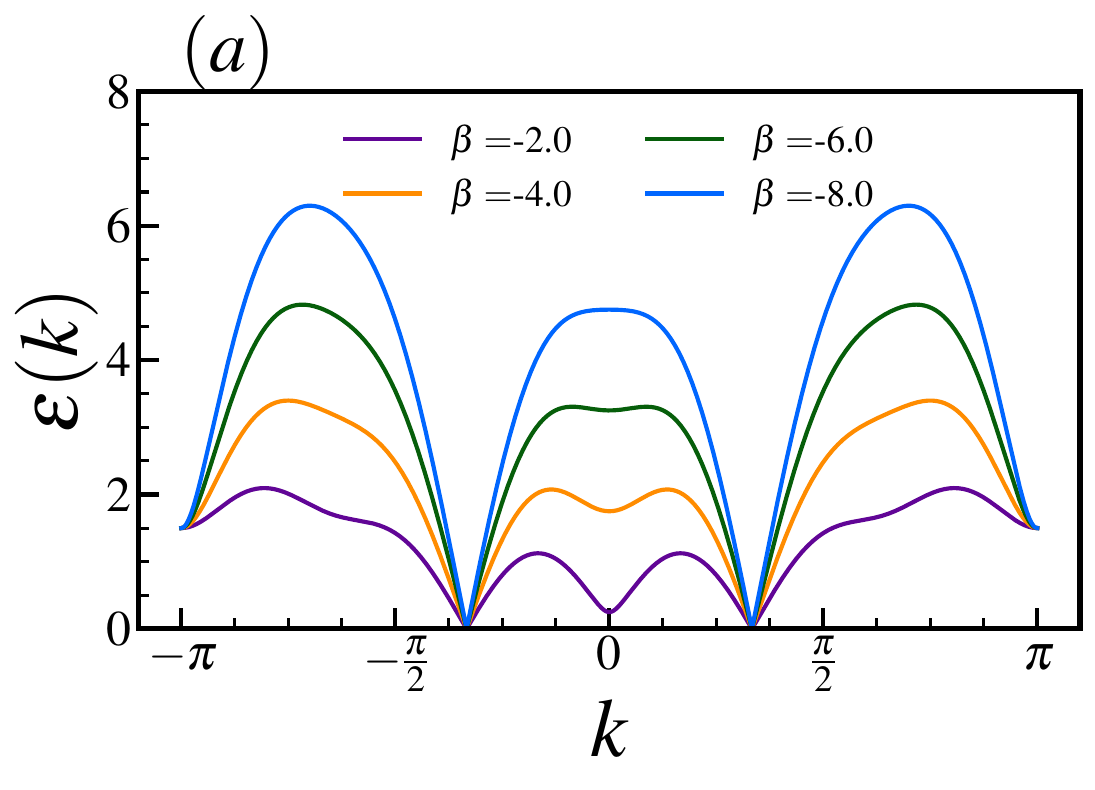} \includegraphics[width=0.5\linewidth,height=0.42\linewidth]{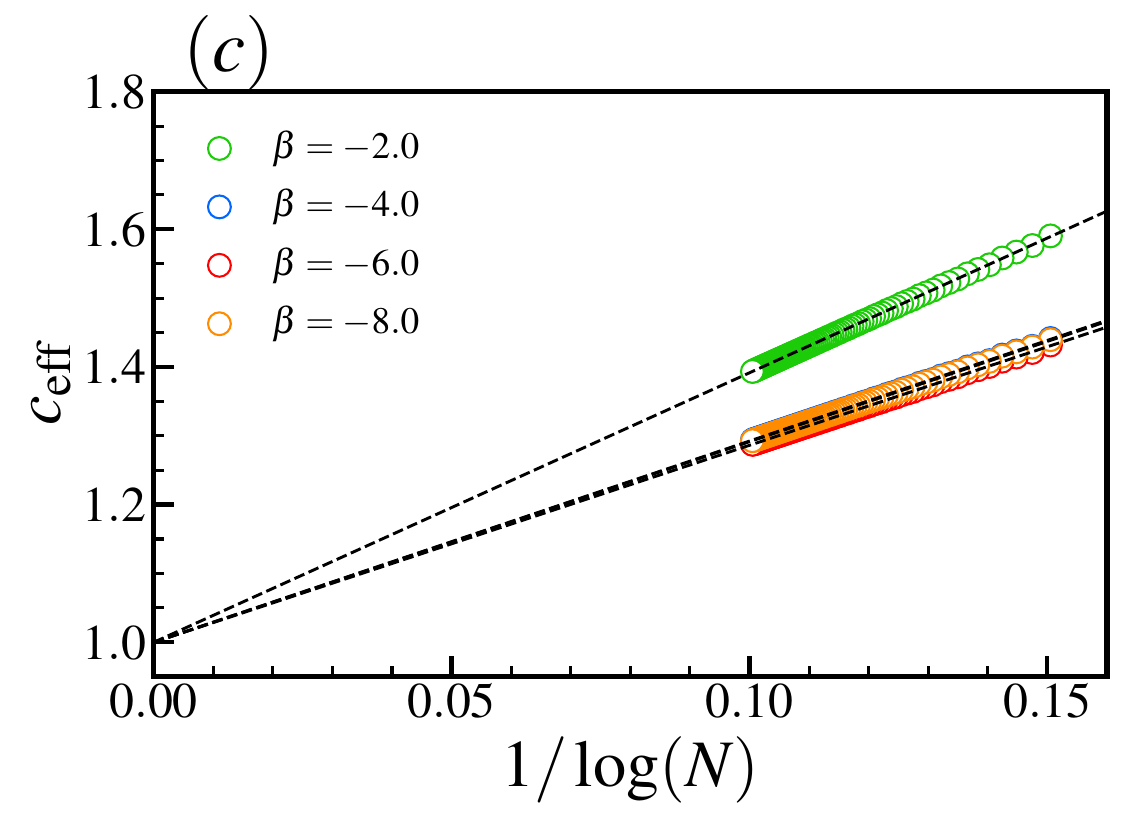} 
		
	}
	\centerline{\includegraphics[width=0.5\linewidth,height=0.42\linewidth]{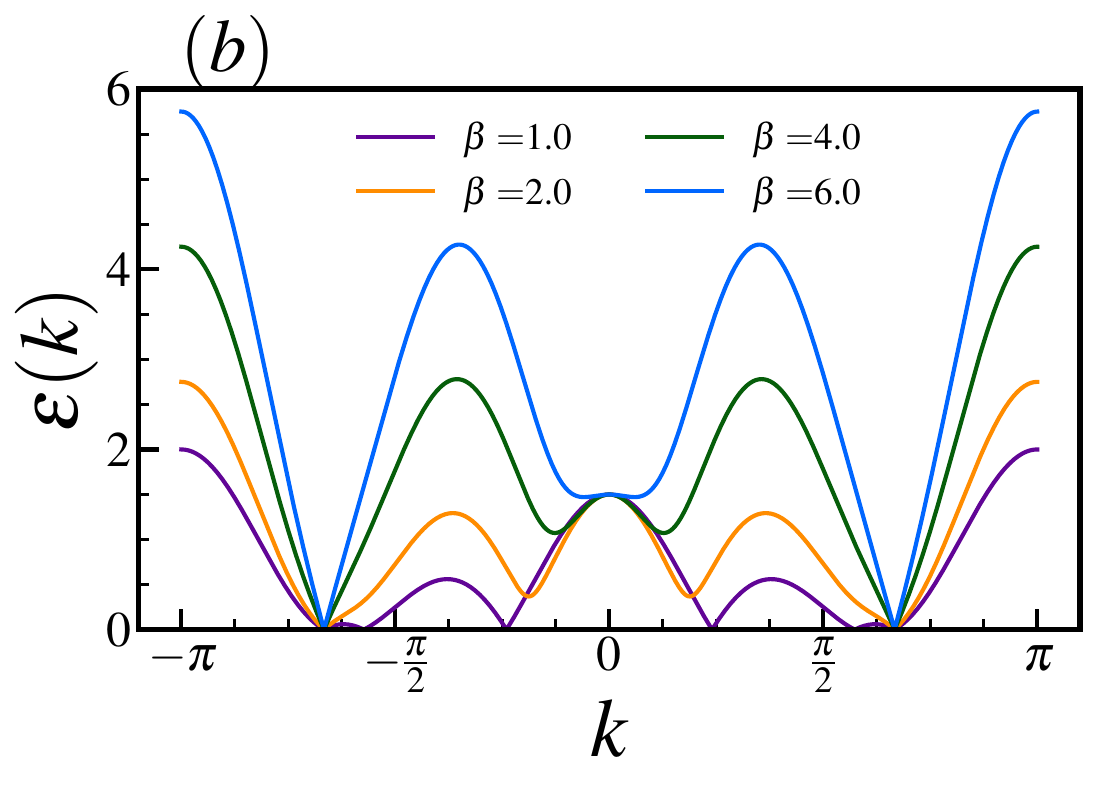} \includegraphics[width=0.5\linewidth,height=0.42\linewidth]{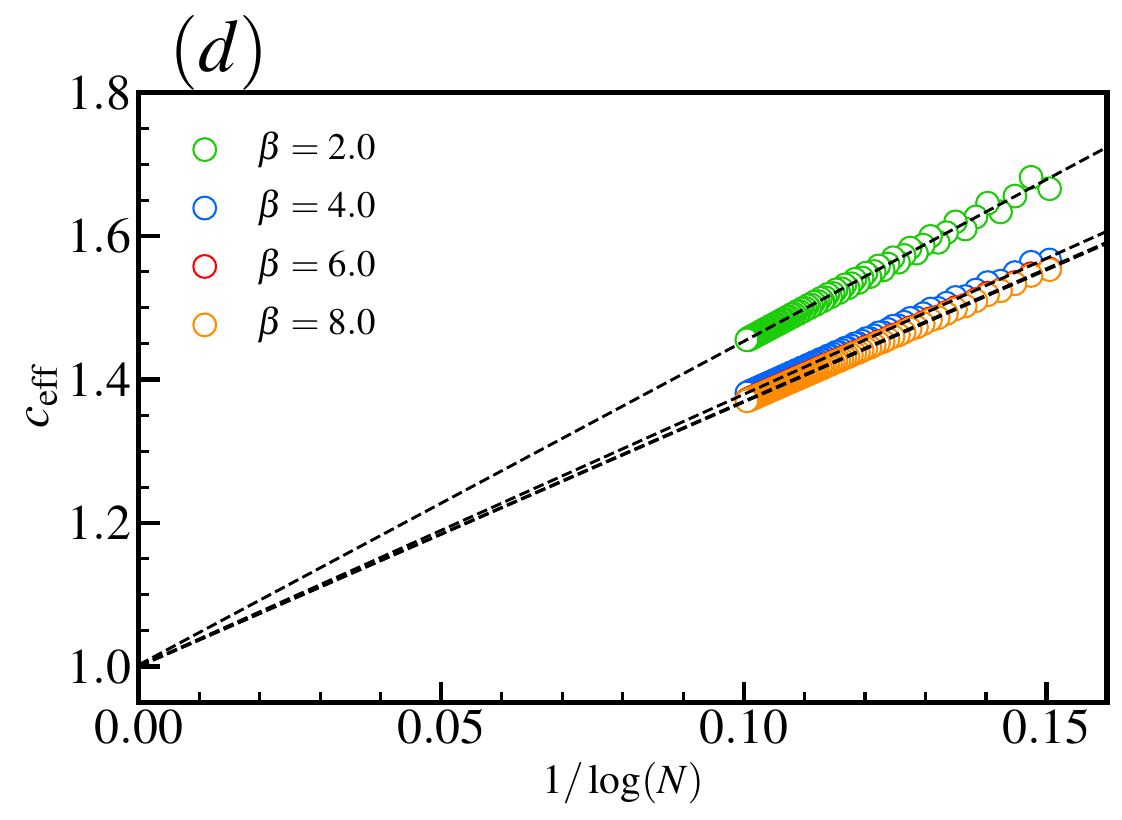} 
		
	}
	\caption{ (a,b) Quasiparticle spectrum and (c,d) Effective central charges extracted from entanglement-entropy scaling, along the critical lines $h_{c_3}$ and $h{c_4}$, respectively.}
	\label{fig8}
	
\end{figure}

We finally consider the remaining field-induced critical branches, $h_{c_3}$ and $h_{c_4}$, whose gapless excitations occur at finite momenta. Along the $h_{c_3}$ branch, shown in Fig.~\ref{fig8}~(a), the quasiparticle gap closes only at the symmetry-related Fermi points $k_F=\pm\pi/3$. Throughout the entire critical line, both the location and the number of gapless points remain unchanged, and the low-energy dispersion is linear,
$\varepsilon(k)\propto |k-k_F|$, corresponding to the conventional dynamical exponent $z=1$. The associated continuum theory is therefore governed by a single massless Dirac fermion, as derived in Appendix~\ref{app:hc34}.

A more intriguing situation occurs along the $h_{c_4}$ branch. As shown in Fig.~\ref{fig8}~(b), the spectrum is gapless at $k_F=\pm2\pi/3$ and again exhibits linearly dispersing excitations with $z=1$. However, unlike $h_{c_3}$, this critical branch crosses the isotropic line $\beta=1$, where the Fermi-surface topology undergoes a singular reconstruction. At this point the number of Fermi points changes through the sequence
$2\rightarrow6\rightarrow2$, while the quasiparticle dispersion remains linear throughout. The corresponding nonanalyticity is also reflected in the second derivative of the ground-state energy, identifying the crossing as an unconventional Lifshitz multicritical point. In contrast to the Lifshitz transition discussed previously along $\beta=1$, the dynamical exponent does not change; instead, the transition is driven solely by a reconstruction of the Fermi-point topology. Remarkably, this topological reconstruction leaves the number of independent continuum sectors unchanged, explaining why the effective central charge remains fixed despite the transient proliferation of Fermi points. The detailed low-energy expansion and ground-state-energy analysis are presented in Appendix~\ref{app:hc34_cft}.

The effective central charge extracted from finite-size scaling is shown in Figs.~\ref{fig8}~(c) and \ref{fig8}(d). For both $h_{c_3}$ and $h_{c_4}$, the numerical data converge to $c_{\rm eff}=1$, demonstrating that both critical branches belong to the same $c_{\rm eff}=1$ conformal universality class despite their different Fermi-point configurations. The continuum analysis reveals that the two symmetry-related Fermi points combine into a single massless Dirac sector, yielding one independent conformal degree of freedom. The corresponding field-theoretical derivation is given in Appendix~\ref{app:hc34_cft}. The real-space and momentum-space bulk entanglement spectra provide independent confirmation of this continuum description. Their detailed analysis, including the evolution of the conformal towers and momentum-space entanglement minima along both critical branches, is presented in Appendix~\ref{app:hc34_cft}.
%%%%%%%%%%%%%%%%%%%%%%%%%%%%%%%%%%%%%%%%%%%%% last%%%%%%%%%

\begin{figure}[h]
	\centerline{\includegraphics[width=0.5\linewidth,height=0.42\linewidth]{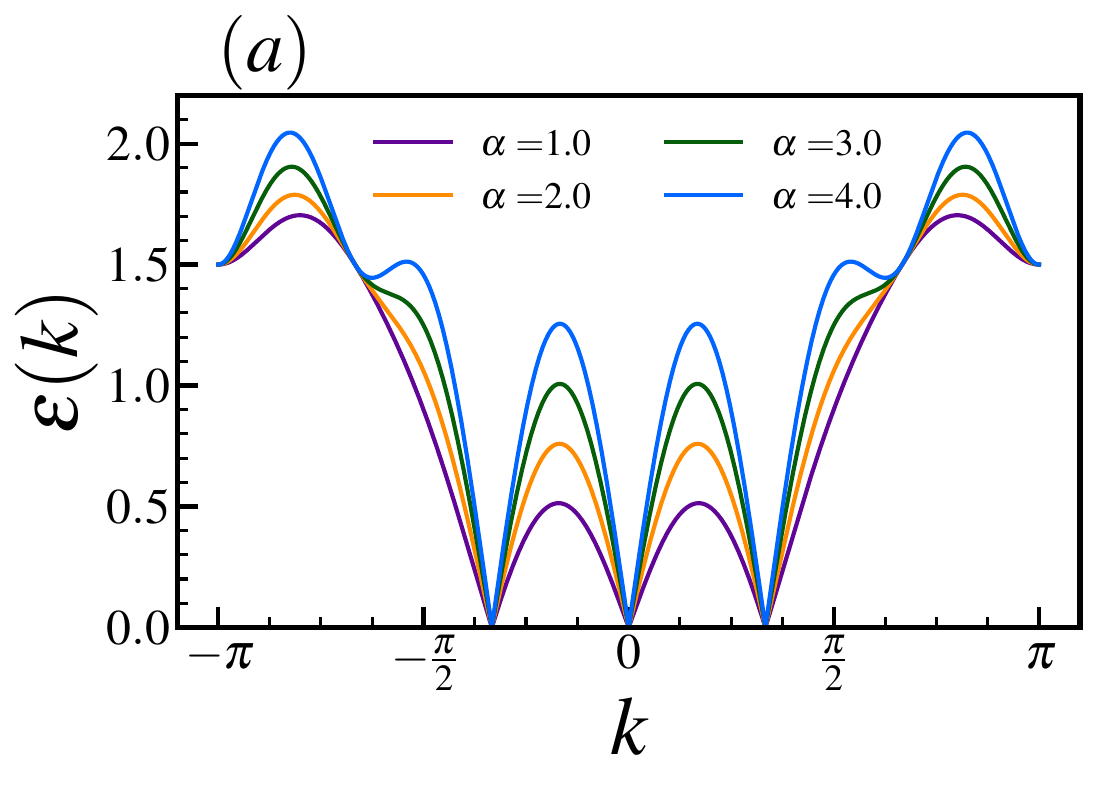} \includegraphics[width=0.5\linewidth,height=0.42\linewidth]{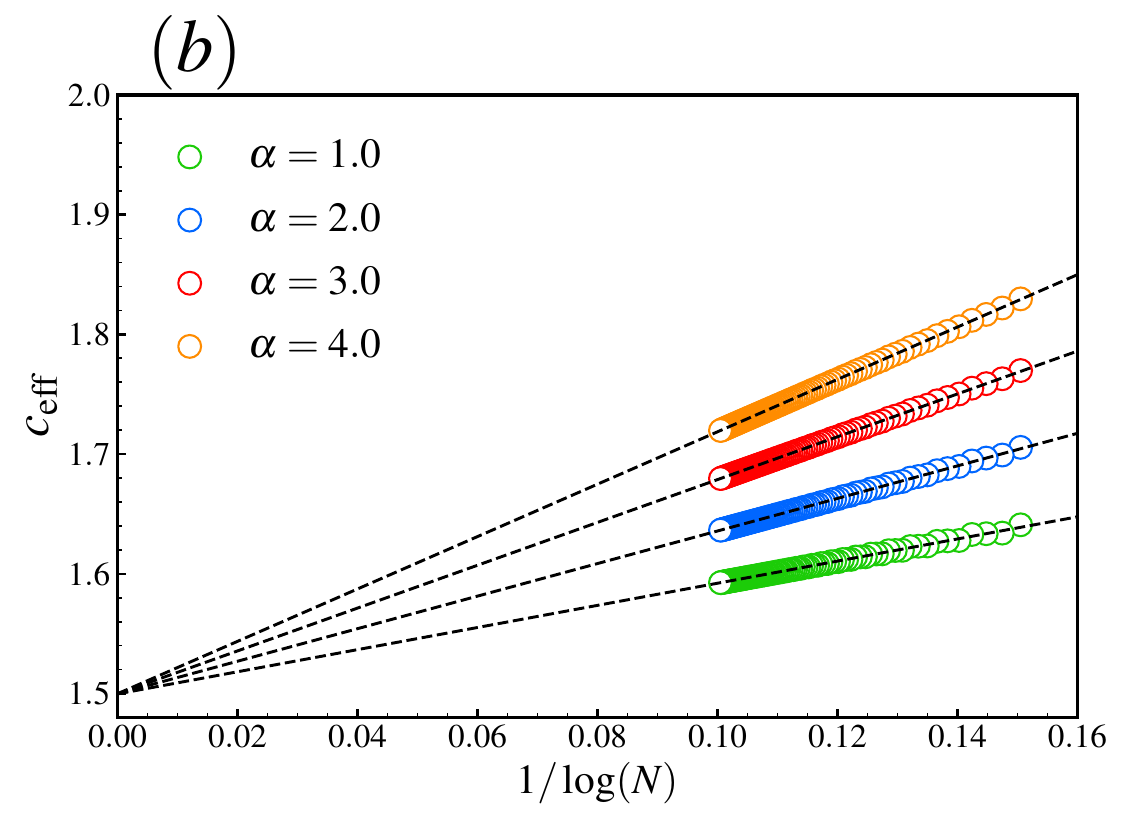} 
		
	}
	\centerline{\includegraphics[width=0.5\linewidth,height=0.42\linewidth]{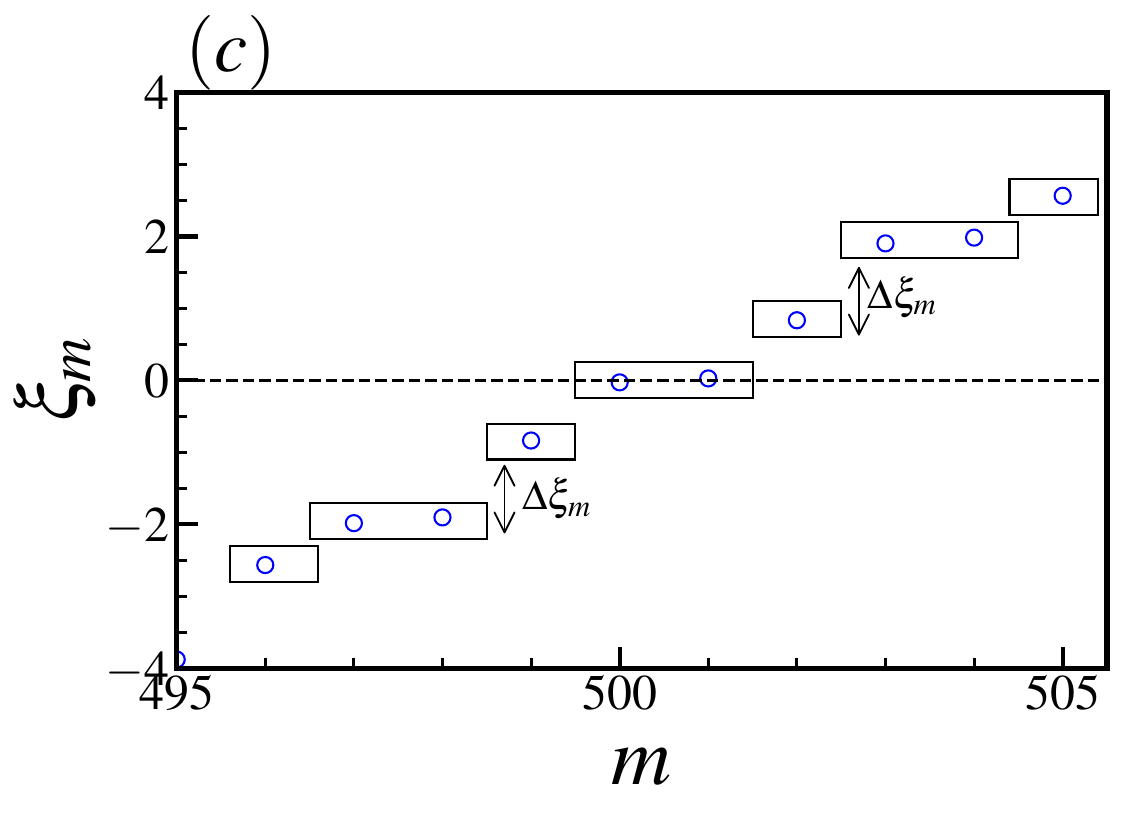} \includegraphics[width=0.5\linewidth,height=0.42\linewidth]{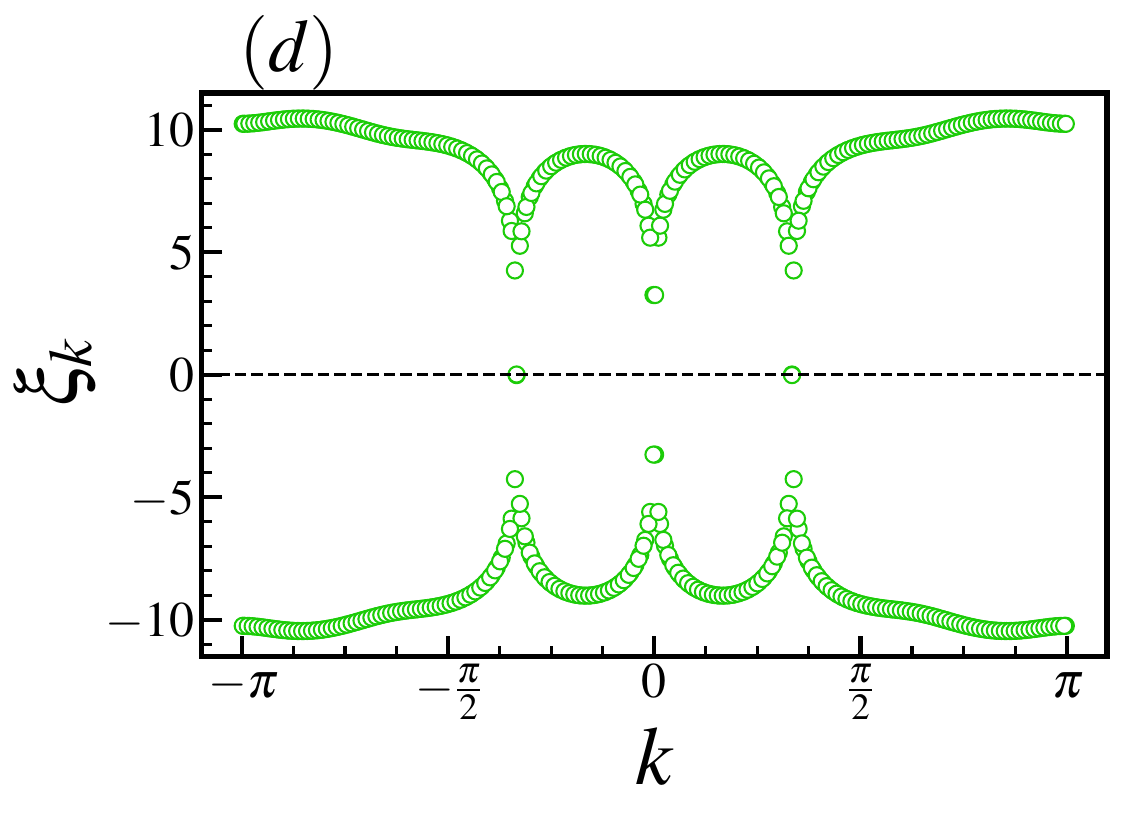} 
		
	}

	\caption{Quasiparticle spectrum, (b) Effective central charge extracted from entanglement-entropy scaling. (c) Real-space (RES) and momentum-space (MES),  (d) Momentum-space (MES) bulk entanglement spectra at the multicritical intersection point.}
	\label{fig9}
	
\end{figure}

A qualitatively distinct critical structure emerges at the intersection of the two branches $h_{c_2}$ and $h_{c_3}$. The intersection occurs at
\begin{equation}
	\beta_m=-1-\frac{2}{\alpha},\qquad h_m=\frac{3}{4},
\end{equation}
and is distinguished by the simultaneous presence of the three Fermi points $k_F=0, k_F=\pm\pi/3$. Thus, the critical modes associated separately with the $h_{c_2}$ and $h_{c_3}$ branches coexist at the intersection. As shown in Fig.~\ref{fig9}~(a), the quasiparticle spectrum remains linear around each gapless momentum, demonstrating that all three critical modes retain Dirac-like low-energy dispersions with dynamical exponent $z=1$. The corresponding analytical low-energy expansions are presented in Appendices~\ref{app:hc12} and \ref{app:hc34}.

The intersection therefore represents considerably more than a simple crossing of two critical lines. As the point is approached along either branch, the Fermi-point topology undergoes a singular reconstruction. Along $h_{c_2}$, the spectrum contains a single gapless momentum on both sides of the intersection, while exactly at the crossing three gapless momenta coexist. Conversely, along $h_{c_3}$, two gapless momenta exist away from the intersection and become three only at the crossing itself. This change in the Fermi-point topology is accompanied by a pronounced singularity in the second derivative of the ground-state energy (see Appendix~\ref{app:MIP}), identifying the intersection as an unconventional Lifshitz multicritical point. Unlike the conventional Lifshitz transition discussed previously, the quasiparticle excitations remain linearly dispersing throughout the reconstruction, while the topology of the Fermi-point configuration along both critical directions changes discontinuously.

The multicritical nature of this point is further reflected in its entanglement scaling. As shown in Fig.~\ref{fig9}~(b), the effective central charge extrapolates to $c_{\mathrm{eff}}={3}/{2}$. This value follows naturally from the continuum description. The symmetry-related pair of modes at $k_F=\pm\pi/3$ forms one massless Dirac fermion, corresponding to the conventional $c_{\rm eff}=1$ Luttinger-liquid sector, while the additional critical mode at $k_F=0$ contributes a single Majorana fermion with central charge $c_{\rm eff}=1/2$, corresponding to the Ising universality class. Consequently,
\begin{equation}
	c_{\mathrm{eff}}
	=
	1+\frac12
	=
	\frac32.
\end{equation}

The agreement between the entanglement scaling and the continuum theory provides strong evidence that the direct product of Dirac ($c_{\rm eff}=1$) and Ising ($c_{\rm eff}=1/2$) conformal field theories governs the multicritical point. This decomposition also clarifies the physical meaning of the multicritical point. The modes at $k_F=\pm\pi/3$ form a particle-hole-related pair and together generate the gapless Luttinger-liquid $c_{\rm eff}=1$ (Dirac) sector, whereas the critical mode at $k_F=0$ supplies the additional $c_{\rm eff}=1/2$ Majorana degree of freedom. The coexistence of a $c_{\rm eff}=1$ Luttinger-liquid sector and a $c_{\rm eff}=1/2$ Ising sector naturally reproduces the total central charge $c_{\rm eff}=3/2$. This establishes the low-energy conformal content of the multicritical point in a model-independent manner. The resulting $c_{\rm eff}=3/2$ theory is therefore not associated with a single conventional Dirac crossing, but with the simultaneous criticality of two distinct low-energy sectors whose gap closings occur at different momenta.

The $c_{\rm eff}=3/2$ value is also the central charge of the second $\mathcal{N}=2$ superconformal minimal model, which can be represented as a compact bosonic sector coupled to a critical Ising sector. Our results exhibit precisely this $c_{\rm eff}=1\oplus c_{\rm eff}=1/2$ critical content. However, the central charge alone does not establish the full $\mathcal{N}=2$ superconformal structure. In the present model, the relevant microscopic symmetry is the particle-hole symmetry of the Bogoliubov–de Gennes Hamiltonian, which is intrinsic to the superconducting fermionic description and constrains the quasiparticle spectrum. This symmetry should not be identified with the $\mathcal{N}=2$ supersymmetry of the continuum theory discussed in the supersymmetric lattice model of Ref.~\cite{bauer2012supersymmetric}. Establishing an emergent superconformal symmetry would require additional evidence beyond the value of $c_{\mathrm{eff}}$, such as the organization of scaling operators and the corresponding symmetry relations in the low-energy theory. We therefore identify the intersection as an unconventional Lifshitz multicritical point governed by coexisting Dirac ($c_{\rm eff}=1$) and Ising ($c_{\rm eff}=1/2$) critical sectors. The simultaneous appearance of these two independent conformal sectors, together with the reconstruction of the Fermi-point topology, provides a unified microscopic characterization of the observed $c_{\rm eff}=3/2$ criticality. A detailed derivation of the low-energy effective Hamiltonians is provided in Appendix~\ref{app:multicritical_cft}.
\\
Figure~\ref{fig9}~(c) shows the bulk entanglement spectrum at the multicritical intersection of the $h_{c_2}$ and $h_{c_3}$ critical branches. The low-lying entanglement levels exhibit a characteristic mixed multiplet structure: successive levels alternate between approximately nondegenerate states and near-degenerate doublets. This pattern is qualitatively distinct from the more uniform degeneracy structures observed along the ordinary critical branches and provides a direct spectroscopic signature of the coexistence of multiple critical sectors at the multicritical point.

The observed structure is naturally understood from the continuum decomposition established above. At the intersection point, the gap closes simultaneously at $k_F=0$ and at the symmetry-related pair $k_F=\pm\pi/3$. The resulting low-energy theory contains a Majorana sector with central charge $c_{\rm eff}=1/2$ and a Dirac, or equivalently Dirac, sector with central charge $c_{\rm eff}=1$. The bulk entanglement spectrum therefore probes the combined conformal tower of two distinct critical sectors rather than that of a single elementary gapless mode. The resulting superposition produces a nontrivial organization of the low-lying entanglement levels, in which states associated with different combinations of the Majorana and Dirac excitations appear with alternating singlet and doublet multiplicities. This mixed degeneracy pattern is consequently consistent with the identification of the multicritical point as an Ising$\times$ Dirac critical theory with
\begin{equation}
	c_{\rm eff}=\frac{1}{2}+1=\frac{3}{2}.
\end{equation}
In particular, the entanglement spectrum provides information complementary to the finite-size scaling of the effective central charge: while the latter directly quantifies the number of low-energy degrees of freedom, the former reveals the characteristic organization of the corresponding low-energy states. The simultaneous appearance of nondegenerate and doubly degenerate entanglement levels is therefore a natural fingerprint of the hybrid conformal structure generated by the coexistence of the $c_{\rm eff}=1/2$ Majorana and $c_{\rm eff}=1$ Dirac sectors.
\\
The degeneracy pattern should be understood as a finite-size realization of this underlying conformal structure. At finite system size, irrelevant operators, marginal interactions, and other microscopic corrections can split levels that become degenerate in the scaling limit. Nevertheless, the persistence of the alternating singlet--doublet organization among the lowest entanglement levels suggests that the observed structure is not an accidental finite-size feature. Rather, it is the entanglement-spectrum manifestation of the same multicritical low-energy theory responsible for the extrapolated value $c_{\mathrm{eff}}\to3/2$.
\\
Figure~\ref{fig9}~(d) shows the momentum-resolved bulk entanglement spectrum at the multicritical intersection of the $h_{c_2}$ and $h_{c_3}$ critical branches. The spectrum develops three pronounced minima at $k=0$ and $k=\pm{\pi}/{3}$, precisely at the momenta where the physical quasiparticle gap closes. Around these momenta, the entanglement branches approach one another most closely, indicating that the lowest entanglement excitations are concentrated in the same momentum regions that control the low-energy critical theory.

This momentum-space structure provides a direct entanglement-spectrum manifestation of the simultaneous gap closing at the multicritical point. The central minimum at $k=0$ is associated with the Majorana  ($c_{\rm eff}=1/2$) sector of the low-energy theory, while the symmetry-related minima at $k=\pm\pi/3$ reflect the pair of finite-momentum modes that together form the gapless  Luttinger-liquid ($c_{\rm eff}=1$) sector. The three minima should therefore not be interpreted as three independent conformal sectors; rather, they resolve the momentum-space constituents of the two-sector critical theory,
\begin{equation}
	\mathrm{Ising}\times\mathrm{Dirac},
	\qquad
	c_{\rm eff}=\frac{1}{2}+1=\frac{3}{2}.
\end{equation}
\\
The momentum-resolved bulk entanglement spectrum thus provides information complementary to the real-space entanglement spectrum. The latter reveals the mixed multiplet structure associated with the coexistence of the two critical sectors, whereas the momentum-resolved spectrum identifies where these critical degrees of freedom are concentrated in momentum space. The simultaneous appearance of minima at $0$ and $\pm\pi/3$, together with the independent extrapolation of the effective central charge to $c_{\mathrm{eff}}=3/2$, provides a consistent characterization of the multicritical point as a state in which the zone-center Majorana mode and the finite-momentum  Luttinger-liquid sector become critical simultaneously.

\subsection{Global structure of the conformal phase diagram}

\begin{figure}[h]
	\centerline{\includegraphics[width=0.5\linewidth,height=0.42\linewidth]{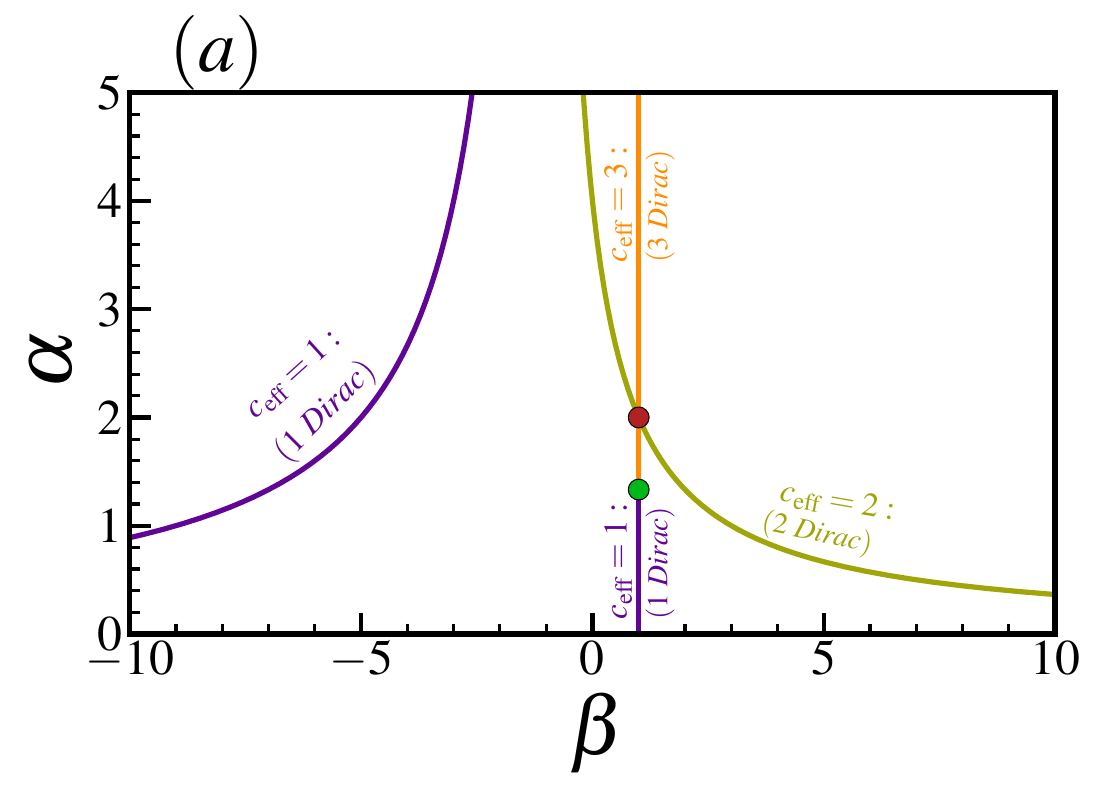} \includegraphics[width=0.5\linewidth,height=0.42\linewidth]{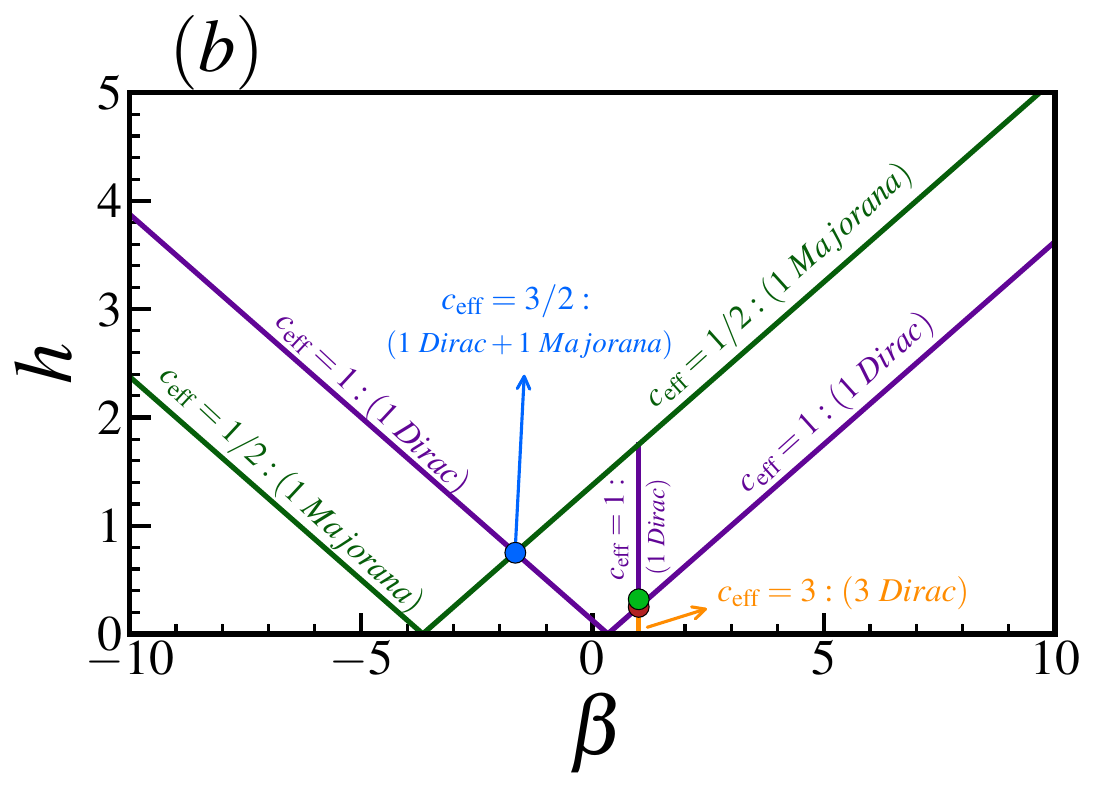} 
		
	}

	\caption{Schematic conformal phase diagrams in the absence and presence of a transverse magnetic field. (a) Zero-field phase diagram in the $(\alpha,\beta)$ plane. (b) Finite-field phase diagram in the $(h,\beta)$ plane. Critical boundaries are labeled by their effective central charges, highlighting the hierarchy of Ising, Luttinger-liquid, and multiple-Dirac critical sectors, as well as the unconventional $c_{\rm eff}=3/2$ multicritical point.}
	\label{fig10}
	
\end{figure}

To provide a global overview of the critical behavior discussed throughout this work, Fig.~\ref{fig10} summarizes the conformal phase diagrams in both the absence and presence of a transverse magnetic field. Panel~(a) presents the $(\alpha,\beta)$ phase diagram at zero field, while panel~(b) shows the corresponding $(h,\beta)$ phase diagram for finite magnetic field. Each critical boundary is labeled by its effective central charge, as determined from the entanglement-entropy scaling and corroborated by the corresponding continuum field-theory analysis. Together, the two diagrams reveal the complete hierarchy of conformal criticalities realized by the model, ranging from the Ising universality class ($c_{\rm eff}=1/2$) to phases containing up to three independent Dirac sectors ($c_{\rm eff}=3$), as well as the unconventional multicritical point with $c_{\rm eff}=3/2$. They also provide a unified visualization of how the transverse magnetic field enriches the topology of the critical manifold by generating additional field-induced quantum critical lines and modifying the organization of the underlying conformal sectors.

An important conclusion emerging from the global phase diagrams is that, throughout the present model, the effective central charge is determined by the number of independent low-energy continuum sectors rather than by the raw number of lattice Fermi crossings. For all critical phases whose Fermi points lie inside the first Brillouin zone, the continuum theory contains one massless Dirac sector for each symmetry-related pair of Fermi points, yielding the simple relation
\begin{equation}
	c_{\rm eff}=\frac{N_f}{2},
\end{equation}
where $N_f$ denotes the total number of Fermi points. This correspondence breaks down when gapless points occur at the Brillouin-zone boundary, where the two crossings at $k=\pm\pi$ are identified by lattice periodicity and collapse into a single continuum sector. Consequently, the effective central charge is controlled by the number of independent continuum sectors after this identification, rather than by the total number of microscopic Fermi crossings.

\section{Conclusion} \label{conc}

In this work, we investigated the quantum critical behavior of an exactly solvable extended quantum spin chain with competing nearest-neighbor and cluster interactions. By combining exact diagonalization, finite-size scaling, low-energy continuum analysis, and bulk entanglement spectra in real and momentum space, we established a unified microscopic correspondence between Fermi-point topology, emergent conformal criticality, and quantum entanglement.

The model exhibits a remarkably rich hierarchy of conformal critical phases with effective central charges $c_{\rm eff}=1/2$, $1$, $3/2$, $2$, and $3$, generated through a sequence of interaction- and field-driven Lifshitz transitions. In particular, the interplay between cluster interactions and the transverse magnetic field produces successive reconstructions of the Fermi-point structure, including an unconventional multicritical point characterized by $c_{\rm eff}=3/2$. The low-energy continuum analysis shows that this point consists of coexisting Ising ($c_{\rm eff}=1/2$) and Luttinger-liquid ($c_{\rm eff}=1$) sectors, providing a microscopic explanation of its effective central charge.

A central result of this work is that the conformal central charge cannot, in general, be inferred from a simple count of gapless Fermi points. Instead, it is determined by the number and conformal content of the independent low-energy continuum sectors that emerge from the Fermi-point structure after accounting for lattice symmetries, reciprocal-lattice identifications, and mode equivalences. In particular, Fermi points related through the periodic identification of the Brillouin zone need not correspond to independent conformal degrees of freedom, whereas genuinely distinct continuum sectors contribute separately to the critical theory. This organizing principle consistently accounts for the different central charges realized throughout the phase diagram, including those associated with field-driven Lifshitz transitions and the multicritical point.

Our results further demonstrate that a Lifshitz transition does not necessarily change the conformal central charge. A Fermi-point reconstruction can either generate a new independent continuum sector, thereby modifying the conformal anomaly, or merely reorganize the momentum-space representation of existing low-energy modes without changing the underlying conformal field theory. Thus, changes in Fermi-point topology need not imply changes in the conformal anomaly: what matters is whether the reconstruction alters the number or conformal content of the independent continuum sectors.

The entanglement spectra provide complementary microscopic signatures of this correspondence. The real-space bulk entanglement spectrum provides characteristic level structures associated with the underlying conformal sectors, while the momentum-space entanglement spectrum resolves the momentum-space organization and reconstruction of the critical modes across the Lifshitz transitions. Combined with the central charge extracted from the universal scaling of the entanglement entropy, these quantities provide complementary probes of both the universal conformal content and its microscopic momentum-space origin. Thus, the entanglement spectra go beyond serving as estimators of the central charge and provide a microscopic bridge between the lattice Fermi-point structure and the emergent continuum conformal field theory.

Beyond the specific model considered here, our results suggest a broader framework for understanding how interaction-driven spectral reconstruction can shape emergent conformal criticality in one-dimensional quantum systems. The organizing principle developed here can potentially be applied to extended spin chains, long-range or interacting fermionic systems, and topological superconducting models, where symmetry, mode equivalence, and Fermi-point reconstruction may similarly control the number and conformal content of independent low-energy sectors. This framework may therefore provide a useful route toward identifying higher-central-charge criticality and unconventional multicritical behavior in a broader class of low-dimensional quantum systems.

\appendix

\section{Exact Fermionic Solution}
\label{app:exact_solution}

In this appendix, we present the exact solution of the extended cluster XX chain by deriving its fermionic Bogoliubov--de Gennes representation and quasiparticle spectrum. Starting from the microscopic Hamiltonian
\begin{equation}
	\mathcal H=\mathcal H_{XX}+\mathcal H_{CI},
\end{equation}
with
\begin{align}
	\mathcal H_{XX}
	&=
	J\sum_{n=1}^{N}
	\left(
	S_n^xS_{n+1}^x+
	S_n^yS_{n+1}^y
	\right)
	-
	Jh\sum_{n=1}^{N}S_n^z,
	\\
	\mathcal H_{CI}
	&=
	J'
	\sum_{n=1}^{N}
	\left[
	\left(
	S_n^xS_{n+3}^x
	+
	\beta S_n^yS_{n+3}^y
	\right)
	S_{n+1}^zS_{n+2}^z
	\right],
\end{align}
we perform the Jordan--Wigner transformation followed by a Bogoliubov diagonalization to obtain the exact quasiparticle spectrum that forms the basis of the analysis presented in the main text.

\subsection{Jordan--Wigner Representation}

To obtain an exact solution, we employ the Jordan--Wigner transformation, which maps the spin Hamiltonian onto a quadratic model of spinless fermions. Introducing fermionic operators $a_n$ and $a_n^\dagger$, the Hamiltonian takes the form
\begin{align}
	\mathcal H
	&= \frac{1}{2} \sum_{n=1}^{N} \left( a_n^\dagger a_{n+1} + a_{n+1}^\dagger a_n \right) \nonumber\\
	&+ \frac{\alpha}{16}(1+\beta) \sum_{n=1}^{N} \left( a_n^\dagger a_{n+3} + a_{n+3}^\dagger a_n \right) \nonumber\\
	&+ \frac{\alpha}{16}(1-\beta) \sum_{n=1}^{N} \left( a_n^\dagger a_{n+3}^\dagger + a_{n+3}a_n \right) \nonumber\\ &- h \sum_{n=1}^{N}	\left( 	a_n^\dagger a_n-\frac12 \right).
\end{align}

The resulting fermionic Hamiltonian remains quadratic and therefore admits an exact analytical treatment.

\subsection{Bogoliubov Quasiparticles and Momentum-Space Structure}

After performing a Fourier transformation  $a_n = \frac{1}{\sqrt{N}} \sum_k e^{-ikn} a_k$, we introduce the Nambu spinor
\begin{align}
	\Psi_k=\begin{pmatrix} a_k\\ a_{-k}^{\dagger}	\end{pmatrix},
\end{align}
which allows the Hamiltonian to be written in Bogoliubov--de Gennes form,
\begin{align}
	& \mathcal H = \frac{1}{2}\sum_k \Psi_k^\dagger H_{\mathrm{BdG}}(k) \Psi_k,
\end{align}
with
\begin{align}
	H_{\mathrm{BdG}}(k) = \begin{pmatrix} 	\mathcal{A}_k & -i\mathcal{B}_k \\ i\mathcal{B}_k & -\mathcal{A}_k \end{pmatrix},
\end{align}
where
\begin{align}
	& \mathcal{A}_k =  \cos k - h + \frac{\alpha}{8}(1+\beta)\cos(3k),  \nonumber \\
	&\mathcal{B}_k = \frac{\alpha}{8}(1-\beta)\sin(3k).
\end{align}

Diagonalization of the BdG Hamiltonian yields the quasiparticle excitation spectrum
\begin{equation}
	\varepsilon(k) = \sqrt{\mathcal{A}_k^2+\mathcal{B}_k^2}.
\end{equation}

Since the Hamiltonian is exactly diagonalizable, the ground state corresponds to the Bogoliubov vacuum, and the low-energy properties are completely determined by the quasiparticle spectrum.

%\subsection{Pseudospin Representation and Gap Closing}

A particularly transparent representation is obtained by expressing the BdG Hamiltonian in terms of Pauli matrices,
\begin{equation}
	H_{\mathrm{BdG}}(k)	=	\mathcal A_k\sigma_z + \mathcal B_k\sigma_y,
\end{equation}
or, equivalently,
\begin{equation}
	H_{\mathrm{BdG}}(k) = d(k) \cdot \boldsymbol \sigma,
\end{equation}
with the effective BdG vector
\begin{equation}
	d(k) = \left( 0,\, \mathcal{B_k},\, \mathcal{A_k} \right).
\end{equation}

Within this representation, the quasiparticle energy is simply given by $\varepsilon(k)=| d(k)| $, and the gap-closing condition assumes the compact form $\mathcal{A}_k=0$, and $\mathcal{B}_k=0$. Consequently, the quantum critical lines and the associated changes in the low-energy structure can be understood directly from the evolution of the zeros of $d(k)$ in momentum space.

%$\varepsilon(k)=|\mathbf {d(k)|}$, and the gap-closing condition assumes the compact form $\mathcal A_k=0$, and $\mathcal B_k=0$. Consequently, the quantum critical lines and the associated changes in the low-energy structure can be understood directly from the evolution of the zeros of $\mathbf d(k)$ in momentum space. 

\section{Low-Energy Expansion Around the Fermi Points}

\label{app:dispersion}

\subsection{Low-energy theory along the $\beta_{c1}$ critical line}
\label{app:bet1}

We derive the low-energy form of the quasiparticle spectrum along the isotropic line $\beta=1$ and establish the origin of the linear Dirac excitations and the cubic band touching characterizing the Lifshitz critical point discussed in the main text.
\\
For $\beta=1$, the quasiparticle spectrum reduces to

\begin{equation} 
	\varepsilon(k) = \cos k + \frac{\alpha}{4}\cos(3k). \label{eq:A1}
\end{equation}
\\
The Fermi points are determined by the condition $\varepsilon(k_F)=0$. Two solutions are always present at $k_F=\pm\pi/2$, while for $\alpha>4J/3$ two additional pairs of nodes emerge,$	k_F=\pm k_0$ and $k_F=\pm(\pi-k_0)$, where $k_0= \arccos \left( \sqrt{({3}/{4})-({J}/{\alpha})} \right)$.

\mbox{}\\
\paragraph*{Linear dispersion around $k_F=\pm\pi/2$.} 
\mbox{}\\

To determine the low-energy behavior around the persistent Fermi points, we set

\begin{equation}
	k=\pm \frac{\pi}{2}+q, \qquad |q|\ll1.
\end{equation}

Using the expansions

\begin{align}
	\cos\left(\pm \frac{\pi}{2}+q\right)
	&= \mp \sin q = \mp q \pm \frac{q^3}{6}\pm O(q^5),
	\\
	\cos\left(\pm \frac{3\pi}{2}+3q\right)
	&= \pm \sin(3q) = \pm 3q \mp \frac{9}{2}q^3\pm O(q^5),
\end{align}

Eq.~(\ref{eq:A1}) becomes

\begin{equation}
	\varepsilon(q)
	= \left( \pm \frac{3\alpha}{4} \mp 1 \right)q + \left( \pm \frac{J}{6} \mp \frac{9\alpha}{8} \right) q^3 \pm O(q^5).
	\label{eq:A5}
\end{equation}

Therefore, away from the Lifshitz point, the leading contribution is linear,

\begin{equation}
	\varepsilon(q) \simeq \pm v_F q,
\end{equation}

with Fermi velocity

\begin{equation}
	v_F= \frac{3\alpha}{4}-1. \label{eq:A6}
\end{equation}

\mbox{}\\
\paragraph*{Cubic band touching at the Lifshitz point.} 
\mbox{}\\

At the critical coupling

\begin{equation}
	\alpha_c=\frac{4}{3},
\end{equation}

the Fermi velocity vanishes identically,

\begin{equation}
	v_F(\alpha_c)=0,
\end{equation}

and the linear term in Eq.~(\ref{eq:A5}) disappears. Retaining the leading nonvanishing contribution yields

\begin{equation}
	\varepsilon(q) = \mp \frac{4}{3}q^3 \pm O(q^5).
	\label{eq:A7}
\end{equation}

\begin{figure}[h]
	\centerline{\includegraphics[width=0.5\linewidth,height=0.42\linewidth]{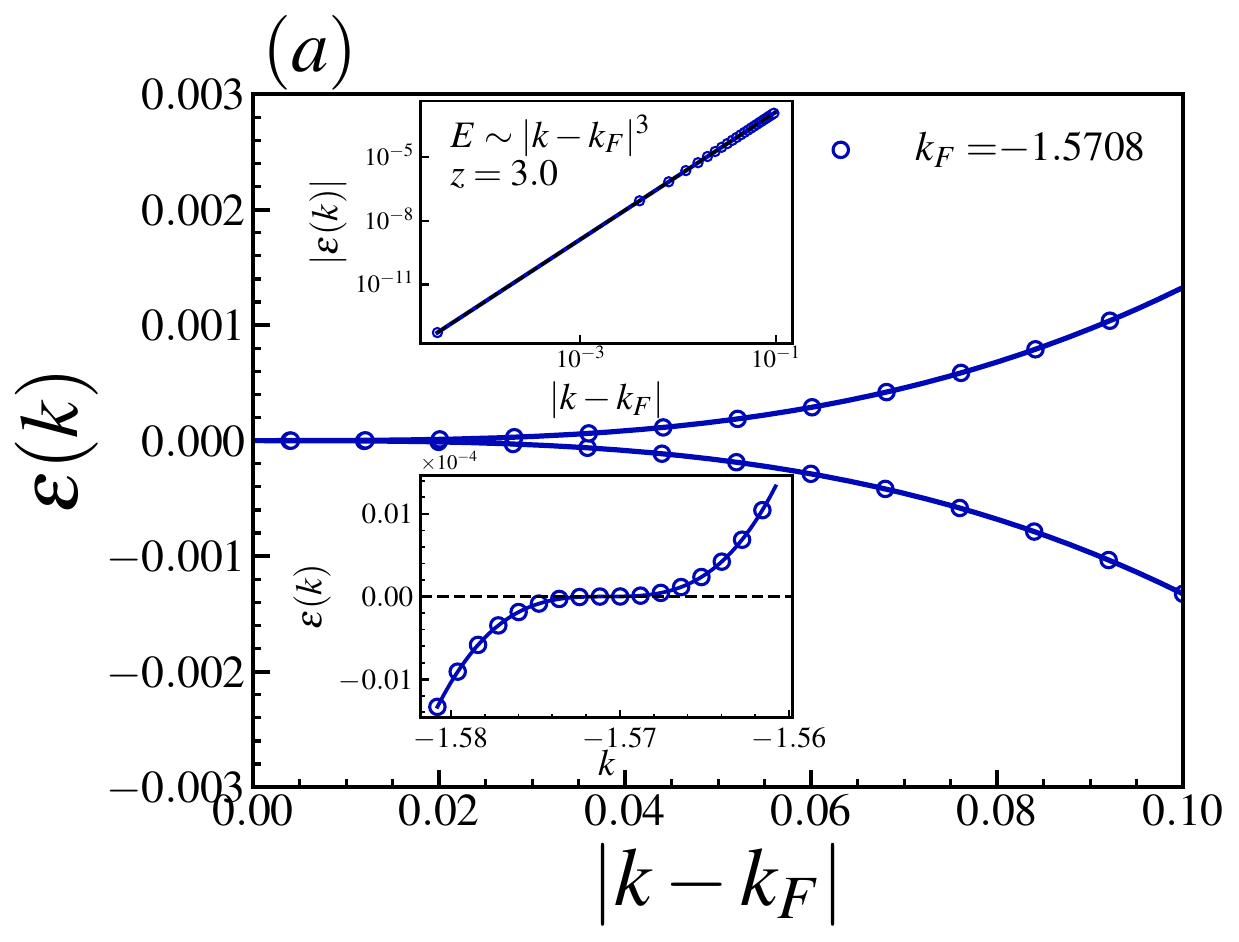} \includegraphics[width=0.5\linewidth,height=0.42\linewidth]{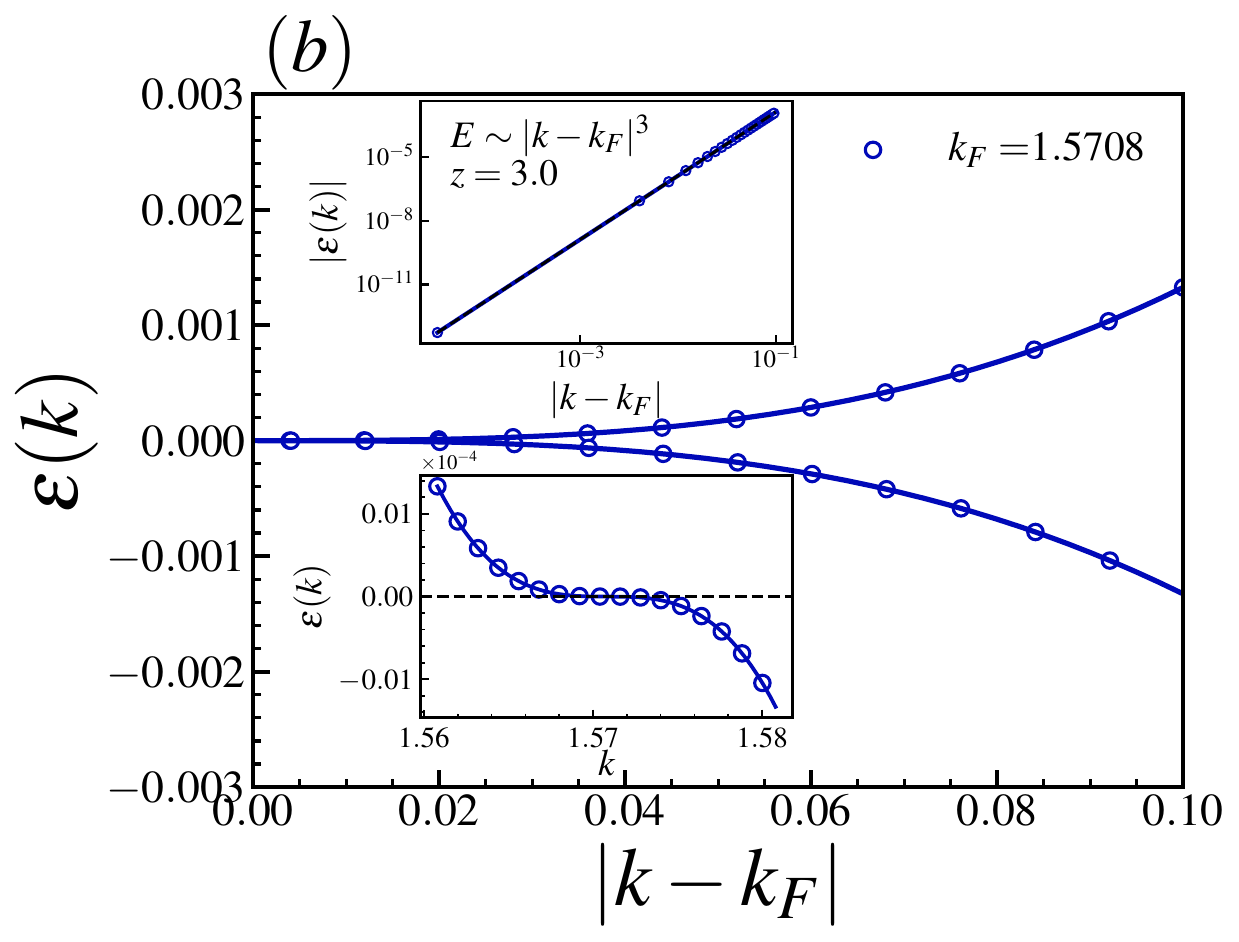} 
		
	}

	\caption{ Low-energy spectrum along the isotropic line $\beta=1$ near the critical momenta $k_F=\pm\pi/2$.}
	\label{fig_A1_2}
	
\end{figure}
Figures~\ref{fig_A1_2}~(a) and \ref{fig_A1_2}~(b) display the low-energy spectrum near the critical momenta $k_F=\pm\pi/2$ exactly at $\alpha_c=4/3$. In contrast to the linear Dirac cones observed away from the transition, the log-log analysis shown in the insets reveals the scaling behavior $ \varepsilon(k)\propto |k-k_F|^3$, yielding a dynamical exponent $ z=3$. Hence, the quasiparticle spectrum near $k_F=\pm\pi/2$ obeys

\begin{equation}
	\varepsilon(q) \propto |q|^3,
\end{equation}
corresponding to the dynamical critical exponent $z=3$. The cubic band touching therefore represents the multicritical mode controlling the Lifshitz transition.

\mbox{}\\
\paragraph*{Emergent Dirac nodes for $\alpha>4/3$.} 
\mbox{}\\

For $\alpha>4/3$, additional Fermi points appear at $k_F=\pm k_0$ and $k_F=\pm(\pi-k_0)$ where $k_0= \arccos \left( \sqrt{ (3/4)-(1/\alpha) } \right)$.
Expanding Eq.~(\ref{eq:A1}) around a generic node,

\begin{equation}
	k= \pm k_F+q, \qquad |q|\ll1,
\end{equation}
gives
\begin{equation}
	\varepsilon(\pm k_F+q) = \varepsilon'(\pm k_F)q \pm O(q^2),
	\label{eq:A8}
\end{equation}
where
\begin{equation}
	\varepsilon'(k) = -\sin k - \frac{3\alpha}{4}\sin(3k).
	\label{eq:A9}
\end{equation}
Using 
\begin{equation}
	\sin(3k) = 3\sin k-4\sin^3k \nonumber
\end{equation} 
together with
\begin{equation}
	\sin^2k_0= \frac14+\frac{1}{\alpha}, \nonumber
\end{equation}
one finds
\begin{equation}
	\varepsilon'(\pm k_0) = \pm \sin k_0 \left( 2-\frac{3\alpha}{2} \right). \label{eq:A10}
\end{equation}
Since $\alpha>4/3$, the derivative remains finite,
\begin{equation}
	\varepsilon'(\pm k_0)\neq0,
\end{equation}
implying
\begin{equation}
	\varepsilon(\pm k_0+q) \simeq \pm \varepsilon'(k_0)q.
\end{equation}
The same result holds for the symmetry-related nodes at
$\pm(\pi-k_0)$,
whose slopes differ only by sign. Therefore all six Fermi points possess linear low-energy dispersions and correspond to massless Dirac excitations with dynamical exponent $z=1$.

\begin{figure}[h]
	\centerline{\includegraphics[width=0.8\linewidth,height=0.6\linewidth]{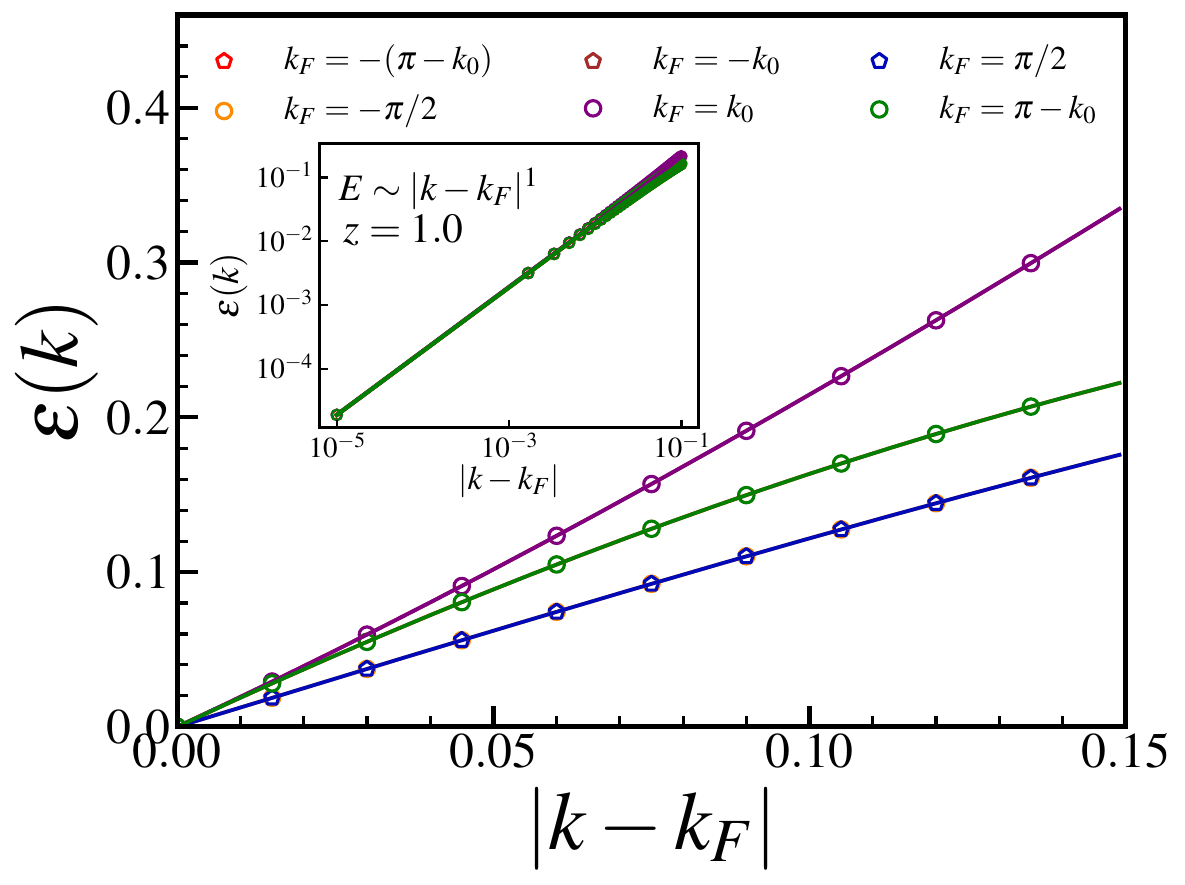} 
		
	}

	\caption{Low-energy spectrum along the isotropic line $\beta=1$ for all six Fermi points in the $\alpha>\alpha_c$ phase.}
	\label{fig_A1_1}
	
\end{figure}

Figure~\ref{fig_A1_1} shows the quasiparticle dispersion in the vicinity of all six Fermi points in the $\alpha>\alpha_c$ phase, where the quasiparticle energy is plotted as a function of $|k-k_F|$ around all Fermi points. Independent of the particular node, the dispersion exhibits a linear behavior, $\varepsilon(k)\propto |k-k_F|$, indicating a dynamical exponent $z=1$. The inset, shown on logarithmic scales, confirms the unit slope expected for relativistic Dirac excitations. Therefore, despite the proliferation of gapless crossings beyond the Lifshitz transition, all six nodes support conventional massless Dirac modes.
\\
Consequently, the low-energy excitations are relativistic Dirac modes characterized by the dynamical exponent $z=1$.
\\

In summary, the Lifshitz transition at $\alpha_c=4/3$ separates a phase containing a single Dirac pair from a six-node phase composed of three Dirac pairs. Exactly at the transition, the Dirac velocity vanishes and the spectrum develops a cubic band touching with $z=3$, which governs the nucleation of the additional Fermi points discussed in the main text.

\subsection{Low-energy theory along the $\beta_{c_2}$ critical line}
\label{app:betc2}

We analyze the low-energy quasiparticle spectrum along the critical lines $\beta_{c_2}$ and $\beta_{c_3}$ and establish the Dirac nature of their gapless excitations through an explicit continuum expansion. For the quadratic Bogoliubov--de Gennes Hamiltonian, the quasiparticle spectrum is

\begin{equation}
	\varepsilon(k)=\sqrt{\mathcal{A}_k^2+\mathcal{B}_k^2},
\end{equation}

where

\begin{align}
	\mathcal{A}_k &= \cos k+\frac{\alpha}{8}(1+\beta)\cos3k,\\
	\mathcal{B}_k &= \frac{\alpha}{8}(1-\beta)\sin3k,
\end{align}
\\
Along the critical line $\beta_{c2}=-\left(1+{8}/{\alpha}\right)$, the quasiparticle gap closes at three symmetry-related momenta, $ k_F=0$, $k_F=\pm\pi$ .
Using the relations
\begin{equation}
	1+\beta_{c_2}=-\frac{8}{\alpha},
	\qquad
	1-\beta_{c_2}=\frac{2(\alpha+4)}{\alpha},
\end{equation}
the spectrum can be expanded around each Fermi point. For the node at $k_F=0$, we write $k=q$ with $|q|\ll1$. Using the Taylor expansions

\begin{align}
	&\cos q=1-\frac{q^2}{2}+O(q^4),\nonumber \\
	& \cos3q=1-\frac{9q^2}{2}+O(q^4), \nonumber \\
	& \sin3q=3q+O(q^3),
\end{align} 
ne finds

\begin{align}
	\mathcal{A}_q &=4q^2+O(q^4),\\
	\mathcal{B}_q &=\frac{3}{4}(\alpha+4)q+O(q^3),
\end{align}
which yields

\begin{equation}
	\varepsilon(q)
	=
	\sqrt{(4q^2)^2+
		\left[\frac34(\alpha+4)q\right]^2}
	\simeq
	\frac34(\alpha+4)|q|.
\end{equation}
The continuum expansion around the zone-boundary Fermi point is obtained by writing
\begin{equation}
	k=\pi+q,\qquad |q|\ll1,
\end{equation}
together with
\begin{align}
	\cos(\pi+q)
	&=-1+\frac{q^2}{2}+O(q^4),\\
	\cos(3\pi+3q)
	&=-1+\frac{9q^2}{2}+O(q^4),\\
	\sin(3\pi+3q)
	&=3q+O(q^3).
\end{align}
Substituting these expressions into $\mathcal{A}_k$ and $\mathcal{B}_k$ gives
\begin{align}
	\mathcal{A}_{\pi+q} &= \left(-1+\frac{q^2}{2}\right) - \left(-1+\frac{9q^2}{2}\right)
	= -4q^2+O(q^4),\\ 
	\mathcal{B}_{\pi+q} &= \frac34(\alpha+4)q+O(q^3),
\end{align}
and therefore
\begin{equation}
	\varepsilon(\pi+q)
	=
	\sqrt{(4q^2)^2+
		\left[\frac34(\alpha+4)q\right]^2}
	\simeq
	\frac34(\alpha+4)|q|.
\end{equation}

The expansion around the remaining Fermi point, $k_F=-\pi$, proceeds analogously by setting $k=-\pi+q$. Owing to the identities $\cos(-x)=\cos x$ and $\sin(-x)=-\sin x$, the resulting continuum coefficients are identical to those obtained at $k_F=\pi$. Consequently, all three gapless nodes are characterized by the same leading-order dispersion,
\begin{equation}
	\varepsilon(k)\simeq v_F|k-k_F|,
\end{equation}
with the common Fermi velocity
\begin{equation}
	v_F=\frac34(\alpha+4).
\end{equation}

\begin{figure}[h]
	\centerline{\includegraphics[width=0.8\linewidth,height=0.6\linewidth]{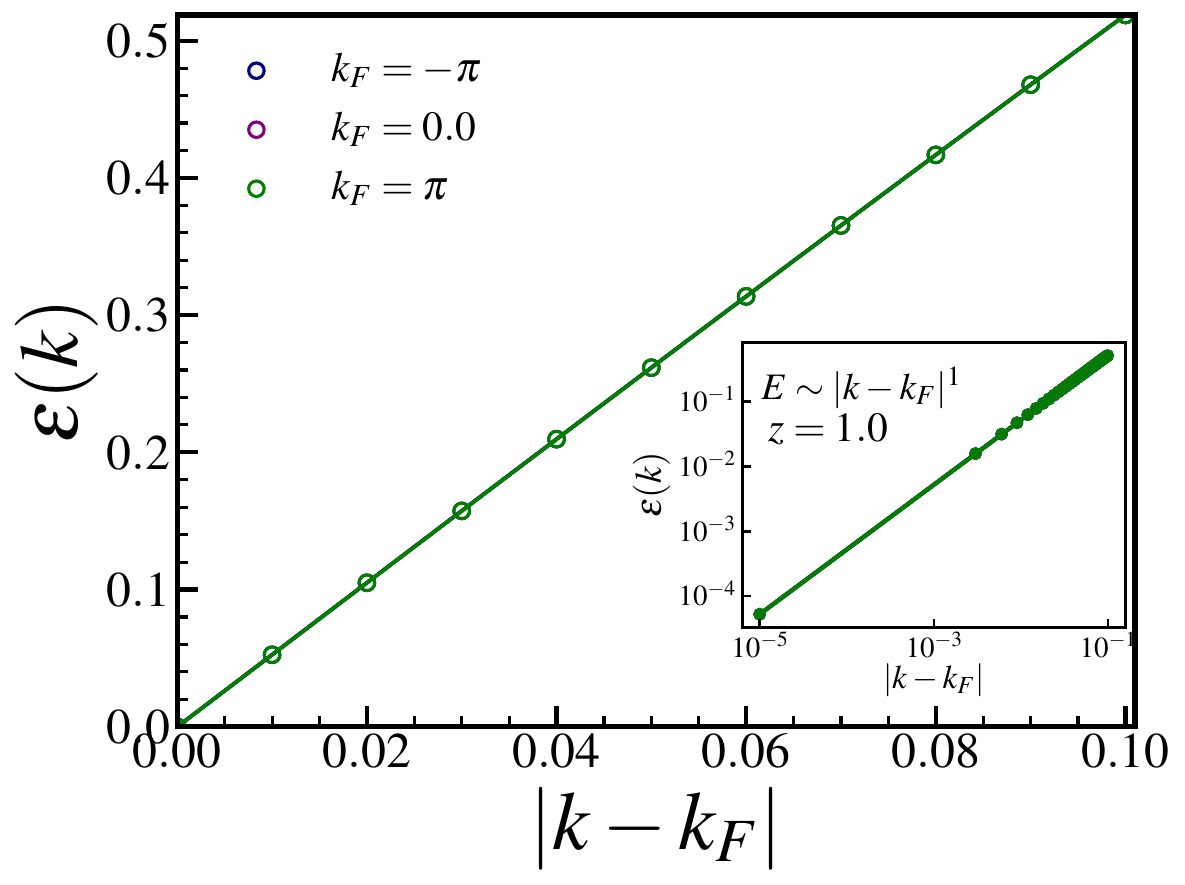} 
		
	}

	\caption{ Low-energy spectrum along the line $\beta_{c_2}$ near the critical momenta $k_F=0,\pm\pi$.}
	\label{fig_A1_3}
	
\end{figure}

Figure~\ref{fig_A1_3} illustrates the quasiparticle spectrum in the vicinity of all three gapless points. The numerical results are in excellent agreement with the analytical expansion and demonstrate that every node is characterized by a linear dispersion. Consequently, the low-energy quasiparticles constitute massless Dirac excitations with a dynamical critical exponent $ z=1$. The logarithmic representation shown in the inset further confirms the unit power-law scaling, $\varepsilon(k)\propto|k-k_F|$, expected for relativistic Dirac quasiparticles.

\subsection{Low-energy theory along the $\beta_{c_3}$ critical line}
\label{app:betc3}
The thired critical line is
\begin{equation}
	\beta_{c_3}=\frac{4}{\alpha}-1,
\end{equation}
for which the quasiparticle spectrum becomes gapless at the four symmetry-related momenta
\begin{equation}
	k_F=\pm\frac{\pi}{3},\qquad
	k_F=\pm\frac{2\pi}{3}.
\end{equation}
Using
\begin{equation}
	1+\beta_{c_3}=\frac{4}{\alpha},
	\qquad
	1-\beta_{c_3}
	=\frac{2(\alpha-2)}{\alpha},
\end{equation}
the BdG coefficients reduce to
\begin{align}
	\mathcal A_k
	&=
	\cos k+\frac{1}{2}\cos3k,\\
	\mathcal B_k
	&=
	\frac{\alpha-2}{4}\sin3k.
\end{align}
We first expand around the representative Fermi point
\begin{equation}
	k=\frac{\pi}{3}+q,
	\qquad |q|\ll1.
\end{equation}
The required trigonometric expansions are
\begin{align}
	\cos\!\left(\frac{\pi}{3}+q\right)
	&=
	\frac12-\frac{\sqrt3}{2}q-\frac14q^2+O(q^3),\\
	\cos(\pi+3q)
	&=
	-1+\frac92q^2+O(q^4),\\
	\sin(\pi+3q)
	&=
	-3q+O(q^3).
\end{align}
Substituting these expressions into $\mathcal A_k$ and
$\mathcal B_k$ gives
\begin{align}
	\mathcal A_{\pi/3+q}
	&=
	-\frac{\sqrt3}{2}\,q+O(q^2),\\
	\mathcal B_{\pi/3+q}
	&=
	-\frac34(\alpha-2)\,q+O(q^3).
\end{align}
The resulting dispersion is therefore
\begin{equation}
	\varepsilon\!\left(\frac{\pi}{3}+q\right)
	=
	\sqrt{
		\frac34
		+\frac{9}{16}(\alpha-2)^2
	}\,
	|q|
	\equiv
	v_F|q|,
\end{equation}
with
\begin{equation}
	v_F=
	\frac14
	\sqrt{
		12+
		9(\alpha-2)^2 }.
\end{equation}
The remaining Fermi points are obtained analogously.
For $k=-\pi/3+q$,
\begin{align}
	\cos\!\left(-\frac{\pi}{3}+q\right)
	&=
	\frac12+\frac{\sqrt3}{2}q-\frac14q^2+O(q^3),\\
	\cos(-\pi+3q)
	&=
	-1+\frac92q^2+O(q^4),\\
	\sin(-\pi+3q)
	&=
	3q+O(q^3),
\end{align}
leading to
\begin{align}
	\mathcal A_{-\pi/3+q}
	&=
	+\frac{\sqrt3}{2}\,q+O(q^2),\\
	\mathcal B_{-\pi/3+q}
	&=
	+\frac34(\alpha-2)\,q+O(q^3).
\end{align}

Similarly, expanding around
$k=\pm2\pi/3+q$ yields
\begin{align}
	\mathcal A_{\pm2\pi/3+q}
	&=
	\mp\frac{\sqrt3}{2}\,q+O(q^2),\\
	\mathcal B_{\pm2\pi/3+q}
	&=
	+\frac34(\alpha-2)\,q+O(q^3),
\end{align}
so that all four gapless points possess the same leading-order spectrum,
\begin{equation}
	\varepsilon(k)
	\simeq
	v_F|k-k_F|,
\end{equation}
with the common Fermi velocity
\begin{equation}
	v_F=
	\frac14
	\sqrt{
		12J^2+
		9(\alpha-2J)^2 }.
\end{equation}

Therefore, every gapless node on the $\beta_{c_3}$ critical line supports a linearly dispersing Dirac quasiparticle. The differences between the four expansions appear only in the signs of the linear coefficients of $\mathcal A_k$ and $\mathcal B_k$, while the quasiparticle energy depends only on their squared magnitudes. Consequently, all four Fermi points are described by the same universal low-energy Dirac theory.

\begin{figure}[h]
	\centerline{\includegraphics[width=0.5\linewidth,height=0.42\linewidth]{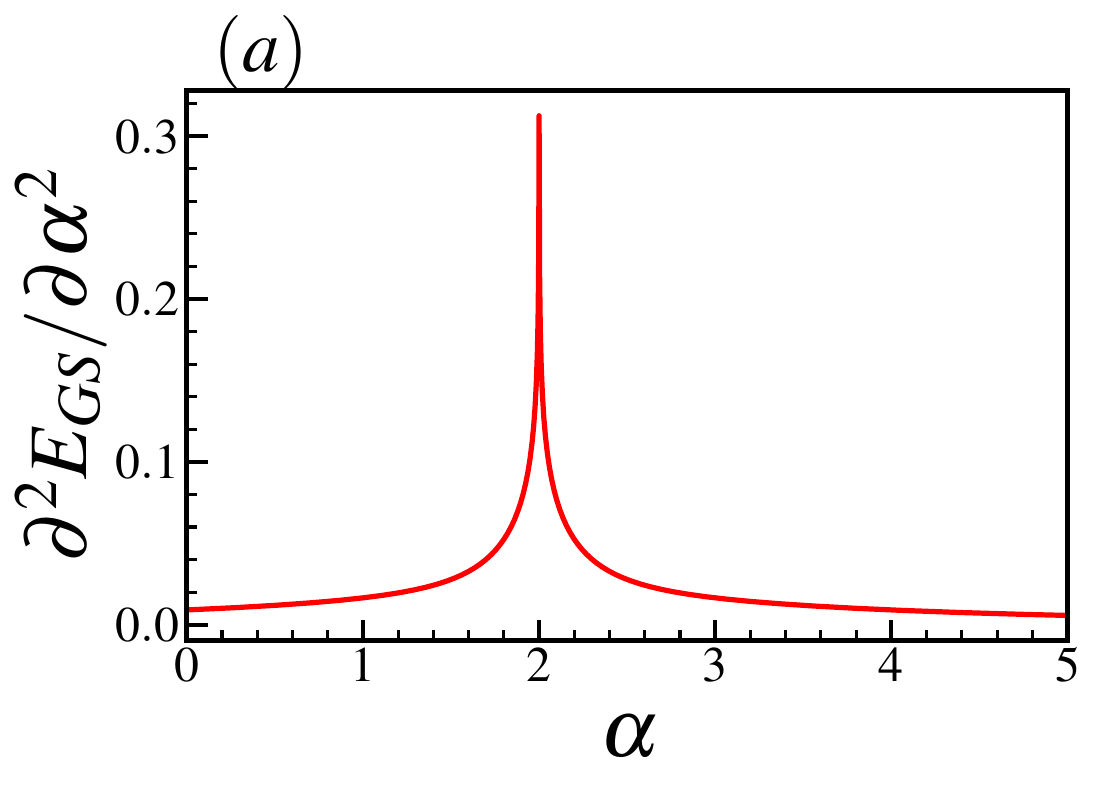} \includegraphics[width=0.5\linewidth,height=0.42\linewidth]{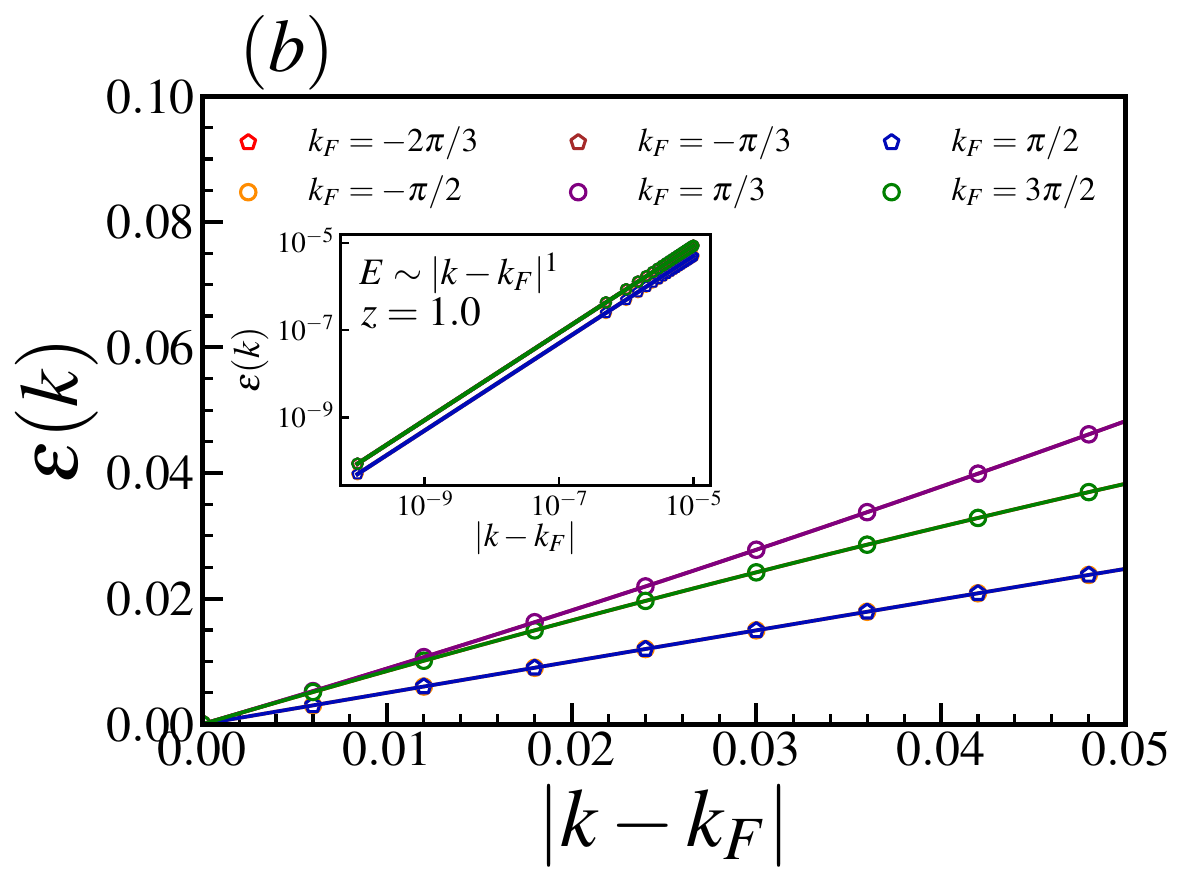} 
		
	}

	\caption{(a) Second derivative of the ground-state energy with respect to $\alpha$, and  (b) Low-energy spectrum along the line $\beta_{c_3}$ for all six Fermi points.}
	\label{fig_A1_4}
	
\end{figure}

A more subtle feature appears at the intersection of the two critical lines, 
$\alpha=2$ and $\beta=1$. Figure~\ref{fig_A1_4}~(a) shows the second derivative of the ground-state energy, which exhibits a clear discontinuity at this point, signaling a quantum phase transition. Simultaneously, the topology of the gapless behavior changes: approaching the intersection along the $\beta_{c_3}$ line, the number of Fermi points increases from four to six at $(2,1)$ and returns to four after crossing the multicritical point. Since the low-energy dispersion remains linear throughout this reconstruction, the dynamical exponent stays fixed at $z=1$. The transition is therefore driven by a change in the topology of the Fermi surface rather than by a modification of the critical scaling. This distinguishes it from the conventional one-dimensional Lifshitz transition discussed in the main text and identifies it as an unconventional Lifshitz reconstruction occurring at the multicritical point. The corresponding numerical dispersion is shown in Fig.~\ref{fig_A1_4}, where the linear behavior around every Fermi point is clearly visible. The agreement between the analytical expansion and the numerical spectrum establishes that the low-energy excitations remain massless Dirac fermions with a dynamical exponent $z=1$. Unlike the Lifshitz transition discussed along the $\beta=1$ critical line, the linear character of the quasiparticle spectrum is preserved throughout the $\beta_{c_3}$ line.

\subsection{Low-energy spectrum at the Lifshitz critical field}
\label{app:betc1h}
Along the isotropic line $\beta=1$, the quasiparticle spectrum is
\begin{equation}
	\varepsilon(k)
	=
	\cos k
	-h
	+\frac{\alpha}{4}\cos3k .
\end{equation}
The Fermi points satisfy
\begin{equation}
	\alpha x^3+
	\left(
	1-\frac{3\alpha}{4}
	\right)x-h=0,
	\qquad
	x=\cos k.
	\label{eq:cubic_lif}
\end{equation}
To illustrate the low-energy structure shown in
Fig.~\ref{fig6}~(a), we consider the representative case
$\alpha=3$.
At the Lifshitz critical field
\begin{equation}
	h=h_{c_L}^{(\mathrm{sp})}
	=\frac{5\sqrt5}{36},
\end{equation}
Eq.~(\ref{eq:cubic_lif}) becomes
\begin{equation}
	3x^3-\frac54x-\frac{5\sqrt5}{36}=0,
\end{equation}
whose physical solutions consist of one double root
\begin{equation}
	x_1=-\frac{\sqrt5}{6},
\end{equation}
and one simple root
\begin{equation}
	x_2=\frac{\sqrt5}{3}.
\end{equation}
The corresponding Fermi points are therefore
\begin{align}
	k_{F}^{(1)}
	&=
	\pm
	\arccos
	\left(
	-\frac{\sqrt5}{6}
	\right),
	\\
	k_{F}^{(2)}
	&=
	\pm
	\arccos
	\left(
	\frac{\sqrt5}{3}
	\right).
\end{align}

We first consider the pair of Fermi points associated with the double root,
\begin{equation}
	k_F^{(1)}
	=
	\pm
	\arccos
	\left(
	-\frac{\sqrt5}{6}
	\right),
\end{equation}
and expand the spectrum around
\begin{equation}
	k
	=
	k_F^{(1)}+q,
	\qquad
	|q|\ll1.
\end{equation}
Substituting the Fermi-point condition into the dispersion immediately gives
\begin{equation}
	\varepsilon(k_F^{(1)})=0.
\end{equation}
Furthermore, because these momenta originate from the double root of the cubic equation, the first derivative also vanishes,
\begin{equation}
	\left.
	\frac{\partial\varepsilon(k)}
	{\partial k}
	\right|_{k=k_F^{(1)}}
	=0,
\end{equation}
whereas the second derivative remains finite,
\begin{equation}
	\left.
	\frac{\partial^2\varepsilon(k)}
	{\partial k^2}
	\right|_{k=k_F^{(1)}}
	=
	-\frac{3\sqrt5}{4}.
\end{equation}
The Taylor expansion of the spectrum therefore becomes
\begin{equation}
	\varepsilon(k_F^{(1)}+q)
	=
	\frac12
	\left.
	\frac{\partial^2\varepsilon}
	{\partial k^2}
	\right|_{k_F^{(1)}}
	q^2
	+
	O(q^3)
	=
	-\frac{3\sqrt5}{8}q^2
	+
	O(q^3),
\end{equation}
which immediately implies
\begin{equation}
	|\varepsilon(k)|
	\propto
	|q|^2.
\end{equation}
Hence, this pair of symmetry-related Fermi points exhibits a quadratic low-energy dispersion characterized by the dynamical critical exponent $z=2$.

We next consider the second pair of Fermi points,
\begin{equation}
	k_F^{(2)}
	=
	\pm
	\arccos
	\left(
	\frac{\sqrt5}{3}
	\right),
\end{equation}
and again write
\begin{equation}
	k=k_F^{(2)}+q,
	\qquad
	|q|\ll1.
\end{equation}
At these momenta,
\begin{equation}
	\varepsilon(k_F^{(2)})=0,
\end{equation}
while the first derivative remains finite,
\begin{equation}
	\left.
	\frac{\partial\varepsilon(k)}
	{\partial k}
	\right|_{k=k_F^{(2)}}
	=v_F\neq0.
\end{equation}
The Taylor expansion is therefore
\begin{equation}
	\varepsilon(k_F^{(2)}+q)
	=
	v_F q
	+
	O(q^2),
\end{equation}
so that
\begin{equation}
	|\varepsilon(k)|
	\propto
	|q|.
\end{equation}
This is the conventional linear Dirac dispersion with dynamical critical exponent $z=1$.

Therefore, the Lifshitz critical field contains two distinct types of low-energy excitations. The Fermi points
\begin{equation}
	k_F
	=
	\pm
	\arccos
	\left(
	-\frac{\sqrt5}{6}
	\right)
\end{equation}
support quadratic quasiparticles with $z=2$, whereas
\begin{equation}
	k_F
	=
	\pm
	\arccos
	\left(
	\frac{\sqrt5}{3}
	\right)
\end{equation}
remain linearly dispersing Dirac points with $z=1$. The coexistence of these two types of critical excitations constitutes the defining feature of the Lifshitz multicritical point discussed in the main text.

\begin{figure}[h]
	\centerline{\includegraphics[width=0.5\linewidth,height=0.42\linewidth]{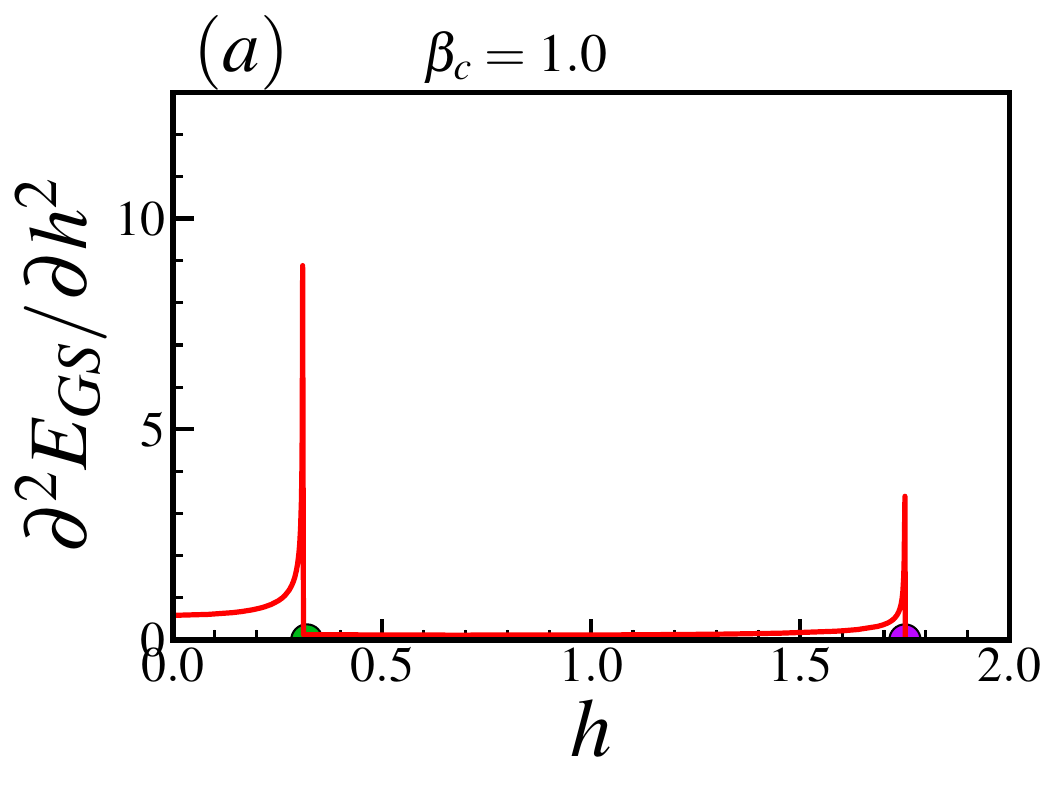} \includegraphics[width=0.5\linewidth,height=0.42\linewidth]{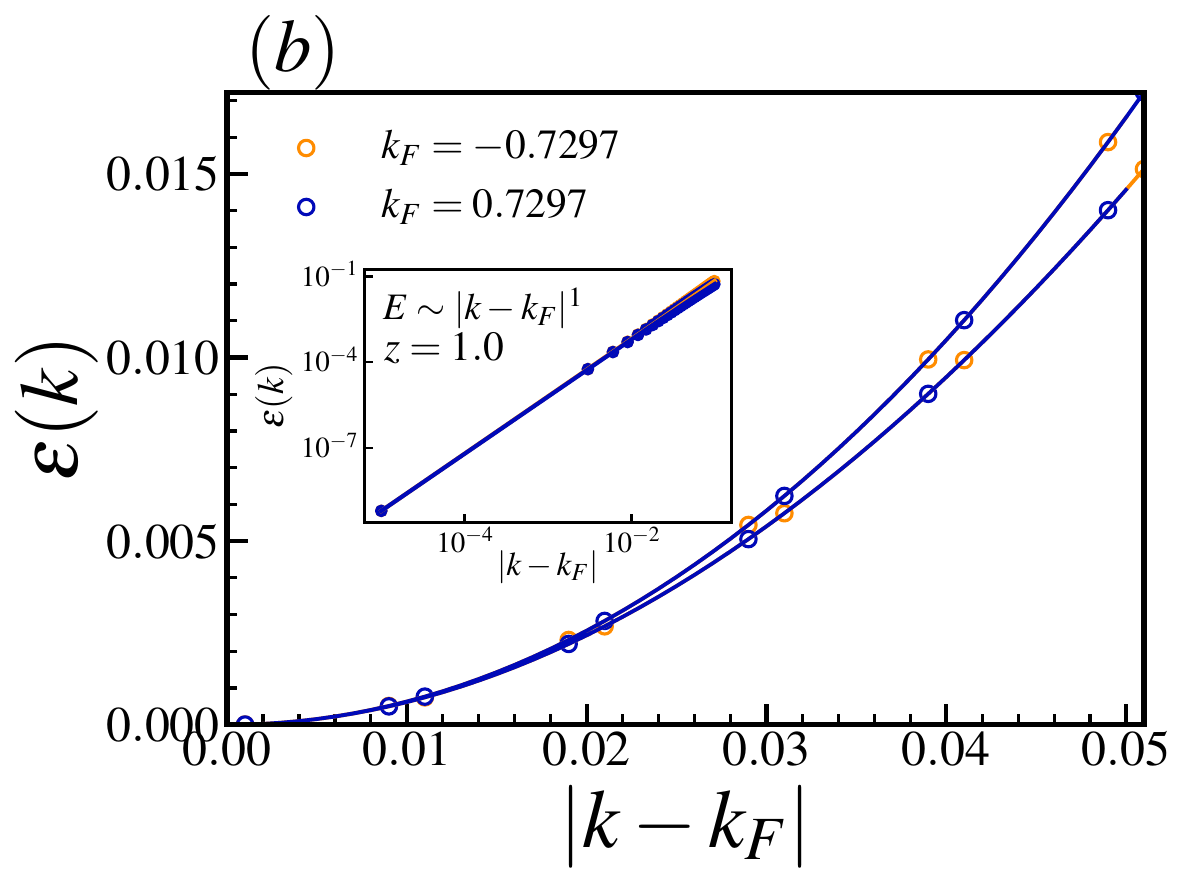} 
		
	}
	
	\centerline{\includegraphics[width=0.5\linewidth,height=0.42\linewidth]{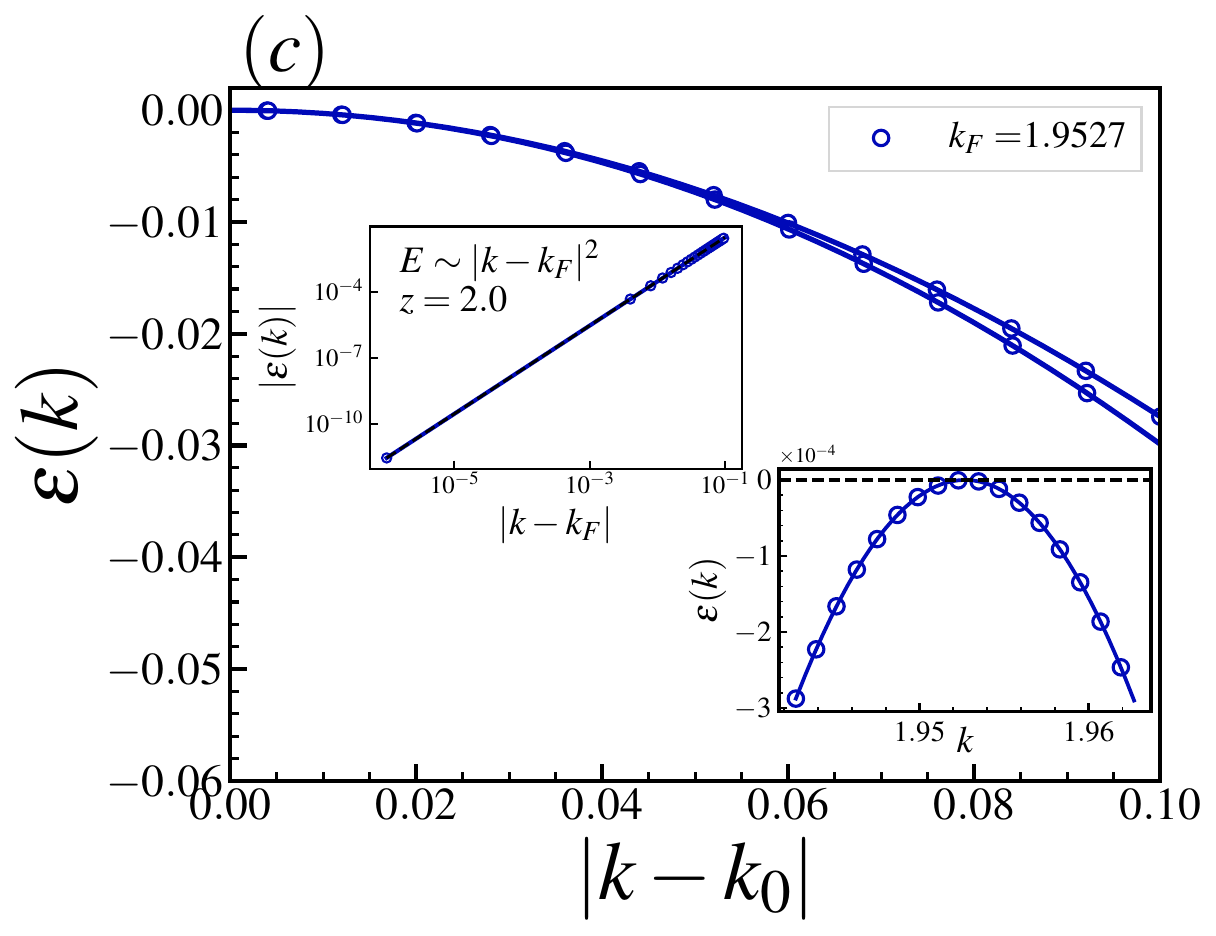} \includegraphics[width=0.5\linewidth,height=0.42\linewidth]{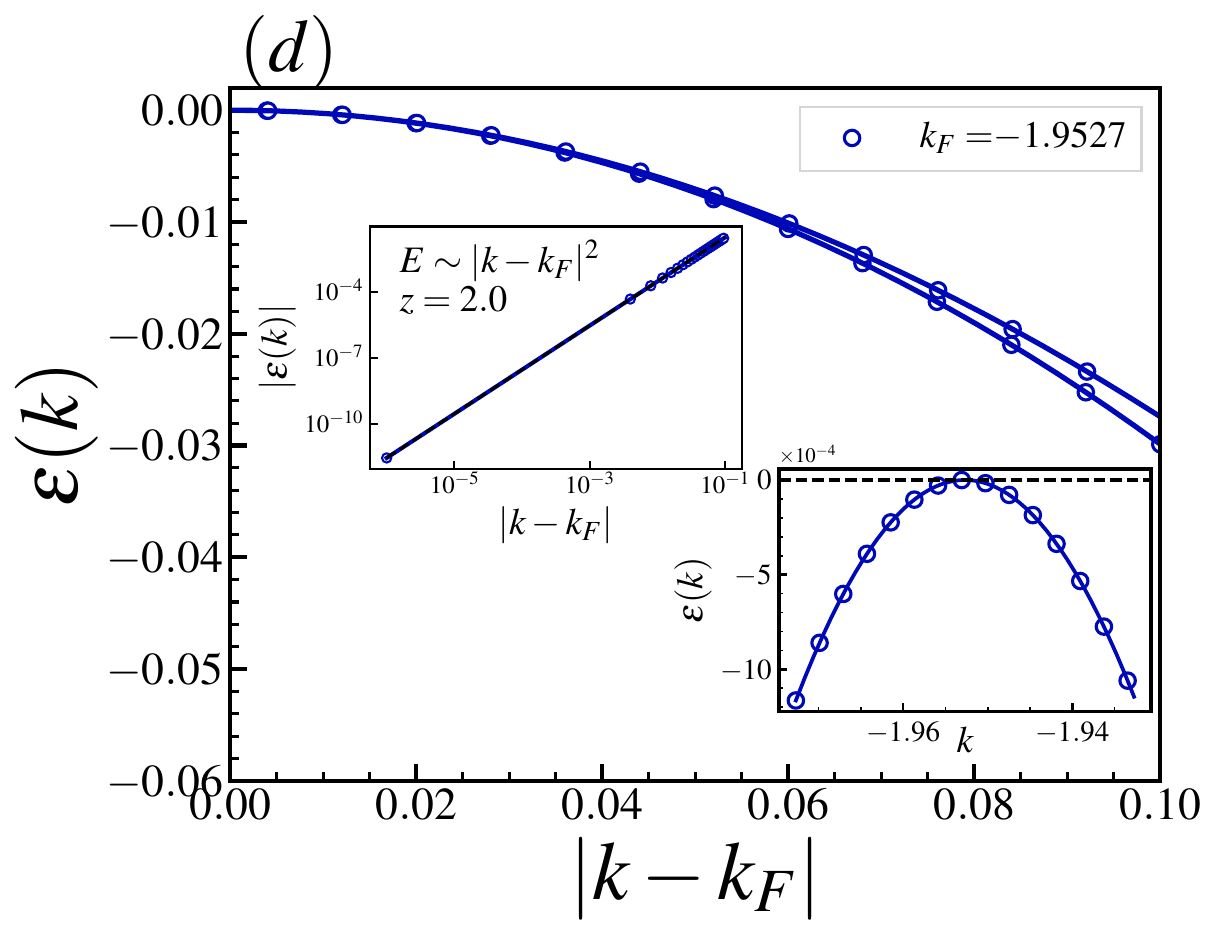} 
		
	}

	\caption{(a) Second derivative of the ground-state energy with respect to $h$. (b) Low-energy spectrum along the isotropic line $\beta=1$ near the critical momenta $k_F^{(2)}$. (c,d) Low-energy spectrum along the isotropic line $\beta=1$ near the critical momenta $k_F^{(1)}$.}
	\label{fig_A1_5}
	
\end{figure}
Figure~\ref{fig_A1_5}~(a) shows the second derivative of the ground-state energy with respect to the transverse field. Pronounced nonanalyticities appear at both critical fields, $h_{c_L}^{(\mathrm{sp})}$ and $h_{c}^{(\mathrm{sp})}$, confirming the continuous quantum phase transitions obtained from the Fermi-point analysis.
\\
Figure~\ref{fig_A1_5}~(b) illustrates the evolution of the Fermi-point structure along the isotropic line. At the Lifshitz critical field $h_{c_L}^{(\mathrm{sp})}$, two Fermi points merge to form quadratic band-touching points, while the remaining pair remains linearly dispersing. Upon further increasing the field, the quadratic nodes disappear, leaving a single pair of Fermi points that is ultimately annihilated at $h_{c}^{(\mathrm{sp})}$, where the system enters the gapped phase.

The corresponding low-energy scaling is examined in Figs.~\ref{fig_A1_5}~(c) and \ref{fig_A1_5}~(d), where the dispersion is plotted as a function of $|k-k_F|$ around the two quadratic Fermi points. The numerical results are accurately described by $\varepsilon(k)\propto |k-k_F|^2$, demonstrating the dynamical exponent $z=2$. This quadratic band touching is the defining characteristic of the conventional Lifshitz critical point on the isotropic line and contrasts with the unconventional multicritical point discussed in the main text, where all gapless excitations remain linearly dispersing with $z=1$.

\subsection{Low-energy theory along the field-driven critical lines $h_{c_1}$ and $h_{c_2}$}
\label{app:hc12}

In this appendix we derive the low-energy continuum description along the field-driven critical lines $h_{c_1}$ and $h_{c_2}$. Throughout this section the quasiparticle spectrum is
\begin{equation}
	\varepsilon(k)=\sqrt{\mathcal A_k^2+\mathcal B_k^2},
\end{equation}
with

\begin{align}
	\mathcal A_k
	&=
	\cos k-h+\frac{\alpha}{8}(1+\beta)\cos3k,\\
	\mathcal B_k
	&=
	\frac{\alpha}{8}(1-\beta)\sin3k.
\end{align}
The gap closes at different high-symmetry momenta on the two critical branches. Expanding the spectrum around the corresponding Fermi points shows that the low-energy quasiparticles remain linearly dispersing, implying a dynamical exponent $z=1$ throughout both critical lines.
\paragraph*{Critical line $h_{c_1}$.}
\mbox{}\\
Along
\begin{equation}
	h_{c_1}
	=
	-1-\frac{\alpha}{8}(1+\beta),
\end{equation}
the quasiparticle gap closes at the Brillouin-zone boundary, $k_F=\pm\pi$. Expanding around the Fermi point by writing
\begin{equation}
	k=\pi+q,\qquad |q|\ll1,
\end{equation}
one obtains
\begin{align}
	\mathcal A_{\pi+q}
	&=
	\left[
	\frac{1}{2}
	+\frac{9\alpha}{16}(1+\beta)
	\right]q^2
	+O(q^4),\\
	\mathcal B_{\pi+q}
	&=
	-\frac{3\alpha}{8}(1-\beta)q
	+O(q^3).
\end{align}
The quasiparticle spectrum therefore becomes
\begin{equation}
	\varepsilon^2(\pi+q)
	=
	\left[
	\frac{1}{2}
	+\frac{9\alpha}{16}(1+\beta)
	\right]^2q^4
	+
	\left[
	\frac{3\alpha}{8}(1-\beta)
	\right]^2q^2
	+O(q^5).
\end{equation}
Since the linear contribution originating from $\mathcal B_k$ dominates for $q\rightarrow0$ ($\beta\neq1$), the leading-order dispersion is
\begin{equation}
	\varepsilon(\pi+q)
	\simeq
	\frac{3\alpha}{8}|1-\beta|\,|q|.
\end{equation}
An identical expansion around the symmetry-related point $k=-\pi+q$ gives the same result owing to the even symmetry $\varepsilon(-k)=\varepsilon(k)$. Consequently,
\begin{equation}
	\varepsilon(k)
	\simeq
	v_F|k-k_F|,
	\qquad
	k_F=\pm\pi,
\end{equation}
with
\begin{equation}
	v_F=\frac{3\alpha}{8}|1-\beta|.
\end{equation}
Thus, the critical excitations along the $h_{c_1}$ branch are Dirac-like with dynamical exponent $z=1$.
\\
\paragraph*{Critical line $h_{c_2}$.}
\mbox{}\\

Along
\begin{equation}
	h_{c_2} = 1+\frac{\alpha}{8}(1+\beta),
\end{equation}
the excitation gap closes at $k_F=0$. Writing
\begin{equation}
	k=q,\qquad |q|\ll1,
\end{equation}
and expanding the trigonometric functions gives
\begin{align}
	\mathcal A_q
	&=
	-
	\left[
	\frac{1}{2}
	+\frac{9\alpha}{16}(1+\beta)
	\right]q^2
	+O(q^4),\\
	\mathcal B_q
	&=
	\frac{3\alpha}{8}(1-\beta)q
	+O(q^3).
\end{align}
Accordingly,
\begin{equation}
	\varepsilon^2(q)
	=
	\left[
	\frac{1}{2}
	+\frac{9\alpha}{16}(1+\beta)
	\right]^2q^4
	+
	\left[
	\frac{3\alpha}{8}(1-\beta)
	\right]^2q^2
	+O(q^6),
\end{equation}
which yields the leading-order dispersion
\begin{equation}
	\varepsilon(k)
	\simeq
	\frac{3\alpha}{8}|1-\beta|\,|k|,
	\qquad
	k\simeq0.
\end{equation}
Therefore, the quasiparticle spectrum along the $h_{c_2}$ branch is likewise linear,
\begin{equation}
	\varepsilon(k)
	\simeq
	v_F|k-k_F|,
\end{equation}
with
\begin{equation}
	k_F=0,
	\qquad
	v_F=\frac{3\alpha}{8}|1-\beta|,
\end{equation}
demonstrating that the critical excitations remain Dirac-like with dynamical exponent $z=1$. Only in the isotropic limit $\beta=1$ does the linear term vanish, giving rise to the quadratic dispersion discussed separately in Appendix~\ref{app:betc1h}.

\begin{figure}[h]
	\centerline{\includegraphics[width=0.5\linewidth,height=0.42\linewidth]{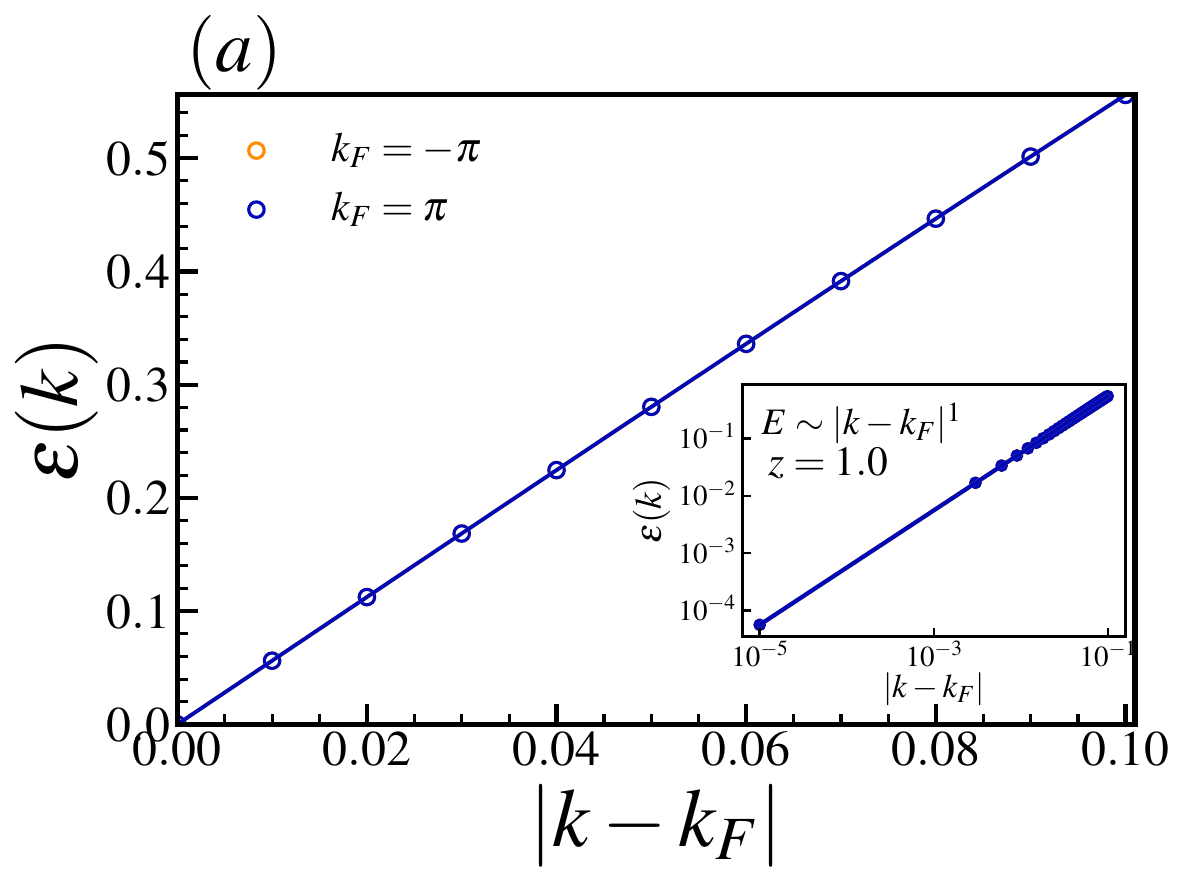} \includegraphics[width=0.5\linewidth,height=0.42\linewidth]{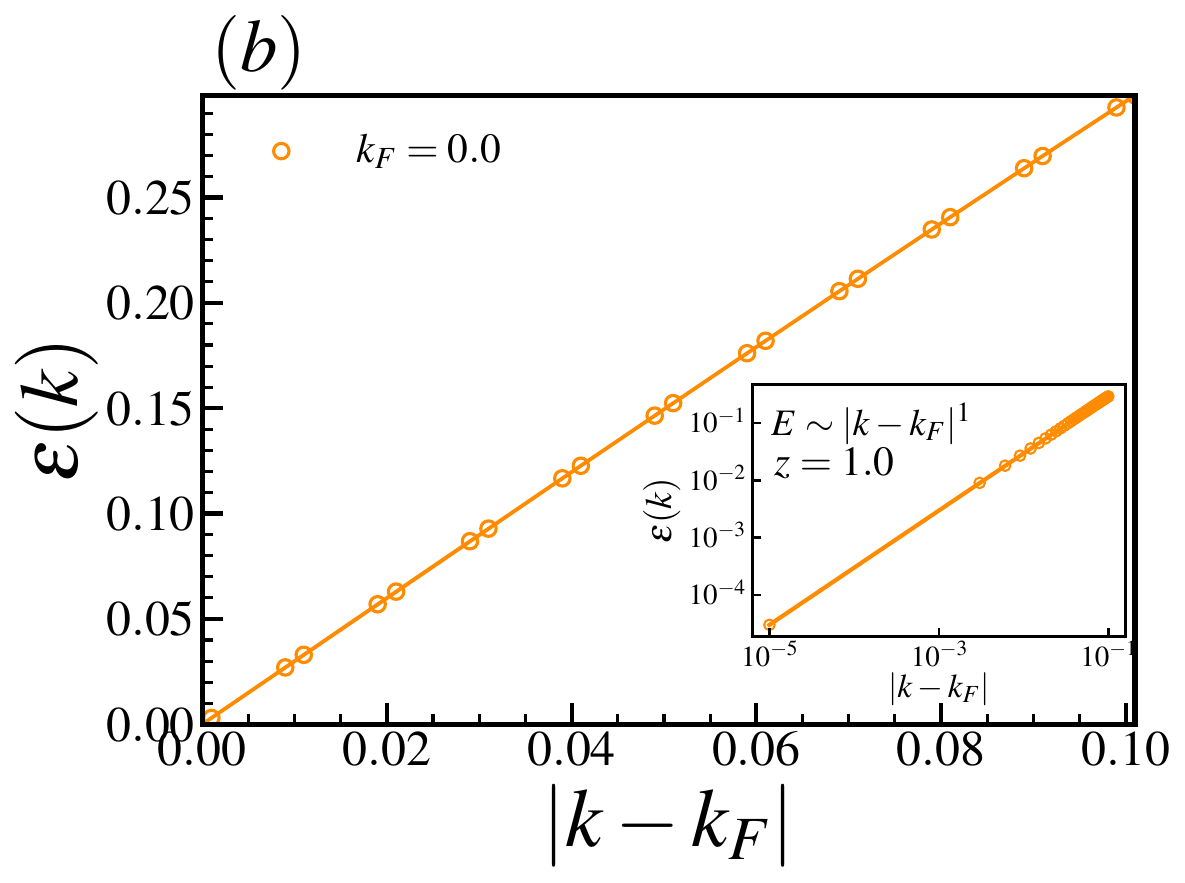} 
		
	}

	\caption{Low-energy spectrum (a) along the line $h_{c_1}$ near the critical momenta $k_F=\pm\pi$ and (b) (a) along the line $h_{c_2}$ near the critical momenta $k_F=0$.}
	\label{fig_A1_6}
	
\end{figure}

Figure~\ref{fig_A1_6}~(a) shows the quasiparticle spectrum in the vicinity of the two symmetry-related Fermi points $k_F=\pm\pi$ along the $h_{c_1}$ critical branch. The numerical spectrum exhibits a linear dispersion near both nodes, in excellent agreement with the continuum expansion derived above. The logarithmic plot in the inset confirms the scaling relation $\varepsilon(k)\propto|k-k_F|$, establishing the Dirac nature of the low-energy excitations.  Figure~\ref{fig_A1_6}~(b) presents the corresponding analysis for the $h_{c_2}$ critical branch. In this case, the spectrum becomes gapless only at the zone center, $k_F=0$, where the quasiparticle dispersion is likewise linear. The inset demonstrates the expected scaling $\varepsilon(k)\propto|k|$, confirming that the critical mode is also characterized by the dynamical exponent $z=1$, consistent with the analytical continuum theory.

\subsection{Low-energy expansions along the critical lines $h_{c_3}$ and $h_{c_4}$}
\label{app:hc34}

In this appendix, we derive the low-energy quasiparticle dispersions along the critical lines $h_{c_3}$ and $h_{c_4}$. The quasiparticle spectrum is given by
\begin{equation}
	\varepsilon(k)=\sqrt{\mathcal A_k^2+\mathcal B_k^2},
\end{equation}
with
\begin{align}
	\mathcal A_k
	&=
	\cos k-h+\frac{\alpha}{8}(1+\beta)\cos3k,\nonumber \\
	\mathcal B_k
	&=
	\frac{\alpha}{8}(1-\beta)\sin3k.
\end{align}
In both cases, the gap closes at a pair of symmetry-related finite-momentum Fermi points. Expanding the spectrum about these momenta shows that the low-energy excitations remain linearly dispersing, corresponding to a dynamical exponent $z=1$.
\\
\paragraph*{Critical line $h_{c_3}$.}
\mbox{}\

Along
\begin{equation}
	h_{c_3}
	=
	\frac{1}{2}
	-
	\frac{\alpha}{8}(1+\beta),
	\qquad
	\beta<1-\frac{4}{\alpha},
\end{equation}
the gap closes at $	k_F=\pm\pi/3$. Writing
\begin{equation}
	k=\pm\frac{\pi}{3}+q,
	\qquad
	|q|\ll1,
\end{equation}
the low-energy expansion of the two components of the Bogoliubov spectrum takes the form
\begin{align}
	\mathcal A_{\pm\pi/3+q}
	&=
	\mp\frac{\sqrt{3}}{2}q
	+
	\left[
	-\frac14
	+
	\frac{9\alpha}{16}(1+\beta)
	\right]q^2
	+
	O(q^3), \nonumber \\
	\mathcal B_{\pm\pi/3+q}
	&=
	-\frac{3\alpha}{8}(1-\beta)q
	+
	O(q^3).
\end{align}
Consequently, the leading low-energy dispersion is
\begin{equation}
	\varepsilon\left(\pm\frac{\pi}{3}+q\right)
	=
	v_F|q|
	+
	O(q^2),
\end{equation}
where
\begin{equation}
	v_F
	=
	\sqrt{
		\frac34
		+
		\frac{9\alpha^2}{64}(1-\beta)^2
	}.
\end{equation}
Equivalently,
\begin{equation}
	\varepsilon(k)
	\simeq
	v_F
	\left|k-k_F\right|,
	\qquad
	k_F=\pm\frac{\pi}{3}.
\end{equation}
Thus, both finite-momentum gapless modes on the $h_{c_3}$ branch exhibit linear dispersion and are characterized by the dynamical exponent $z=1$.

\begin{figure}[h]
	\centerline{\includegraphics[width=0.8\linewidth,height=0.6\linewidth]{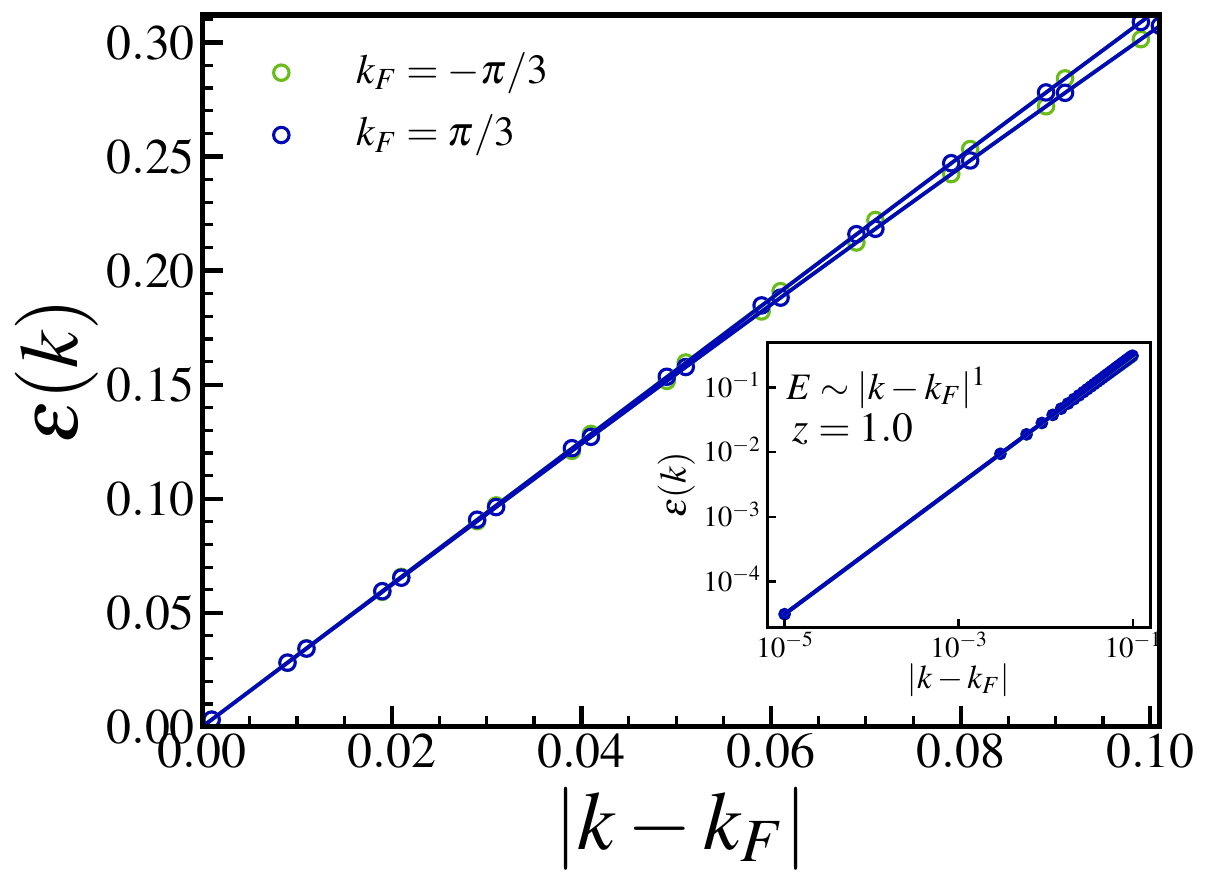} 
		
	}

	\caption{Low-energy spectrum along the line $h_{c_3}$ near the critical momenta $k_F=\pm\pi/3$.}
	\label{fig_A1_7}
	
\end{figure}

Figure~\ref{fig_A1_7} shows the quasiparticle spectrum in the vicinity of the Fermi points $k_F=\pm\pi/3$. The numerical results confirm the analytical low-energy expansion, exhibiting a linear dispersion about both gapless momenta. The inset, plotted on logarithmic scales, verifies the scaling $\varepsilon(k)\propto|k-k_F|$, consistent with a dynamical exponent $z=1$ and linearly dispersing Dirac quasiparticles.

\paragraph*{Critical line $h_{c_4}$.}
\mbox{}\
\\
Along
\begin{equation}
	h_{c_4}
	=
	-\frac{1}{2}
	+
	\frac{\alpha}{8}(1+\beta),
	\qquad
	\beta>\frac{4}{\alpha}-1,
\end{equation}
the gap closes at $k_F=\pm 2\pi/3$. Expanding about the two Fermi points according to
\begin{equation}
	k=\pm\frac{2\pi}{3}+q,
	\qquad
	|q|\ll1,
\end{equation}
gives
\begin{align}
	\mathcal A_{\pm2\pi/3+q}
	&=
	\mp\frac{\sqrt{3}}{2}q
	+
	\left[
	\frac14
	-
	\frac{9\alpha}{16}(1+\beta)
	\right]q^2
	+
	O(q^3),
	\nonumber \\
	\mathcal B_{\pm2\pi/3+q}
	&=
	\frac{3\alpha}{8}(1-\beta)q
	+
	O(q^3).
\end{align}
The corresponding quasiparticle dispersion is therefore
\begin{equation}
	\varepsilon\left(\pm\frac{2\pi}{3}+q\right)
	=
	v_F|q|
	+
	O(q^2),
\end{equation}
with
\begin{equation}
	v_F
	=
	\sqrt{
		\frac34
		+
		\frac{9\alpha^2}{64}(1-\beta)^2
	}.
\end{equation}
Hence,
\begin{equation}
	\varepsilon(k)
	\simeq
	v_F
	\left|k-k_F\right|,
	\qquad
	k_F=\pm\frac{2\pi}{3},
\end{equation}
and the low-energy excitations remain linearly dispersing with $z=1$. The expansions above establish that the critical branches $h_{c_3}$ and $h_{c_4}$ are characterized by stable linearly dispersing finite-momentum gapless modes. The corresponding low-energy expansions provide the analytical basis for the $z=1$ scaling observed numerically in the quasiparticle spectra. In contrast to the $h_{c_3}$ branch, the $h_{c_4}$ line exhibits an additional special reconstruction at its intersection with the isotropic line $\beta=1$, where the number of Fermi points changes from two to six and subsequently returns to two. The associated singularity in the ground-state energy and the corresponding change in Fermi-point topology constitute the unconventional Lifshitz transition.

\begin{figure}[h]
	\centerline{\includegraphics[width=0.5\linewidth,height=0.42\linewidth]{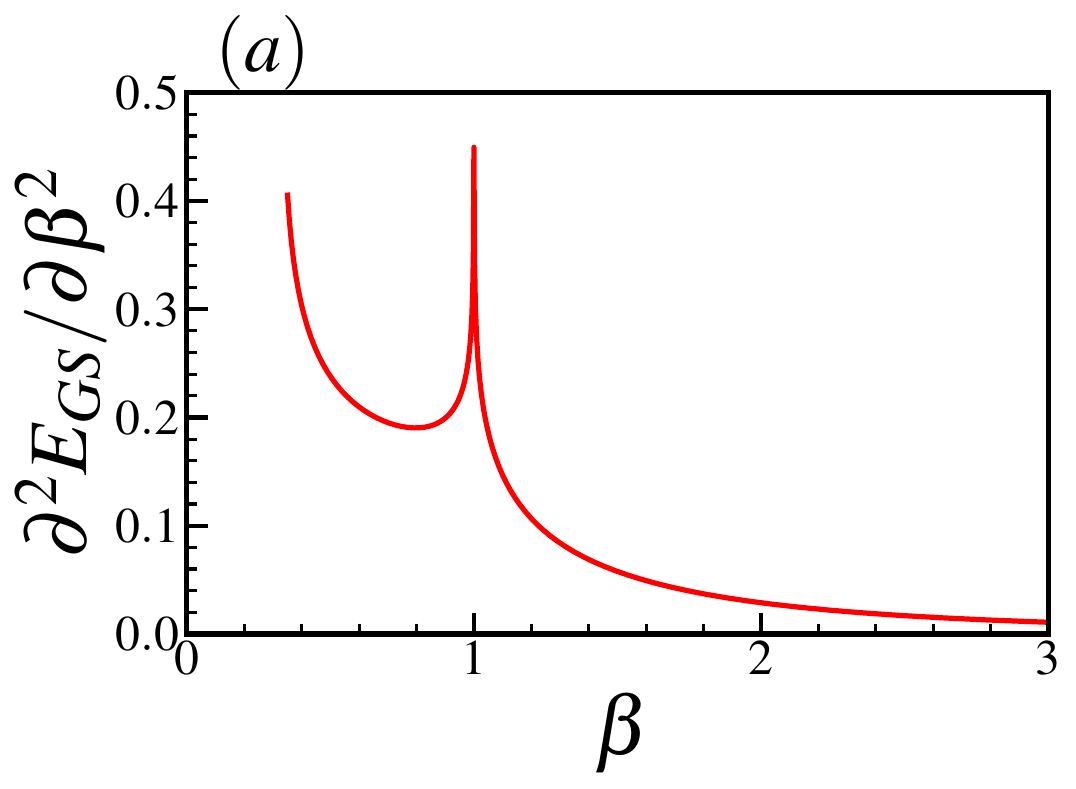} \includegraphics[width=0.5\linewidth,height=0.42\linewidth]{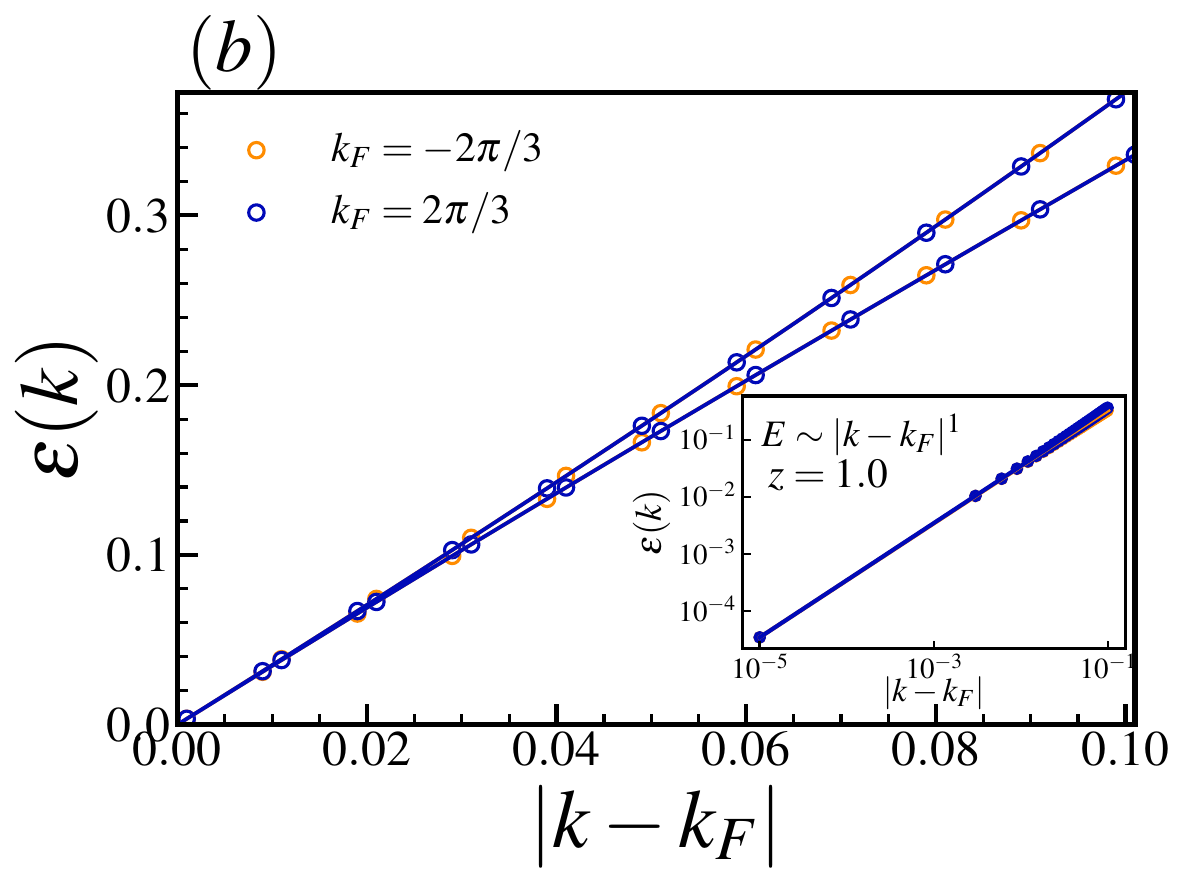} 
		
	}

	\caption{(a) Second derivative of the ground-state energy with respect to $h$ and (b) Low-energy spectrum along the line $h_{c_4}$ near the critical momenta $k_F=\pm2\pi/3$}
	\label{fig_A1_8}
	
\end{figure}

A special reconstruction occurs on the $h_{c_4}$ critical line upon reaching the isotropic point $\beta=1$. Figure~\ref{fig_A1_8}~(a) shows the second derivative of the ground-state energy, which develops a pronounced peak precisely at $\beta=1$, signaling a nonanalytic change in the ground-state properties. At the same point, the topology of the gapless spectrum changes: the number of Fermi points increases from two to six upon crossing the intersection of the $h_{c_4}$ line with $\beta=1$, before returning to two beyond the reconstruction. Because the quasiparticle dispersion remains linear near the gapless momenta, the dynamical exponent remains $z=1$ throughout. This transition is therefore driven by a reconstruction of the Fermi-point topology rather than by a change in the low-energy scaling, identifying the intersection as an unconventional special Lifshitz point. Figure~\ref{fig_A1_8}~(b) further confirms the linear, Dirac-like dispersion in the vicinity of the Fermi points $k_F=\pm2\pi/3$, in agreement with the analytical low-energy expansion.

\subsection{Ground-State Energy Curvature at the Multicritical Intersection Point}
\label{app:MIP}

\begin{figure}[h]
	\centerline{\includegraphics[width=0.8\linewidth,height=0.6\linewidth]{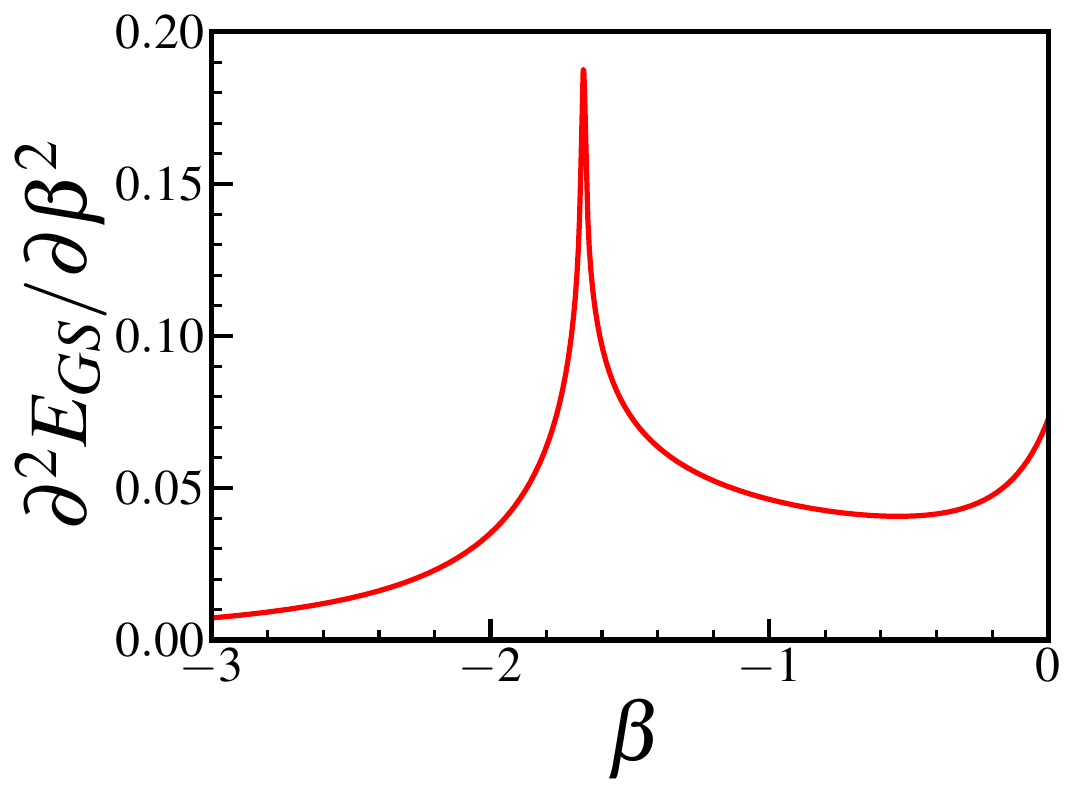} 
		
	}

	\caption{Second derivative of the ground-state energy with respect to $\beta$ along the isotropic line $h_{c_4}$.}
	\label{fig_A1_9}
	
\end{figure}

Figure~\ref{fig_A1_9} shows the second derivative of the ground-state energy with respect to the anisotropy parameter $\beta$ in the vicinity of the intersection point $\beta=\beta_m$. A pronounced singular peak develops precisely at $\beta_m$, signaling a nonanalytic change in the ground-state energy. This thermodynamic singularity coincides with the momentum-space reconstruction discussed in the main text, where the number of gapless Fermi points changes at the intersection of the $h_{c_2}$ and $h_{c_3}$ critical branches. The coincidence of the spectral reconstruction and the singular behavior of the ground-state energy demonstrates that the intersection is a genuine quantum multicritical point rather than a simple geometric crossing of two critical lines. Unlike conventional field-driven critical points, the singularity originates from a reconstruction of the Fermi-point topology while the gapless quasiparticle excitations remain linearly dispersing on both sides of the transition. The second derivative of the ground-state energy therefore provides an independent thermodynamic signature of the unconventional Lifshitz multicritical point identified in the main text.

\section{Correlation-matrix formulation of entanglement measures}
\label{app:entanglement}

Since the fermionic representation of the Hamiltonian is quadratic, the ground state is Gaussian and is therefore completely characterized by two-point correlation functions. Introducing the Nambu basis
\begin{equation}
	\Psi^\dagger=
	(a_1^\dagger,\ldots,a_N^\dagger,
	a_1,\ldots,a_N),
\end{equation}
the full many-body correlation matrix is defined as
\begin{equation}
	\label{eq:GN}
	\mathcal{G}_N=
	\left(
	\begin{array}{cccccc}
		\langle a_{1}^{\dagger}a_{1}\rangle &
		\cdots &
		\langle a_{1}^{\dagger}a_{N}\rangle &
		\langle a_{1}^{\dagger}a_{1}^{\dagger}\rangle &
		\cdots &
		\langle a_{1}^{\dagger}a_{N}^{\dagger}\rangle
		\\
		\langle a_{2}^{\dagger}a_{1}\rangle &
		\cdots &
		\langle a_{2}^{\dagger}a_{N}\rangle &
		\langle a_{2}^{\dagger}a_{1}^{\dagger}\rangle &
		\cdots &
		\langle a_{2}^{\dagger}a_{N}^{\dagger}\rangle
		\\
		\vdots & \vdots & \vdots &
		\vdots & \vdots & \vdots
		\\
		\langle a_{N}^{\dagger}a_{1}\rangle &
		\cdots &
		\langle a_{N}^{\dagger}a_{N}\rangle &
		\langle a_{N}^{\dagger}a_{1}^{\dagger}\rangle &
		\cdots &
		\langle a_{N}^{\dagger}a_{N}^{\dagger}\rangle
		\\
		\langle a_{1}a_{1}\rangle &
		\cdots &
		\langle a_{1}a_{N}\rangle &
		\langle a_{1}a_{1}^{\dagger}\rangle &
		\cdots &
		\langle a_{1}a_{N}^{\dagger}\rangle
		\\
		\langle a_{2}a_{1}\rangle &
		\cdots &
		\langle a_{2}a_{N}\rangle &
		\langle a_{2}a_{1}^{\dagger}\rangle &
		\cdots &
		\langle a_{2}a_{N}^{\dagger}\rangle
		\\
		\vdots & \vdots & \vdots &
		\vdots & \vdots & \vdots
		\\
		\langle a_{N}a_{1}\rangle &
		\cdots &
		\langle a_{N}a_{N}\rangle &
		\langle a_{N}a_{1}^{\dagger}\rangle &
		\cdots &
		\langle a_{N}a_{N}^{\dagger}\rangle
	\end{array}
	\right).
\end{equation}
For a subsystem $A$ consisting of $l$ sites, the reduced correlation matrix $C_A$ is obtained by restricting Eq.~(\ref{eq:GN}) to the corresponding degrees of freedom. The reduced density matrix is
\begin{equation}
	\rho_A=
	{\rm Tr}_B
	\left(
	|\Psi_0\rangle\langle\Psi_0|
	\right),
\end{equation}
where $|\Psi_0\rangle$ denotes the many-body ground state and $B$ is the complementary subsystem. Since Gaussianity is preserved under partial tracing, $\rho_A$ remains Gaussian and is completely determined by the eigenvalues $\{\zeta_i\}$ of $C_A$.

The bipartite von Neumann entanglement entropy is
\begin{equation}
	S_A=-{\rm Tr}\left(\rho_A\log_2\rho_A\right),
\end{equation}
which, in terms of the correlation-matrix eigenvalues, becomes
\begin{equation}
	S_A=-\sum_i\left[\zeta_i\log_2\zeta_i+(1-\zeta_i)\log_2(1-\zeta_i)\right].
	\label{eq:SEE}
\end{equation}

For a critical one-dimensional system with periodic boundary conditions, conformal field theory predicts the Calabrese--Cardy scaling form
\begin{equation}
	S(l)	=	\frac{c}{3}	\log_2	\left[	\frac{N}{\pi}	\sin	\left(	\frac{\pi l}{N} 	\right) \right] +s_0 ,
	\label{eq:CC}
\end{equation}
where $c$ is the central charge and $s_0$ is a nonuniversal constant. Throughout this work, the effective central charge $c_{\rm eff}$ is obtained by fitting the numerical entropy to Eq.~(\ref{eq:CC}). For half-chain bipartitions, Eq.~(\ref{eq:CC}) reduces to
\begin{equation}
	S(N/2) = \frac{c_{\rm eff}}{3}\log_2 N+ s_0,
\end{equation}
The reduced density matrix can be formally expressed as
\begin{equation}
	\rho_A=e^{-H_E},
\end{equation}
where $H_E$ denotes the entanglement Hamiltonian. For free-fermion Gaussian states,
\begin{equation}
	H_E= \sum_i \xi_i f_i^\dagger f_i,
\end{equation}
with $\xi_i$ the single-particle entanglement energies. Their relation to the eigenvalues of the correlation matrix is
\begin{equation}
	\zeta_i= \frac{1}{e^{\xi_i}+1},
\end{equation}
which yields
\begin{equation}
	\xi_i= \ln \left( \frac{1-\zeta_i}{\zeta_i} \right).
	\label{eq:ES}
\end{equation}
\\
The set $\{\xi_i\}$ constitutes the real-space entanglement spectrum.
\\
To characterize momentum-space correlations, we partition the Hilbert space into positive- and negative-momentum sectors,
\begin{equation}
	\mathcal H = \mathcal H_{k>0} \otimes \mathcal H_{k<0},
\end{equation}
and trace out one sector,
\begin{equation}
	\rho_{k>0} = {\rm Tr}_{k<0} \left( |\Psi_0\rangle \langle\Psi_0| \right).
\end{equation}
\\
The resulting momentum-space reduced density matrix is likewise Gaussian and is completely specified by the corresponding momentum-space correlation matrix. Denoting its eigenvalues by $\zeta_i^{(k)}$, the momentum-space entanglement energies are

\begin{equation}
	\xi_i^{(k)} = \ln \left( \frac{1-\zeta_i^{(k)}} {\zeta_i^{(k)}} \right).
	\label{eq:kES}
\end{equation}

Equation~(\ref{eq:kES}) has the same functional form as Eq.~(\ref{eq:ES}); the distinction lies solely in the choice of subsystem. Whereas the real-space entanglement spectrum probes spatial quantum correlations, the momentum-space entanglement spectrum measures correlations between quasiparticle modes.

\section{Continuum Bogoliubov--de Gennes Field Theory}
\label{app:continuum}

\subsection{General Continuum Formulation}
\label{app:general_continuum}

We derive the continuum low-energy field theory associated with the microscopic Bogoliubov--de Gennes Hamiltonian introduced in the main text. Our goal is to obtain a general effective description valid near criticality, independent of the particular phase under consideration. The derivation proceeds by expanding the lattice fermion operators around all symmetry-related gapless momenta and retaining only the slowly varying long-wavelength degrees of freedom. The resulting continuum Bogoliubov--de Gennes theory provides the general low-energy description of the model and forms the basis for determining the critical mode content and the corresponding effective central charges in the different parameter regimes considered throughout the paper.

The microscopic lattice Hamiltonian is
\begin{align}
	\mathcal H = \sum_{n} \Big[
	& \frac{J}{2} \left( a_n^\dagger a_{n+1} + a_{n+1}^\dagger a_n \right) - h\,a_n^\dagger a_n \nonumber\\
	&
	+\frac{\alpha(1+\beta)}{16} \left( a_n^\dagger a_{n+3} + a_{n+3}^\dagger a_n \right) \nonumber\\
	&
	+\frac{\alpha(1-\beta)}{16} \left( a_n^\dagger a_{n+3}^\dagger + a_{n+3}a_n \right) \Big].
	\label{eqham}
\end{align}
We assume that the excitation spectrum possesses $N_f$ symmetry-related pairs of gapless momenta $\pm k_i$ ($i=1,\ldots,N_f$). Near criticality, the lattice fermion operator can therefore be decomposed into slowly varying continuum fields  expanded as
\begin{equation}
	a_n = \sum_{i=1}^{N_f} \Big[ \psi_{R,i}(x)e^{ik_i n} + \psi_{L,i}(x)e^{-ik_i n} \Big], \qquad x=na ,
	\label{eqan}
\end{equation}
where $\psi_{R,i}$ and $\psi_{L,i}$ denote right- and left-moving continuum fields associated with the Fermi points $\pm k_i$. For a displacement $\Delta$, the continuum fields are expanded to leading order as
\begin{equation}
	\psi_{R/L,i}(x+\Delta) = \psi_{R/L,i}(x) + \Delta\,\partial_x\psi_{R/L,i}(x) + \mathcal O(\partial_x^2).
	\label{eqpsi}
\end{equation}

Substituting Eq.~(\ref{eqan}) into the hopping and pairing operators generates terms oscillating as $e^{\pm i(k_i-k_j)n}$ and $e^{\pm i(k_i+k_j)n}$. After replacing $\sum_n \rightarrow a^{-1}\int dx$, all rapidly oscillating contributions average to zero in the continuum limit, and only momentum-conserving terms survive. The normal hopping channel therefore reduces to
\begin{align} 
	&
	\sum_n \left( c_n^\dagger c_{n+\Delta} +\mathrm{H.c.} \right) \nonumber \\
	&
	= \sum_i \int dx \Big[ 2\cos(k_i\Delta) \left( \psi_{R,i}^\dagger\psi_{R,i} + \psi_{L,i}^\dagger\psi_{L,i} \right) \nonumber\\
	& + 2i\Delta\sin(k_i\Delta) \left( \psi_{R,i}^\dagger\partial_x\psi_{R,i} -
	\psi_{L,i}^\dagger\partial_x\psi_{L,i} \right) \Big].
	\label{eqcsc}
\end{align}
Similarly, the anomalous pairing channel becomes
\begin{align}
	&
	\sum_n \left( c_n^\dagger c_{n+\Delta}^\dagger + c_{n+\Delta}c_n \right) \nonumber \\
	&
	= \sum_i \int dx \Big[  2i\sin(k_i\Delta) \left( \psi_{R,i}^\dagger\psi_{L,i}^\dagger
	- \psi_{L,i}\psi_{R,i} \right) \nonumber \\
	& + 2\Delta\cos(k_i\Delta) \left( \psi_{R,i}^\dagger \partial_x\psi_{L,i}^\dagger + \psi_{L,i}\partial_x\psi_{R,i} \right) \Big].
	\label{eqcscs}
\end{align}
Evaluating Eqs.~(\ref{eqcsc}) and (\ref{eqcscs}) for the nearest-neighbor hopping ($\Delta=1$) and third-neighbor normal and pairing processes ($\Delta=3$) and collecting all contributions yields

\begin{align}
	H_{\rm eff} = \sum_{i=1}^{N_f} \int dx \Big[
	& M_i \left( \psi_{R,i}^\dagger\psi_{R,i} + \psi_{L,i}^\dagger\psi_{L,i} \right) \nonumber \\
	&
	+i v_{F,i}	\left(	\psi_{R,i}^\dagger\partial_x\psi_{R,i}	-	\psi_{L,i}^\dagger\partial_x\psi_{L,i}	\right)	\nonumber\\
	&
	+i \Delta_i	\left(	\psi_{R,i}^\dagger\psi_{L,i}^\dagger	-	\psi_{L,i}\psi_{R,i} 	\right) 	\nonumber\\
	&
	+\tilde v_i	\left(	\psi_{R,i}^\dagger\partial_x\psi_{L,i}^\dagger	+ 	\psi_{L,i}\partial_x\psi_{R,i} 	\right) 	\Big].
	\label{eqheff}
\end{align}
The continuum couplings are directly determined by the microscopic BdG structure,
\begin{align}
	M_i	&=	\cos k_i -	h +	\frac{\alpha(1+\beta)}{8} \cos(3k_i), \\
	\Delta_i &= \frac{\alpha(1-\beta)}{8} \sin(3k_i), \\
	v_{F,i}
	&= -\left. \frac{d\mathcal A_k}{dk} \right|_{k_i} = \sin k_i	+ \frac{3\alpha(1+\beta)}{8} \sin(3k_i), \\
	\tilde v_i	&=	\left.	\frac{d\mathcal B_k}{dk}	\right|_{k_i} = \frac{3\alpha(1-\beta)}{8} \cos(3k_i),
\end{align}
where
\begin{align}
	\mathcal A_k &= \cos k - h	+ \frac{\alpha(1+\beta)}{8} \cos(3k), \\
	\mathcal B_k &= \frac{\alpha(1-\beta)}{8} \sin(3k).
\end{align}
Equation~(\ref{eqheff}) constitutes the continuum Bogoliubov--de Gennes field theory associated with the microscopic lattice Hamiltonian. The low-energy content of the theory is determined by the set of gapless momenta satisfying $\mathcal A_k=\mathcal B_k=0$, which fixes the number of Fermi-point pairs retained in the continuum expansion of Eq.~(\ref{eqan}). Different critical regimes of the lattice model correspond to different solutions of these conditions and therefore to different realizations of the effective low-energy theory. The analysis presented in the main text follows from appropriate specializations of this general continuum framework.

\subsection{Specialization to the $\beta=1$, $h=0$ Line: Lifshitz Transition and Central-Charge Reconstruction}
\label{app:bet1_continuum}

The general continuum formulation derived in the previous subsection simplifies considerably along the exactly solvable line $\beta=1$ in the absence of an external field $h=0$. In this case the pairing amplitudes vanish identically, $\Delta_i=\tilde v_i=0$, and the lattice Hamiltonian reduces to a purely hopping problem. Consequently, the low-energy theory is completely determined by the structure of the Fermi surface.

The single-particle dispersion is given by
\begin{equation}
	\mathcal A_k = \cos k+\frac{\alpha}{4}\cos(3k),
\end{equation}
and the Fermi points are determined by the condition $	\mathcal A_k=0$.
Using the trigonometric identity $\cos(3k)=4\cos^3k-3\cos k$, Eq.~$\mathcal A_k=0$ can be rewritten as
\begin{equation}
	\cos k 	\left[	1-\frac{3\alpha}{4}	+\alpha\cos^2k	\right]	=0.
	\label{eqroot}
\end{equation}
The solutions of Eq.~(\ref{eqroot}) determine the topology of the Fermi surface and therefore the number of gapless continuum modes.
\\
\paragraph*{Weak-coupling regime: $\alpha<\alpha_c$.} 
\mbox{}\\
\\
For
\begin{equation}
	\alpha<\alpha_c\equiv\frac{4}{3},
\end{equation}
the bracket in Eq.~(\ref{eqroot}) has no real solution. The only Fermi points are therefore located at $k_F=\pm{\pi}/{2}$. Expanding the dispersion around either Fermi point,
\begin{equation}
	k=\pm\frac{\pi}{2}+q, \qquad |q|\ll1,
\end{equation}
gives
\begin{equation}
	\mathcal A_k\simeq \pm v_F q,
\end{equation}
with Fermi velocity
\begin{equation}
	v_F=-\left.\frac{d\mathcal A_k}{dk}\right|_{k=\pm \pi/2}= \pm 1 \mp \frac{3\alpha}{4}.
\end{equation}
Substituting this result into the general continuum Hamiltonian yields
\begin{equation}
	H_{\rm eff} = i v_F \int dx \left( \psi_R^\dagger\partial_x\psi_R - \psi_L^\dagger\partial_x\psi_L \right).
	\label{eqc1}
\end{equation}
Equation~(\ref{eqc1}) describes a single massless Dirac fermion and therefore realizes the conventional $c_{\rm eff}=1$ Luttinger-liquid universality class.
\\
\paragraph*{Lifshitz critical point: $\alpha=\alpha_c$.}
\mbox{}\\
\\
At
\begin{equation}
	\alpha=\alpha_c=\frac{4}{3},
\end{equation}
the Fermi points remain fixed at $\pm\pi/2$, but the linear coefficient vanishes,$	v_F=0$.

The leading contribution to the dispersion is then obtained from the next nonvanishing term in the momentum expansion, $\mathcal A_k\propto q^3$.
The continuum theory is therefore governed by a Lifshitz critical point with dynamical exponent $z=3$. Importantly, no additional Fermi-point pairs emerge exactly at the transition, so the number of gapless branches remains unchanged. This explains why the effective central charge remains compatible with $c_{\rm eff}=1$ at the critical point.
\\
\paragraph*{Strong-coupling regime: $\alpha>\alpha_c$.}
\mbox{}\\
\\
For $\alpha>\alpha_c$, Eq.~(\ref{eqroot}) admits two additional real solutions,
\begin{equation}
	\cos^2k_0=\frac{3}{4}-\frac{1}{\alpha},
\end{equation}
which generate three distinct pairs of Fermi points, $\pm{\pi}/{2}$,$\pm k_0$, and  $\pm(\pi-k_0)$.

Linearizing the dispersion around each Fermi point produces an independent massless Dirac sector. The low-energy Hamiltonian therefore becomes
\begin{equation}
	H_{\rm eff}	=	i	\sum_{a=1}^{3}	v_a	\int dx	\left( \psi_{R,a}^{\dagger}\partial_x\psi_{R,a}- \psi_{L,a}^{\dagger}\partial_x\psi_{L,a} \right),
	\label{eqc3}
\end{equation}
where $v_a$ denotes the corresponding Fermi velocity evaluated at each Fermi point.

Equation~(\ref{eqc3}) represents three decoupled gapless Dirac fermions. Since the central charge is additive for independent conformal sectors,
\begin{equation}
	c_{\rm eff}	=	\sum_{a=1}^{3} c_a,
\end{equation}
and each massless Dirac fermion contributes $c_a=1$, one obtains
\begin{equation}
	c_{\rm eff}=3.
\end{equation}

The Lifshitz transition at $\alpha=\alpha_c$ therefore corresponds to a reconstruction of the Fermi surface from one to three pairs of gapless Fermi points. Within the continuum description, this reconstruction changes the number of independent Dirac sectors from one to three, providing a direct field-theoretic explanation for the observed jump of the effective central charge,
\begin{equation}
	c_{\rm eff}:1\rightarrow3.
\end{equation}

\subsection{Low-Energy Effective Field Theory on the Critical Line $\beta_{c_2}$}
\label{app:betc2_cft}

We derive the continuum theory associated with the critical line
\begin{equation}
	\beta_{c_2}=-\left(1+\frac{8}{\alpha}\right),
\end{equation}
and demonstrate that its low-energy critical behavior is governed by a single massless Dirac fermion with central charge $c_{\rm eff}=1$.

Setting $h=0$, the Bogoliubov--de Gennes coefficients become
\begin{align}
	\mathcal A_k
	&=\cos k+\frac{\alpha}{8}(1+\beta)\cos3k
	=\cos k-J\cos3k,\\
	\mathcal B_k
	&=\frac{\alpha}{8}(1-\beta)\sin3k
	=\left(\frac{\alpha}{4}+\right)\sin3k.
\end{align}

The quasiparticle spectrum,
\begin{equation}
	\varepsilon(k)=\sqrt{\mathcal A_k^2+\mathcal B_k^2},
\end{equation}
vanishes at three symmetry-related momenta, $k_F=0$, $k_F=\pm\pi$.

Although three gapless points appear in the lattice spectrum, only two independent continuum sectors need to be retained. The two zone-boundary points are related by the reciprocal lattice vector $G=2\pi$ and therefore represent the same continuum degree of freedom.

This follows directly from the lattice expansion of the fermionic operator near the Brillouin-zone boundary,
\begin{equation}
	c_n
	\sim
	\psi_{R,\pi}(x)e^{i\pi n}
	+
	\psi_{L,\pi}(x)e^{-i\pi n}.
\end{equation}
Since
\begin{equation}
	e^{i\pi n}
	=
	e^{-i\pi n}
	=
	(-1)^n,
\end{equation}
one obtains
\begin{equation}
	c_n
	\sim
	\left[
	\psi_{R,\pi}(x)
	+
	\psi_{L,\pi}(x)
	\right]
	(-1)^n,
\end{equation}
which implies the identification
\begin{equation}
	\psi_{R,\pi}(x)
	\equiv
	\psi_{L,\pi}(x)
	\equiv
	\Psi_\pi(x).
	\label{eq:collapse}
\end{equation}
Consequently, the two lattice momenta $k=\pm\pi$ collapse into a single independent continuum sector.

The continuum theory therefore contains two flavor sectors, corresponding to the zone center ($k_F=0$) and the Brillouin-zone boundary ($k_F=\pi$). Evaluating the continuum couplings appearing in Eq.~(\ref{eqheff}) gives
\begin{align}
	M_0&=0,
	&
	\Delta_0&=0,
	&
	v_{F,0}&=0,
	&
	\tilde v_0
	=
	-\frac{3\alpha}{4}-3,
	\\
	M_\pi&=0,
	&
	\Delta_\pi&=0,
	&
	v_{F,\pi}&=0,
	&
	\tilde v_\pi
	=
	\frac{3\alpha}{4}+3.
\end{align}
Accordingly, both the conventional mass term and the ordinary kinetic contribution vanish identically, leaving the derivative pairing term as the leading contribution to the continuum Hamiltonian,
\begin{equation}
	H_{\rm eff}
	=
	2i
	\sum_{a=0,\pi}
	\tilde v_a
	\int dx
	\left(
	\psi^\dagger_{R,a}
	\partial_x
	\psi^\dagger_{L,a}
	-
	\psi_{L,a}
	\partial_x
	\psi_{R,a}
	\right).
	\label{eq:heff_c2}
\end{equation}

Equation~(\ref{eq:heff_c2}) contains two continuum sectors, associated with the zone center and the Brillouin-zone boundary. As established in Eq.~(\ref{eq:collapse}), however, the two lattice momenta $k=\pm\pi$ are related by a reciprocal lattice vector and therefore correspond to the same continuum field. Consequently, the zone-boundary contribution represents only one independent low-energy flavor despite the presence of two gap-closing lattice momenta.

Since the effective theory is formulated in the Bogoliubov--de Gennes representation, the continuum quasiparticles obey the intrinsic particle-hole symmetry characteristic of superconducting systems. It is therefore natural to express each complex continuum field in terms of two real Majorana fermions,
\begin{equation}
	\Psi_a(x)
	=
	\frac{\eta_{1,a}(x)+i\eta_{2,a}(x)}
	{\sqrt2},
	\qquad
	\eta_{\mu,a}^\dagger=\eta_{\mu,a},
\end{equation}
where $\nu=0,\pi$ labels the two independent continuum sectors. Here the Majorana fermions should be understood as the real components of the Bogoliubov quasiparticle fields rather than boundary zero-energy excitations. Their appearance follows solely from the particle-hole symmetry of the continuum BdG Hamiltonian and is independent of the choice of boundary conditions.

Because the Brillouin-zone boundary contributes only a single continuum flavor, the low-energy theory contains two independent gapless Majorana fermions, which combine into one massless Dirac fermion,
\begin{equation}
	\Psi_{\rm gapless}(x)
	=
	\frac{\gamma_1(x)+i\gamma_2(x)}
	{\sqrt2}.
\end{equation}
A massless Dirac fermion carries conformal central charge
\begin{equation}
	c_{\rm eff}=1,
\end{equation}
equivalently corresponding to two gapless Majorana fermions, each contributing $c_{\rm eff}=1/2$. Therefore, although the lattice spectrum exhibits three gap-closing momenta, the continuum limit contains only one independent Dirac sector because the two Brillouin-zone boundaries are physically equivalent. This analytical result is fully consistent with the finite-size scaling analysis presented in the main text.

\subsection{Low-Energy Effective Field Theory on the Critical Line $\beta_{c_3}$}
\label{app:betc3_cft}

In this Appendix, we derive the low-energy effective continuum field theory for the third critical line, defined by
\begin{equation}
	\beta_{c_3} = \frac{4}{\alpha} - 1,
\end{equation}
and explicitly demonstrate that its low-energy physics is described by two independent copies of massless Dirac fermions, yielding a total effective central charge of $c_{\text{eff}} = 2$.

Substituting the parameter relation of $\beta_{c_3}$ into the microscopic structural coefficients, we obtain the simplified forms:
\begin{align}
	\mathcal{A}_k &= \cos k + \frac{\alpha}{8}\left[1 + \left(\frac{4}{\alpha} - 1\right)\right]\cos 3k = \cos k + \frac{1}{2}\cos 3k, \\
	\mathcal{B}_k &= \frac{\alpha}{8}\left[1 - \left(\frac{4}{\alpha} - 1\right)\right]\sin 3k = \left(\frac{\alpha}{4} - \frac{1}{2}\right)\sin 3k.
\end{align}
The single-particle excitation spectrum $\varepsilon(k)$ vanishes identically when both $\mathcal{A}_k = 0$ and $\mathcal{B}_k = 0$. On this critical line, the quasiparticle gap closes at four distinct Fermi momenta within the first Brillouin zone  $k_F \in \left\{ \pm{\pi}/{3}, \pm{2\pi}/{3} \right\}$.

Exploiting the time-reversal and inversion symmetries of the underlying lattice ($k \to -k$), these four nodes are naturally grouped into $N_f = 2$ symmetric flavor channels: the first valley centered at $k_1 = {\pi}/{3}$ and the second valley centered at $k_2 = {2\pi}/{3}$. We evaluate the exact structural coefficients and their corresponding momentum derivatives at these designated nodes to extract the coupling parameters of the master continuum field theory. For the first flavor channel ($k_1 = {\pi}/{3}$, noting that $3k_1 = \pi$), we find:
\begin{align}
	M_1 &= \mathcal{A}_{\pi/3} = \cos\left(\frac{\pi}{3}\right) + \frac{1}{2}\cos(\pi) = \frac{1}{2} - \frac{1}{2} = 0, \\
	\Delta_1 &= \mathcal{B}_{\pi/3} = \left(\frac{\alpha}{4} - \frac{1}{2}\right)\sin(\pi) = 0, \\
	v_{F,1} &= \sin\left(\frac{\pi}{3}\right) + \frac{3\alpha}{8}(1+\beta_{c_3})\sin(\pi) = \frac{\sqrt{3}}{2}, \\
	\tilde{v}_1 &= \frac{3\alpha}{8}(1-\beta_{c_3})\cos(\pi) = 3\left(\frac{\alpha}{4} - \frac{J}{2}\right)(-1) = -\frac{3\alpha}{4} + \frac{3J}{2}.
\end{align}
Similarly, for the second flavor channel ($k_2 = {2\pi}/{3}$, noting that $3k_2 = 2\pi$), the coefficients evaluate to:
\begin{align}
	M_2 &= \mathcal{A}_{2\pi/3} = \cos\left(\frac{2\pi}{3}\right) + \frac{1}{2}\cos(2\pi) = -\frac{1}{2} + \frac{1}{2} = 0, \\
	\Delta_2 &= \mathcal{B}_{2\pi/3} = \left(\frac{\alpha}{4} - \frac{1}{2}\right)\sin(2\pi) = 0, \\
	v_{F,2} &= \sin\left(\frac{2\pi}{3}\right) + \frac{3\alpha}{8}(1+\beta_{c_3})\sin(2\pi) = \frac{\sqrt{3}}{2}, \\
	\tilde{v}_2 &= \frac{3\alpha}{8}(1-\beta_{c_3})\cos(2\pi) = 3\left(\frac{\alpha}{4} - \frac{1}{2}\right)(1) = \frac{3\alpha}{4} - \frac{3}{2}.
\end{align}
As explicitly demonstrated, across both active sectors, the effective mass terms and regular pairing gaps vanish identically ($M_i = 0$, $\Delta_i = 0$). Crucially, unlike the behavior observed on other critical lines, the standard diagonal Fermi velocity remains finite and dominant here $v_{F,i} = {\sqrt{3}}/{2} \neq 0$. 

Substituting these specific parameter sets into the generalized effective Hamiltonian yields the low-energy continuum action:
\begin{align}
	\mathcal{H}_{\text{eff}} &= \sum_{a \in \{1, 2\}} \int dx \bigg[ i v_{F,a} (\psi^\dagger_{R,a}\partial_x \psi_{R,a} - \psi^\dagger_{L,a}\partial_x \psi_{L,a}) \nonumber \\
	&+ \tilde{v}_a \left( \psi^\dagger_{R,a}\partial_x\psi^\dagger_{L,a} + \psi_{L,a}\partial_x\psi_{R,a} \right) \bigg].
\end{align}

A vital kinematic distinction must be emphasized: because all four Fermi points lie strictly within the interior of the first Brillouin zone and remain isolated from the zone boundary ($k_F \neq \pm\pi$), no field-collapse or chiral constraint occurs. The right-moving fields $\psi_{R,a}(x)$ and left-moving fields $\psi_{L,a}(x)$ for each flavor remain fully independent, unconstrained complex chiral fields. Consequently, each flavor sector $a \in \{1, 2\}$ independently supports a complete, massless Dirac fermion channel. Since a fully unconstrained Dirac channel contributes exactly $c_a = 1$ to the conformal anomaly, the total effective central charge of the system is determined by summing the contributions of these two decoupled, independent conformal sectors:
\begin{equation}
	c_{\text{eff}} = c_1 + c_2 = 1 + 1 = 2.
\end{equation}
Therefore, the low-energy effective field theory on the $\beta_{c_3}$ critical line is rigorously described by a multi-component conformal field theory with a quantized total central charge $c_{\rm eff}=2$, which perfectly matches the numerical finite-size scaling analysis presented in the main text.

\subsection{Field-Induced Reconstruction of the Low-Energy Theory Along the $\beta=1$ Line}
\label{app:bet1_field_cft}

We now specialize the general continuum formulation to the isotropic line
$\beta=1$ in the presence of a transverse magnetic field. In contrast to the
zero-field case, the magnetic field continuously deforms the Fermi surface.
To illustrate this mechanism explicitly, we consider the representative
parameter choice $\alpha=3$ and compare two magnetic fields located on opposite
sides of the field-driven Lifshitz transition.
\\
\paragraph*{Weak-field regime: $h=0.1$.}
\mbox{}\\
\\
For $h=0.1$, the quasiparticle spectrum possesses three symmetry-related pairs
of Fermi points,
\begin{align}
	k_{F,1}&=\pm\arccos(0.68233),\nonumber\\
	k_{F,2}&=\pm\arccos(-0.09463),\nonumber\\
	k_{F,3}&=\pm\arccos(-0.60102).
\end{align}

Each pair defines an independent low-energy valley. Expanding the lattice
fermions around $\pm k_{F,i}$ and projecting onto the corresponding continuum
sector, the general effective Hamiltonian reduces to
\begin{equation}
	H_{\rm eff}	= \sum_{i=1}^{3} H_{\rm eff}^{(i)},
\end{equation}
with
%\begin{align}
%	H_{\rm eff}^{(i)} &= \int dx \left[
%	iv_{F,i} (\psi_{R,i}^{\dagger}\partial_x\psi_{R,i} - \psi_{L,i}^{\dagger}\partial_x\psi_{L,i}) \nonumber
%	\\
%	& + \tilde v_i ( \psi_{R,i}^{\dagger}\partial_x\psi_{L,i}^{\dagger} + \psi_{L,i}\partial_x\psi_{R,i}) \right].
%\end{align}
\begin{align}
	H_{\rm eff}^{(i)} &= \int dx \left[
	iv_{F,i} (\psi_{R,i}^{\dagger}\partial_x\psi_{R,i} - \psi_{L,i}^{\dagger}\partial_x\psi_{L,i}) \right. \nonumber\\
	& \left. + \tilde v_i ( \psi_{R,i}^{\dagger}\partial_x\psi_{L,i}^{\dagger} + \psi_{L,i}\partial_x\psi_{R,i}) \right].
\end{align}

Since the three valleys remain well separated in momentum space, no inter-valley
hybridization occurs and each symmetric pair of Fermi points forms an
independent massless Dirac channel. The low-energy theory therefore consists of
three decoupled conformal sectors,
\begin{equation}
	c_{\rm eff}
	=
	\sum_{i=1}^{3}c_i
	=
	3.
\end{equation}

\paragraph*{Strong-field regime: $h=1$.}
\mbox{}\\

Increasing the magnetic field beyond the Lifshitz transition reconstructs the
Fermi surface, leaving only a single symmetry-related pair of Fermi points,
\begin{equation}
	k_F=\pm\arccos(0.88953).
\end{equation}

The continuum Hamiltonian therefore contains only one low-energy flavor,
\begin{equation}
	H_{\rm eff}
	=
	\int dx
	\left[
	iv_F
	(\psi_R^\dagger\partial_x\psi_R
	-
	\psi_L^\dagger\partial_x\psi_L)
	+
	\tilde v
	(
	\psi_R^\dagger\partial_x\psi_L^\dagger
	+
	\psi_L\partial_x\psi_R)
	\right].
\end{equation}

Diagonalization of this Bogoliubov-de Gennes sector yields a single linearly
dispersing massless Dirac fermion. Consequently, the continuum theory contains
only one independent conformal sector,
\begin{equation}
	c_{\rm eff}=1.
\end{equation}

The magnetic field therefore drives a Lifshitz reconstruction of the Fermi
surface from three pairs of Fermi points to a single pair. Within the continuum
description, this reconstruction reduces the number of independent massless
Dirac sectors from three to one, providing the field-theoretical explanation
for the discontinuous change of the effective central charge,
\begin{equation}
	c_{\rm eff}:3\rightarrow1.
\end{equation}
This result is fully consistent with the finite-size scaling analysis presented
in the main text.

\begin{figure}[h]
	\centerline{\includegraphics[width=0.5\linewidth,height=0.42\linewidth]{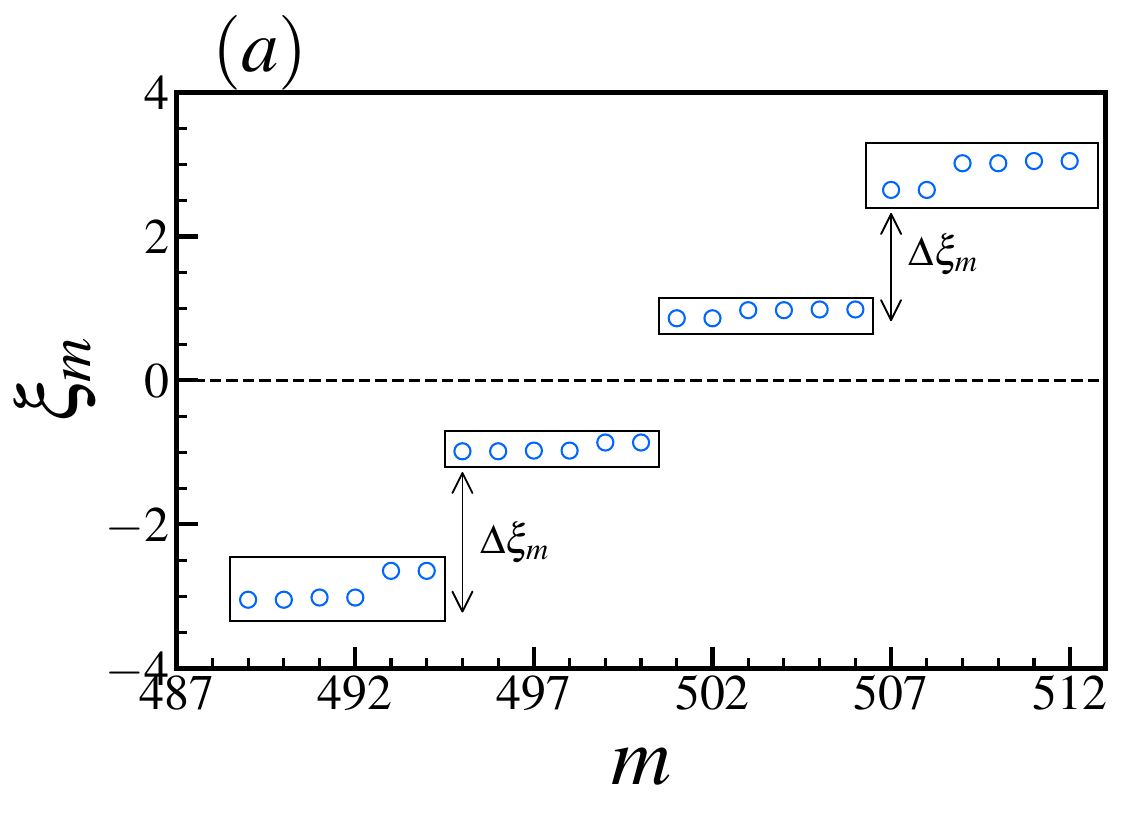} \includegraphics[width=0.5\linewidth,height=0.42\linewidth]{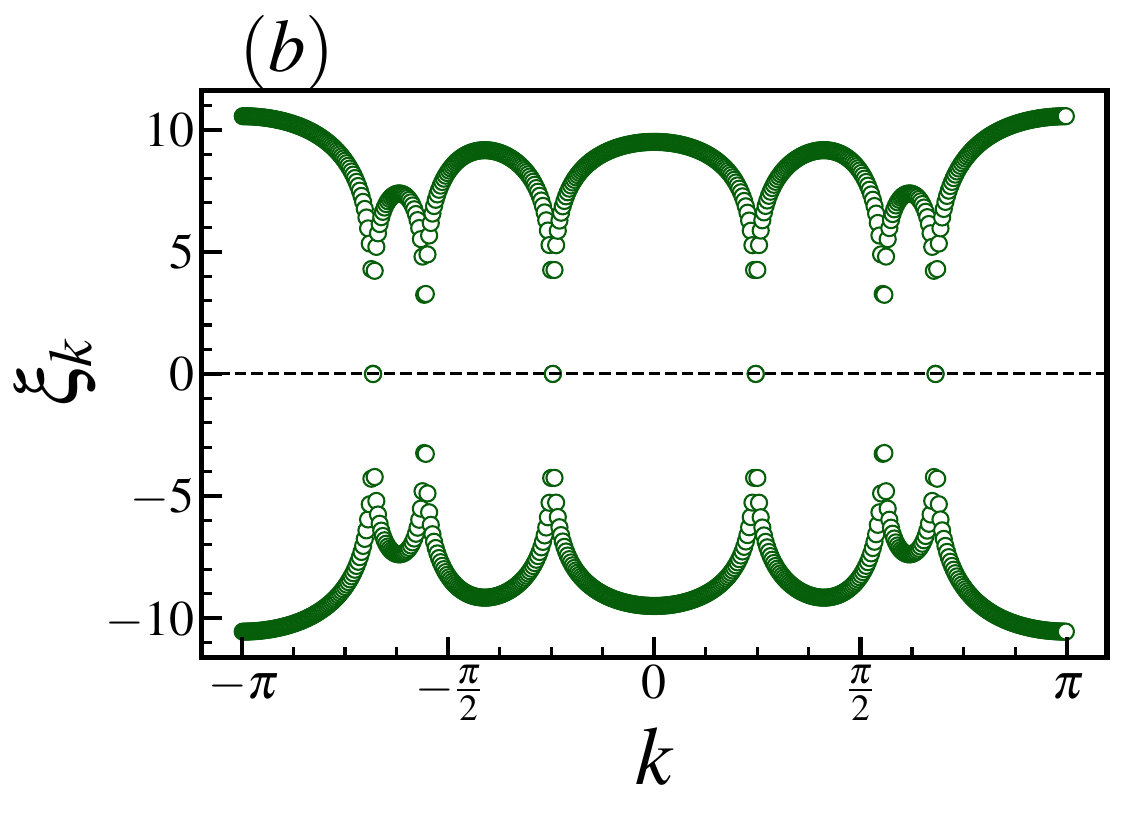} 
		
	}
	
	\centerline{\includegraphics[width=0.5\linewidth,height=0.42\linewidth]{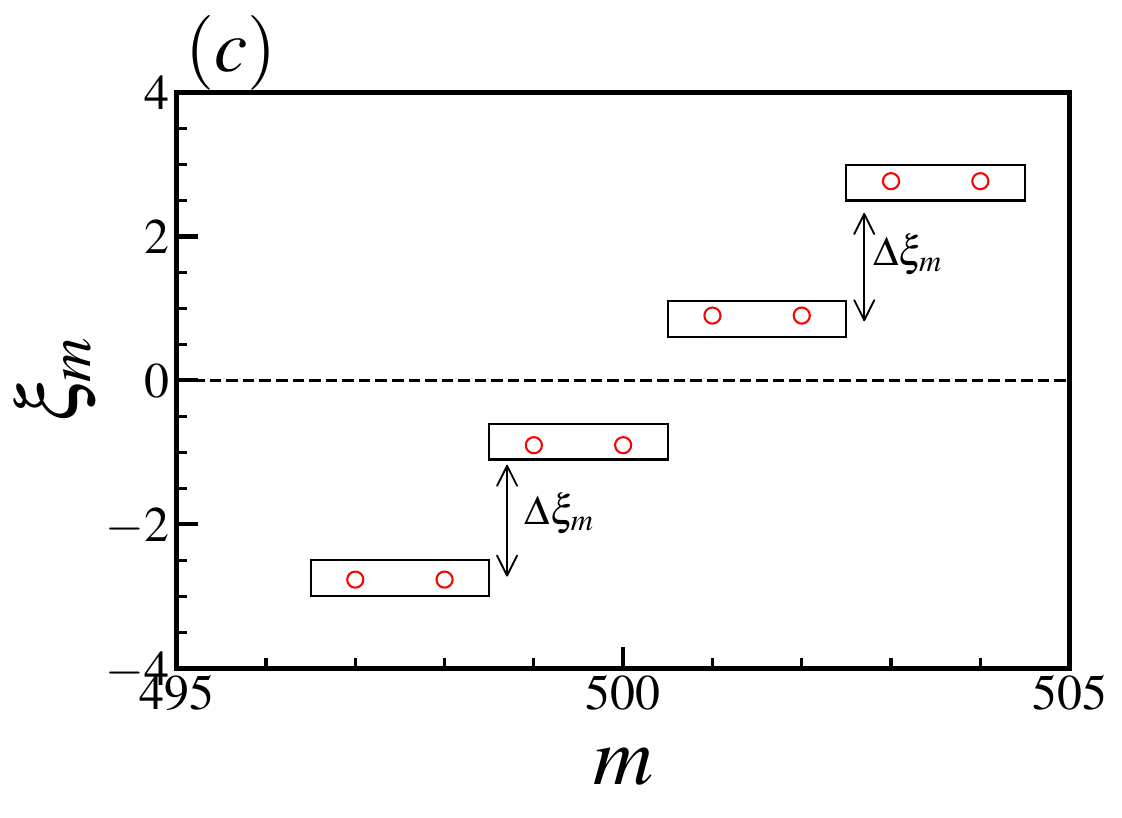} \includegraphics[width=0.5\linewidth,height=0.42\linewidth]{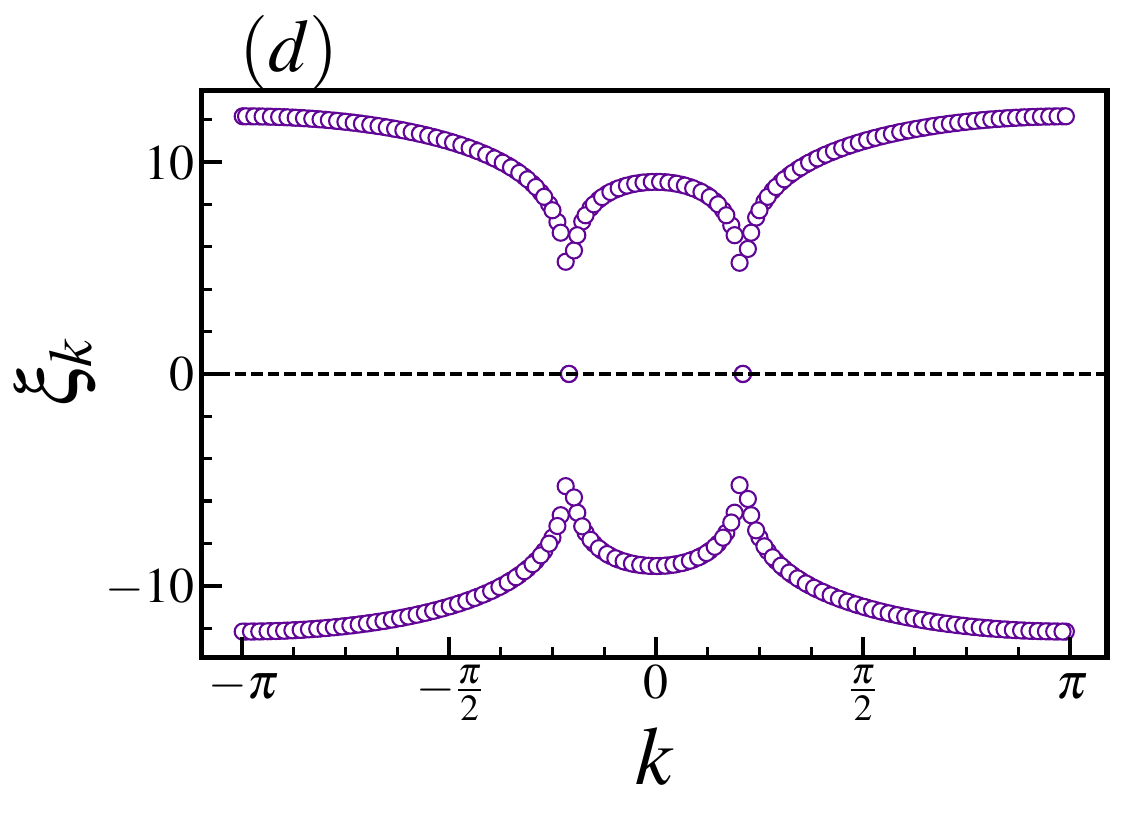} 
		
	}

	\caption{(a,b) Real-space (RES) and momentum-space (MES) bulk entanglement spectra in the phase ($\alpha<4/3$). (c,d) Real-space (RES) and momentum-space (MES) bulk entanglement spectra in the phase ($\alpha>4/3$).}
	\label{fig_c_1}
	
\end{figure}

Figure~\ref{fig_c_1}~(a) shows the real-space bulk entanglement spectrum for the $c_{\rm eff}=3$ phase ($\alpha=3$, $h=0.1$). The low-lying entanglement levels form a sixfold degenerate state, reflecting the presence of three independent pairs of gapless Fermi points.  The corresponding momentum-space bulk entanglement spectrum, shown in Fig.~\ref{fig_c_1}~(b), exhibits three pairs of entanglement cones centered at the Fermi momenta, providing a direct momentum-space signature of the reconstructed Fermi surface.

After the field-driven Lifshitz transition, only a single pair of Fermi points remains. Accordingly, the real-space bulk entanglement spectrum shown in Fig.~\ref{fig_c_1}~(c) reduces to doubly degenerate low-energy levels, consistent with a single massless Dirac sector and $c_{\rm eff}=1$. The corresponding momentum-space spectrum in Fig.~\ref{fig_c_1}~(d) contains only one pair of entanglement cones, confirming that the Lifshitz reconstruction removes two Dirac sectors from the low-energy theory. These bulk entanglement spectra therefore provide an independent entanglement signature of the field-induced transition from the $c_{\rm eff}=3$ phase to the $c_{\rm eff}=1$ phase.

\subsection{Low-Energy Effective Field Theory on the Critical Lines $h_{c_1}$ and $h_{c_2}$}
\label{app:hc12_cft}

The field-driven critical lines $h_{c_1}$ and $h_{c_2}$ are both characterized by Ising criticality with an effective central charge
\begin{equation}
	c_{\rm eff}=\frac12 .
\end{equation}
Although the gap closes at different momenta along the two critical branches, their continuum descriptions are governed by the same low-energy Majorana theory.
\\
Along the $h_{c_1}$ branch,
\begin{equation}
	h_{c_1}=-1-\frac{\alpha}{8}(1+\beta),
\end{equation}
the quasiparticle gap closes only at the Brillouin-zone boundaries,
\begin{equation}
	k_F=\pm\pi,
\end{equation}
where
\begin{equation}
	\mathcal A_{\pm\pi}=0,
	\qquad
	\mathcal B_{\pm\pi}=0,
\end{equation}
while the zone center remains fully gapped. In contrast, along
\begin{equation}
	h_{c_2}=1+\frac{\alpha}{8}(1+\beta),
\end{equation}
the gap closes only at the zone center,
\begin{equation}
	k_F=0,
\end{equation}
with
\begin{equation}
	\mathcal A_{0}=0,
	\qquad
	\mathcal B_{0}=0,
\end{equation}
whereas the Brillouin-zone boundary is gapped.

Within the general continuum formulation of Eq.~\eqref{eqheff}, the relevant couplings evaluated at the corresponding gapless momentum satisfy
\begin{equation}
	M=0,\qquad
	\Delta=0,\qquad
	v_F=0,
\end{equation}
leaving only the anomalous derivative coupling,
\begin{equation}
	\tilde v=
	\pm\frac{3\alpha}{8}(1-\beta),
\end{equation}
where the sign distinguishes the two critical branches but does not affect the universality class.

The resulting low-energy Hamiltonian therefore reduces to
\begin{equation}
	H_{\rm eff}
	=
	\tilde v
	\int dx
	\left(
	\Psi^\dagger\partial_x\Psi^\dagger
	+
	\Psi\partial_x\Psi
	\right).
	\label{eq:hc_majorana}
\end{equation}

The only distinction between the two transitions concerns the continuum field $\Psi(x)$. Along the $h_{c_1}$ branch, the gap closes at the equivalent lattice momenta $k_F=\pm\pi$, which differ by the reciprocal lattice vector $2\pi$. Consequently,
\begin{equation}
	c_n
	\sim
	[\psi_{R,\pi}(x)+\psi_{L,\pi}(x)](-1)^n,
\end{equation}
so that the two lattice modes represent the same continuum degree of freedom,
\begin{equation}
	\psi_{R,\pi}(x)
	\equiv
	\psi_{L,\pi}(x)
	\equiv
	\Psi(x).
\end{equation}

Along the $h_{c_2}$ branch, the gap closes only at the zone center,
\begin{equation}
	c_n
	\sim
	\psi_{R,0}(x)+\psi_{L,0}(x),
\end{equation}
which likewise gives a single continuum flavor. Thus, despite their different momentum-space locations, both critical lines contain only one independent gapless continuum sector.

Finally, exploiting the particle-hole symmetry of the Bogoliubov--de Gennes Hamiltonian, the complex field can be decomposed into two real Majorana fields,
\begin{equation}
	\Psi(x)
	=
	\frac{\eta_1(x)+i\eta_2(x)}{\sqrt2},
	\qquad
	\eta_\mu^\dagger=\eta_\mu,
\end{equation}
which transforms Eq.~(\ref{eq:hc_majorana}) into
\begin{equation}
	H_{\rm eff}
	=
	\int dx
	\left(
	\tilde v\,\eta_1\partial_x\eta_1
	-
	\tilde v\,\eta_2\partial_x\eta_2
	\right).
\end{equation}

The continuum limit therefore consists of a single massless Majorana fermion in $(1+1)$ dimensions, corresponding to the Ising conformal field theory with
\begin{equation}
	c_{\rm eff}=\frac12,
\end{equation}
in complete agreement with the finite-size scaling analysis presented in the main text.

\begin{figure}[h]
	\centerline{\includegraphics[width=0.5\linewidth,height=0.42\linewidth]{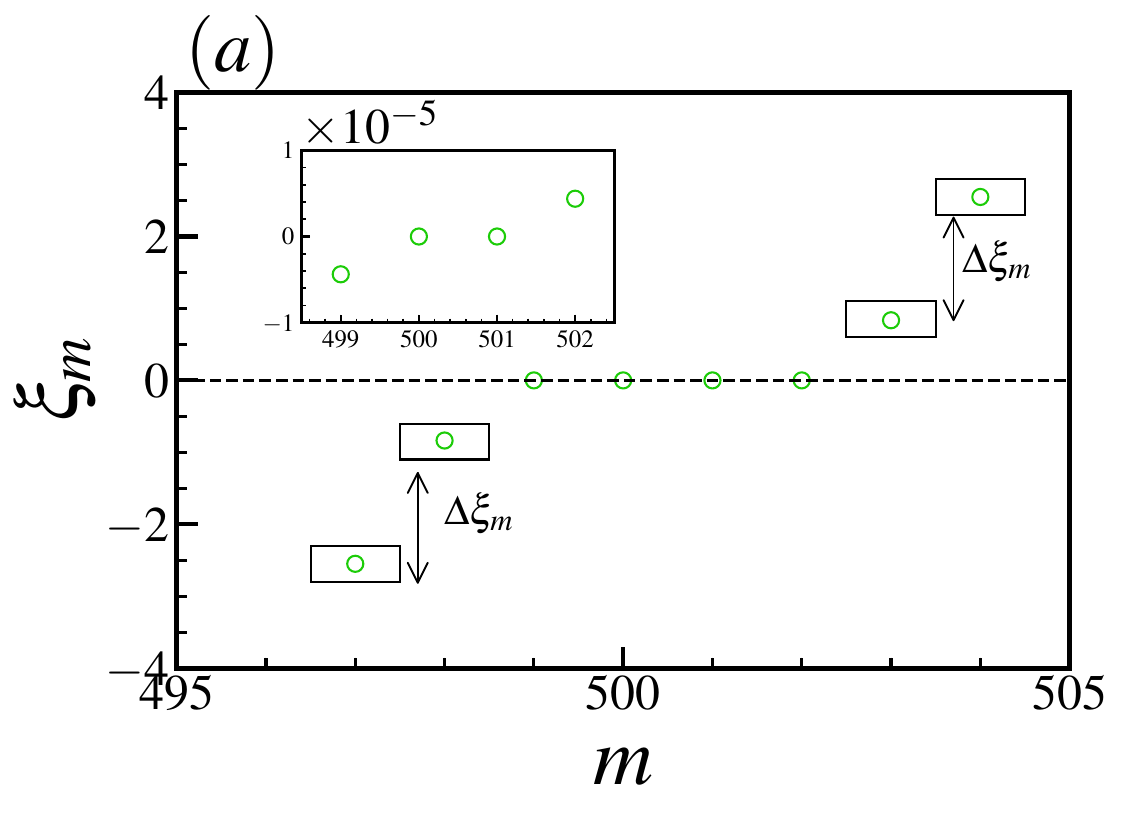} \includegraphics[width=0.5\linewidth,height=0.42\linewidth]{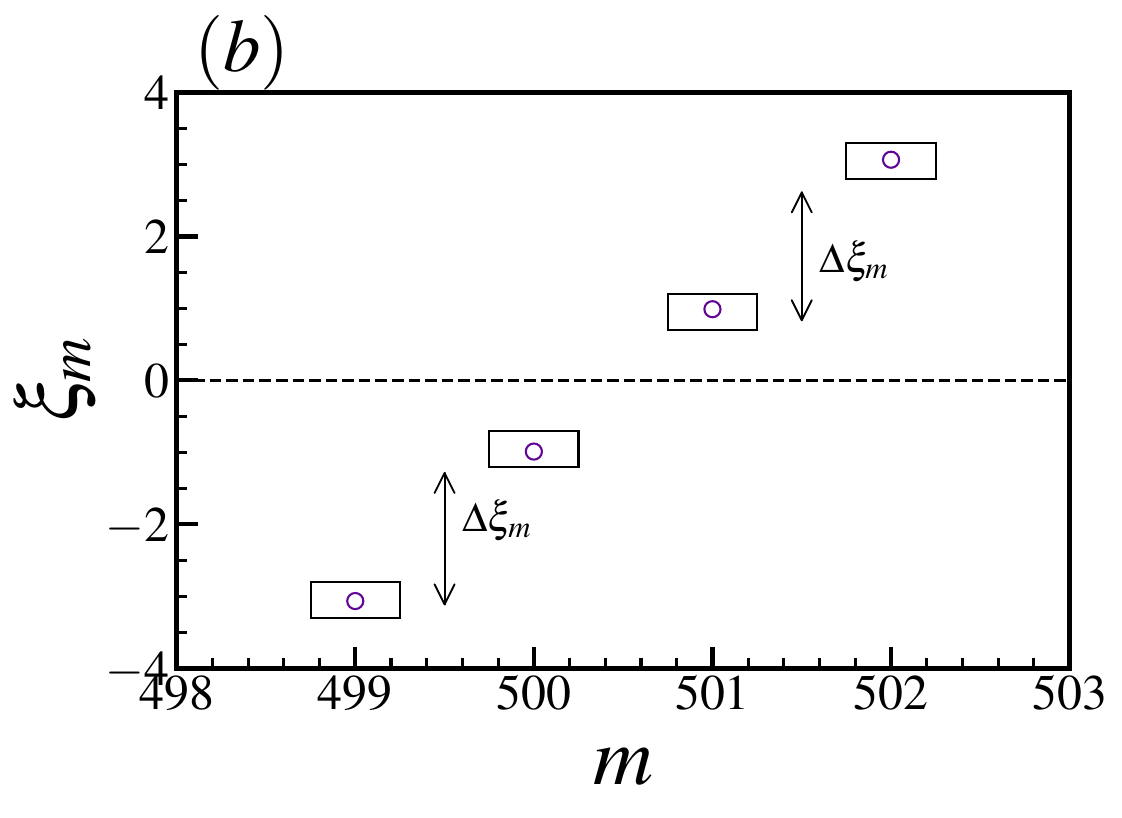} 
		
	}
	
	\centerline{\includegraphics[width=0.5\linewidth,height=0.42\linewidth]{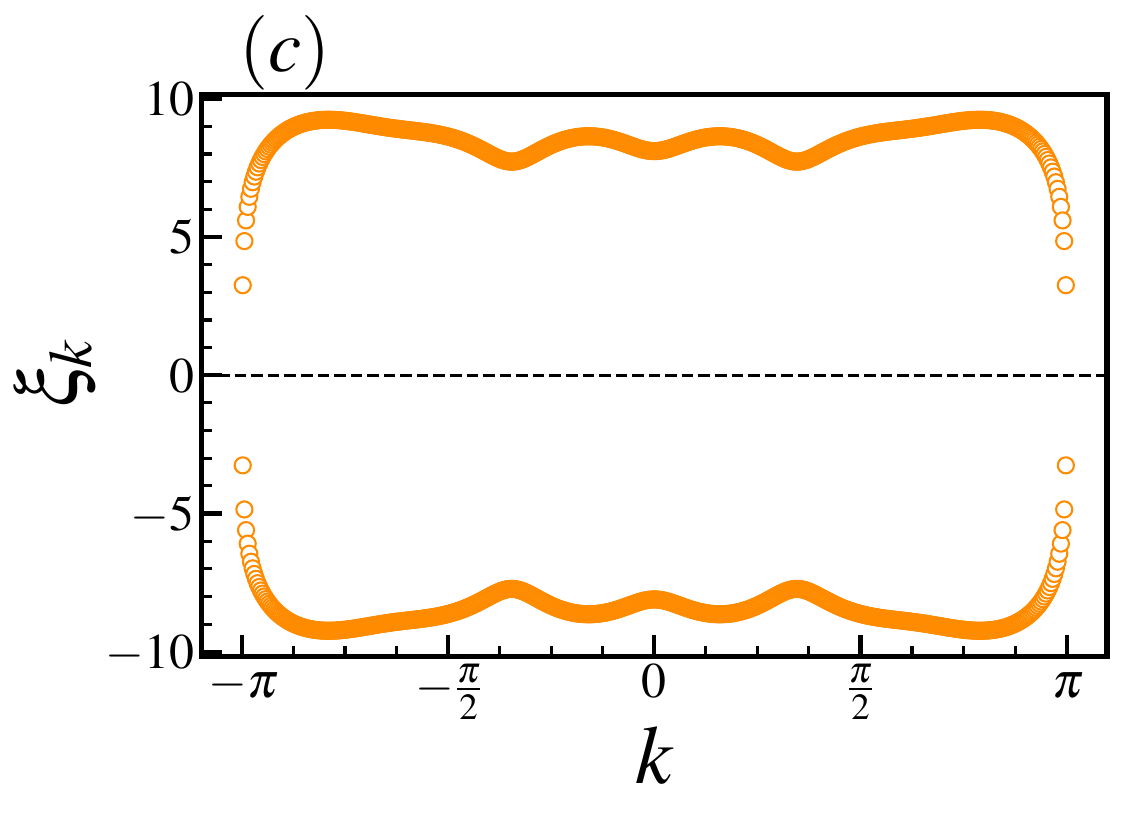} \includegraphics[width=0.5\linewidth,height=0.42\linewidth]{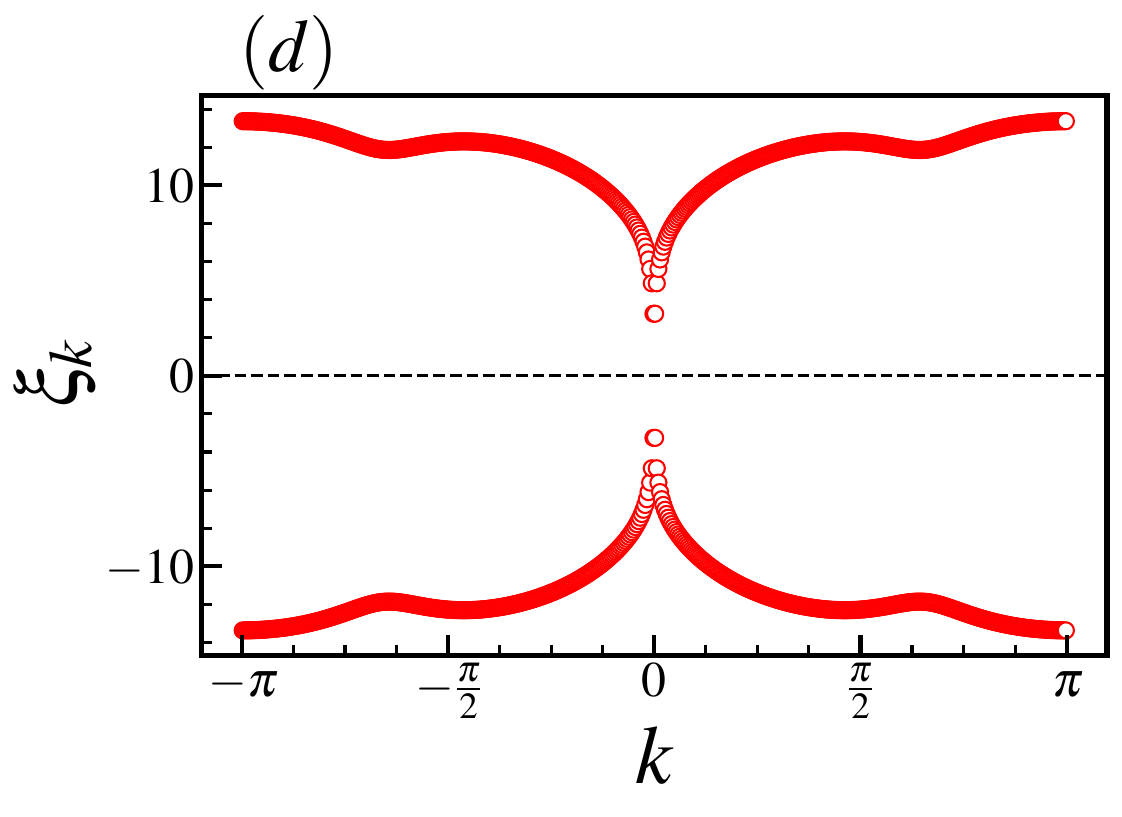} 
		
	}

	\caption{ (a,b) Real-space (RES) bulk entanglement spectra, and (c,d) momentum-space (MES) bulk entanglement spectra for $h_{c_3}$ and $h_{c_2}$, respectively.}
	\label{fig_c_2}
	
\end{figure}
We now present the real-space and momentum-space bulk entanglement spectra associated with the critical lines $h_{c_1}$ and $h_{c_2}$, providing an independent numerical verification of the continuum Majorana theory derived above.

For the critical line $h_{c_1}$, the real-space bulk entanglement spectrum shown in Fig.~\ref{fig_c_2}~(a) forms a single conformal tower without any level degeneracy. In contrast to the $c_{\rm eff}=1,2$, and $3$ critical phases discussed above, no doublets or higher-order multiplets are observed. As the system size increases, the entanglement gaps between successive levels decrease while the nondegenerate level structure is preserved, indicating that the spectrum becomes gapless in the thermodynamic limit. This behavior is fully consistent with a single gapless Majorana mode and the effective central charge $c_{\rm eff}=1/2$. The same behavior is observed along the critical line $h_{c_2}$, as shown in Fig.~\ref{fig_c_2}~(b). Although the gapless excitation is located at the Brillouin-zone center rather than the zone boundary, the bulk entanglement spectrum again consists of a single nondegenerate conformal tower with no evidence of level doubling or higher degeneracies. The identical entanglement structure on the two critical lines demonstrates that the degeneracy pattern is determined by the conformal content of the low-energy theory rather than by the momentum-space location of the gapless mode. Together, Figs.~\ref{fig_c_2}~(a) and \ref{fig_c_2}~(b) establish the absence of multiplet formation as the characteristic entanglement signature of the Ising criticality with $c_{\rm eff}=1/2$.

Figures~\ref{fig_c_2}~(c) and \ref{fig_c_2}~(d) present the momentum-space bulk entanglement spectra along the critical lines $h_{c_1}$ and $h_{c_2}$, respectively. In both cases, the entanglement spectrum develops a single isolated minimum that coincides precisely with the gapless momentum of the quasiparticle spectrum. Along $h_{c_1}$, the minimum is located at the Brillouin-zone boundary, $k=\pm\pi$, whereas along $h_{c_2}$ it shifts to the zone center, $k=0$. Away from these momenta, the entanglement spectrum remains finite, indicating that no additional low-energy entanglement channels emerge. Thus, unlike the higher-central-charge phases, where multiple entanglement minima directly reflect the presence of several independent conformal sectors, the two critical lines exhibit only a single gapless entanglement sector, consistent with the Ising criticality characterized by $c_{\rm eff}=1/2$. The only qualitative distinction between the two spectra is therefore the momentum-space position of the entanglement minimum, while their low-energy conformal structure remains identical.

\subsection{Low-Energy Effective Field Theory on the Critical Lines $h_{c_3}$ and $h_{c_4}$}
\label{app:hc34_cft}

The critical lines $h_{c_3}$ and $h_{c_4}$ are characterized by two symmetry-related gapless momenta located inside the Brillouin zone and exhibit an effective central charge
\begin{equation}
	c_{\rm eff}=1.
\end{equation}
Unlike the field-driven critical lines $h_{c_1}$ and $h_{c_2}$, where the gapless modes occur at high-symmetry points and the continuum theory reduces to a single Majorana sector, the interior Fermi points on $h_{c_3}$ and $h_{c_4}$ support a complete massless Dirac channel.

We first consider the critical line
\begin{equation}
	h_{c_3}
	=
	\frac{1}{2}
	-
	\frac{\alpha}{8}(1+\beta),
	\qquad
	\beta<1-\frac{4}{\alpha}.
\end{equation}
Along this branch, the quasiparticle gap closes at
\begin{equation}
	k_F=\pm\frac{\pi}{3}.
\end{equation}
The two gapless momenta are related by inversion symmetry and therefore constitute the two chiral components of a single continuum flavor rather than two independent Dirac species.

Evaluating the continuum couplings of Eq.~\eqref{eqheff} at $k_F=\pi/3$ gives
\begin{equation}
	M=0,
	\qquad
	\Delta=0,
\end{equation}
while
\begin{equation}
	v_F=\frac{\sqrt{3}}{2},
	\qquad
	\tilde v
	=
	-\frac{3\alpha}{8}(1-\beta).
\end{equation}
Thus, the mass and nonderivative pairing terms vanish on the critical line, whereas the ordinary Fermi velocity remains finite. The resulting low-energy Hamiltonian is
\begin{align}
	H_{\rm eff}
	&=
	\int dx
	\left[
	i v_F
	\left(
	\psi_R^\dagger\partial_x\psi_R
	-
	\psi_L^\dagger\partial_x\psi_L
	\right) \right. \nonumber\\
	& \left. + \tilde v
	\left(
	\psi_R^\dagger\partial_x\psi_L^\dagger
	+
	\psi_L\partial_x\psi_R
	\right)
	\right].
	\label{eq:hc3_dirac}
\end{align}
Because the gapless momenta lie away from $k=0$ and the Brillouin-zone boundary, the right- and left-moving fields remain independent continuum degrees of freedom. They therefore form a single complete massless Dirac sector, yielding
\begin{equation}
	c_{\rm eff}=1.
\end{equation}

The same structure is realized along the critical line
\begin{equation}
	h_{c_4}
	=
	-\frac{1}{2}
	+
	\frac{\alpha}{8}(1+\beta),
	\qquad
	\beta>\frac{4}{\alpha}-1,
\end{equation}
where the gap closes at
\begin{equation}
	k_F=\pm\frac{2\pi}{3}.
\end{equation}
Again, the two nodes are related by inversion symmetry and together define a single continuum flavor. At $k_F=2\pi/3$, the coefficients entering Eq.~\eqref{eqheff} are
\begin{equation}
	M=0,
	\qquad
	\Delta=0,
\end{equation}
with
\begin{equation}
	v_F=\frac{\sqrt{3}}{2},
	\qquad
	\tilde v
	=
	\frac{3\alpha}{8}(1-\beta).
\end{equation}
Consequently, the low-energy Hamiltonian has the same Dirac form as Eq.~\eqref{eq:hc3_dirac}, differing only in the momentum at which the continuum limit is taken and in the sign of the anomalous derivative coupling.

The two critical branches therefore share the same low-energy conformal structure. Their distinct Fermi-point locations, $k_F=\pm\pi/3$ for $h_{c_3}$ and $k_F=\pm2\pi/3$ for $h_{c_4}$, do not produce additional independent continuum flavors. In both cases, the two symmetry-related nodes combine into a single massless Dirac fermion, and hence
\begin{equation}
	c_{\rm eff}=1.
\end{equation}
This continuum result agrees with the finite-size scaling analysis presented in the main text.

\begin{figure}[h]
	\centerline{\includegraphics[width=0.5\linewidth,height=0.42\linewidth]{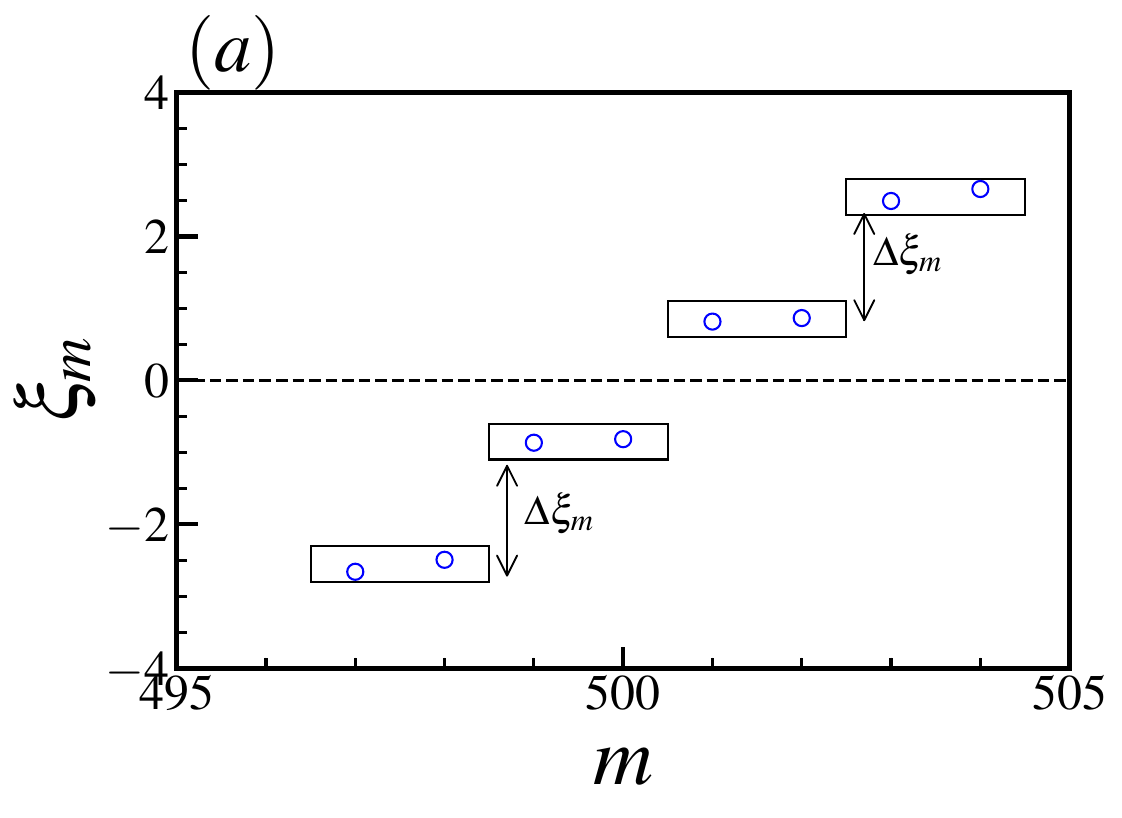} \includegraphics[width=0.5\linewidth,height=0.42\linewidth]{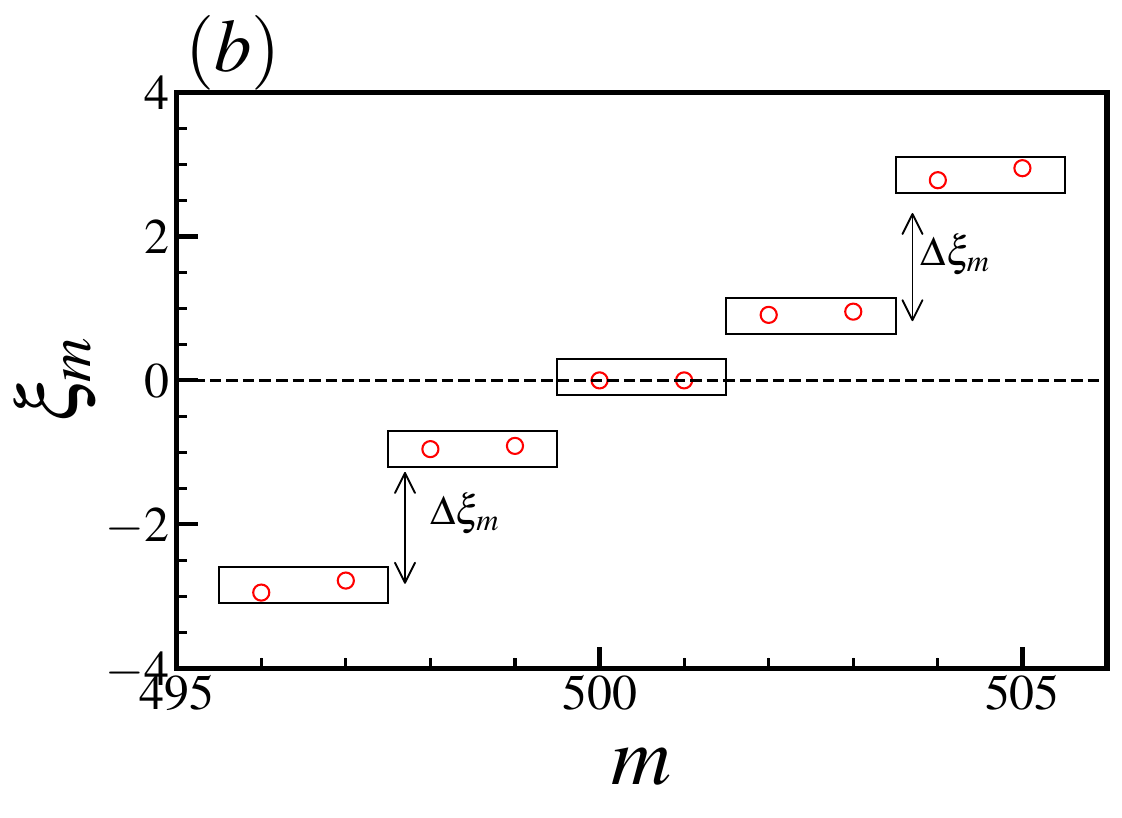} 
		
	}
	
	\centerline{\includegraphics[width=0.5\linewidth,height=0.42\linewidth]{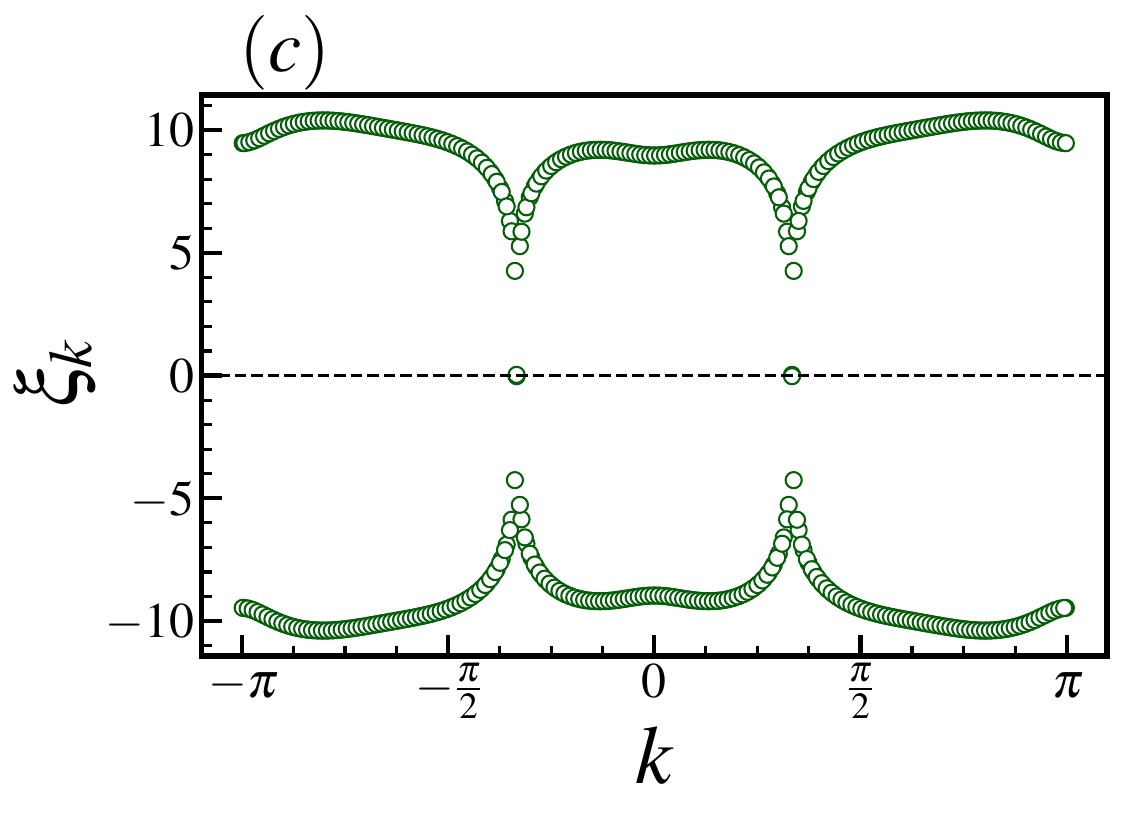} \includegraphics[width=0.5\linewidth,height=0.42\linewidth]{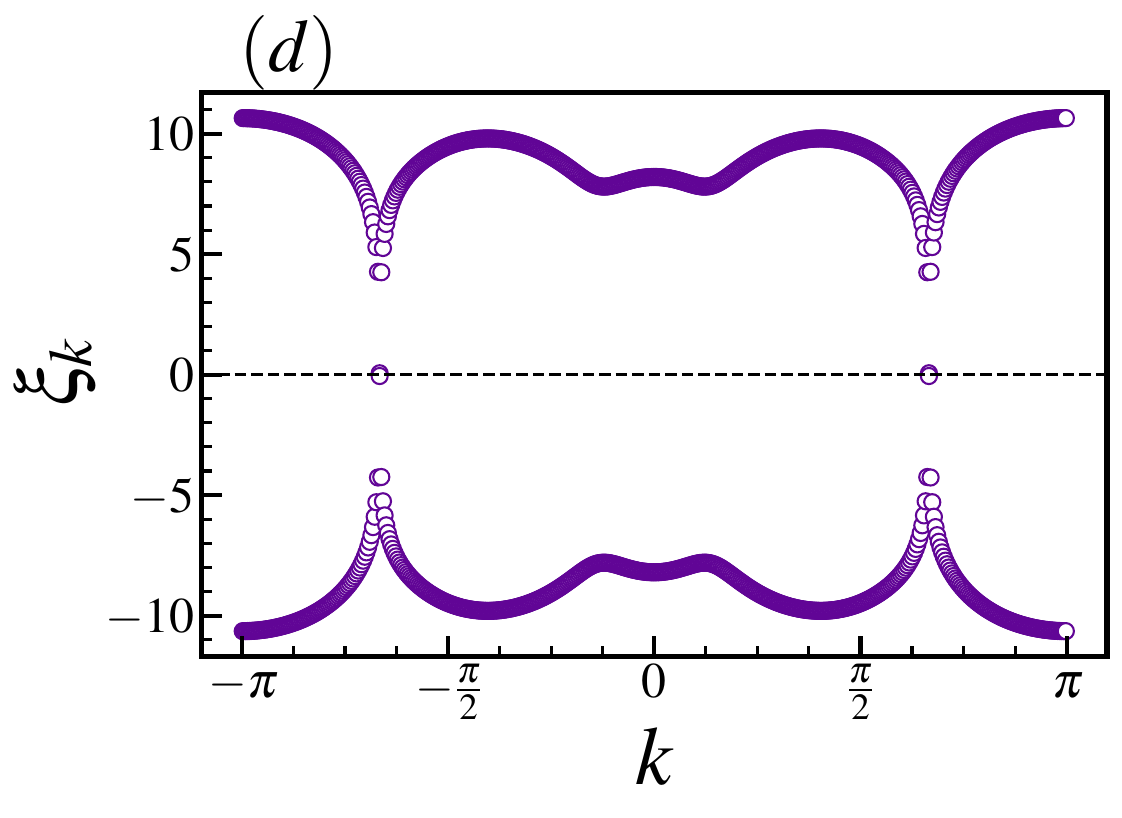} 
		
	}

	\caption{(a,b) Real-space (RES) bulk entanglement spectra, and (c,d) momentum-space (MES) bulk entanglement spectra for $h_{c_3}$ and $h_{c_4}$, respectively.}
	\label{fig_c_3}
	
\end{figure}

Figures~\ref{fig_c_3}~(a) and \ref{fig_c_3}~(b) show the real-space bulk entanglement spectra along the $h_{c_3}$ and $h_{c_4}$ critical lines, respectively. In both cases, the observed double degeneracy is consistent with the presence of a single gapless Dirac channel. The corresponding momentum-space spectra in Figs.~\ref{fig_c_3}~(c) and \ref{fig_c_3}~(d) exhibit minima at the respective Fermi momenta, $k_F=\pm\pi/3$ for $h_{c_3}$ and $k_F=\pm2\pi/3$ for $h_{c_4}$. Together, these results provide a direct momentum-resolved signature of the Dirac modes governing the two $c_{\rm eff}=1$ critical lines.

\subsection{Low-Energy Effective Field Theory at the Multicritical Intersection Point}
\label{app:multicritical_cft}

The intersection of the critical branches $h_{c_2}$ and $h_{c_3}$ exhibits a critical structure that is qualitatively distinct from that of either branch separately. The intersection occurs at
\begin{equation}
	\beta_m=-1-\frac{2}{\alpha},
	\qquad
	h_m=\frac{3}{4},
\end{equation}
and is characterized by the simultaneous closing of the quasiparticle gap at three distinct momenta,
\begin{equation}
	k_F=0,\qquad k_F=\pm\frac{\pi}{3}.
\end{equation}
Thus, the critical modes associated with the $h_{c_2}$ branch and those associated with the $h_{c_3}$ branch become simultaneously gapless. The resulting low-energy theory contains two qualitatively different gapless sectors, whose contributions to the conformal anomaly add to
\begin{equation}
	c_{\mathrm{eff}}=\frac{3}{2}.
\end{equation}
To make the structure of the low-energy theory explicit, we substitute the multicritical parameters into the Bogoliubov--de Gennes coefficients. Up to an overall choice of energy units, the resulting functions take the form
\begin{align}
	\mathcal{A}_k
	&=
	\cos k-\frac{3}{4}-\frac{1}{4}\cos(3k),
	\\
	\mathcal{B}_k
	&=
	\frac{\alpha+1}{4}\sin(3k).
\end{align}
The gap closes whenever both $\mathcal{A}_k$ and $\mathcal{B}_k$ vanish. At the multicritical point this condition is satisfied simultaneously at $k=0$ and at the symmetry-related pair $k=\pm\pi/3$. The crucial feature is that the nature of the linearized theory is not identical at these momenta. In particular, the conventional particle-hole dispersion has vanishing slope at the zone center, whereas it has a finite slope at the two interior Fermi points.
\mbox{}\\
\\
\paragraph{Zone-center sector: an emergent Majorana critical mode.}
\mbox{}\\
\\
We first expand around the zone center,
\begin{equation}
	k=q,\qquad |q|\ll1.
\end{equation}

At $k_F=0$, one finds
\begin{align}
	M_0
	&=
	\mathcal{A}_0
	=0,
	&
	\Delta_0
	&=
	\mathcal{B}_0
	=0,
	\\
	v_{F,0}
	&=
	-\left.\frac{\partial\mathcal{A}_k}{\partial k}\right|_{k=0}
	=0,
	&
	\tilde v_0
	&=
	\left.\frac{\partial\mathcal{B}_k}{\partial k}\right|_{k=0}
	=
	\frac{3(\alpha+1)}{4}.
\end{align}

Consequently, the usual linearized particle-hole term is absent at the zone center. The leading gapless contribution instead originates from the anomalous, derivative pairing channel. In the continuum limit, the corresponding quadratic Hamiltonian can be expressed in terms of the complex low-energy field $\Psi_0$ and its particle-hole conjugate. Upon decomposing $\Psi_0$ into real components,
\begin{equation}
	\Psi_0=\frac{1}{\sqrt{2}}\left(\chi_1+i\chi_2\right),
\end{equation}
the particle-hole constraint identifies the independent low-energy degrees of freedom such that only one real Majorana channel remains as an independent critical mode. The zone-center sector therefore realizes an Ising-type conformal sector with
\begin{equation}
	c_0=\frac{1}{2}.
\end{equation}

This sector is therefore not an ordinary Dirac cone. Its criticality is generated by the vanishing of both the conventional mass and the conventional Fermi velocity, with the derivative pairing term providing the leading linear dispersion. The zone-center mode consequently represents a distinct realization of relativistic low-energy criticality, in which the independent gapless degree of freedom is Majorana rather than Dirac.
\mbox{}\\
\paragraph{Finite-momentum sector: a massless Dirac mode.}
\mbox{}\\
\\
We next expand around the pair of symmetry-related Fermi points,
\begin{equation}
	k=\pm\frac{\pi}{3}+q,
	\qquad |q|\ll1.
\end{equation}

At $k_F=\pi/3$, and equivalently at $k_F=-\pi/3$, the low-energy parameters are
\begin{align}
	M_{\pi/3}
	&=
	\mathcal{A}_{\pi/3}
	=0,
	&
	\Delta_{\pi/3}
	&=
	\mathcal{B}_{\pi/3}
	=0,
	\\
	v_{F,\pi/3}
	&=
	-\left.
	\frac{\partial\mathcal{A}_k}{\partial k}
	\right|_{k=\pi/3}
	=
	\frac{\sqrt{3}}{2},
	&
	\tilde v_{\pi/3}
	&=
	\left.
	\frac{\partial\mathcal{B}_k}{\partial k}
	\right|_{k=\pi/3}
	=
	-\frac{3(\alpha+1)}{4}.
\end{align}

In contrast to the zone-center sector, the ordinary particle-hole velocity is finite. The two symmetry-related Fermi points therefore combine into a conventional complex low-energy fermionic channel. Equivalently, the pair of modes at $\pm\pi/3$ forms a complete massless Dirac sector,
\begin{equation}
	c_{\pm\pi/3}=1.
\end{equation}

The distinction between the two sectors is important. The gap closing at $k=0$ does not simply provide an additional copy of the finite-momentum Dirac mode. Rather, the vanishing of the conventional Fermi velocity changes the structure of the low-energy expansion and leaves a single independent Majorana critical channel. The pair of finite-momentum nodes, on the other hand, remains a conventional complex fermionic sector. The multicritical point therefore combines an Ising-like Majorana mode with a  Dirac mode.

The resulting low-energy Hamiltonian may consequently be organized as
\begin{equation}
	H_{\mathrm{eff}}
	=
	H_{\mathrm{Ising}}
	+
	H_{\mathrm{Luttinger-liquid}},
\end{equation}
or, equivalently,
\begin{equation}
	H_{\mathrm{eff}}
	=
	H_{\mathrm{eff}}^{(0)}
	+
	H_{\mathrm{eff}}^{(\pm\pi/3)}.
\end{equation}
The corresponding conformal anomaly is additive,
\begin{equation}
	c_{\mathrm{eff}}
	=
	c_0+c_{\pm\pi/3}
	=
	\frac{1}{2}+1
	=
	\frac{3}{2}.
\end{equation}
Thus, the simultaneous criticality of the zone-center Majorana sector and the finite-momentum Dirac sector provides a microscopic explanation for the value $c_{\rm eff}=3/2$ observed at the intersection of $h_{c_2}$ and $h_{c_3}$. Within the continuum description, the multicritical point is naturally interpreted as the coexistence of an Ising ($c_{\rm eff}=1/2$) sector and a Dirac (Luttinger-liquid) ($c_{\rm eff}=1$) sector.

The same result also clarifies why the intersection is more than a simple crossing of two phase-boundary curves. The two branches correspond to distinct gap-closing mechanisms and therefore to distinct low-energy momentum structures. Along $h_{c_2}$, the critical excitation is associated with the zone-center momentum, whereas along $h_{c_3}$ the critical modes occur at the symmetry-related finite momenta $\pm\pi/3$. At their intersection, all three momenta become simultaneously gapless. Consequently, the low-energy theory combines the critical content of both branches, with the zone-center excitation contributing a Majorana sector and the finite-momentum pair forming a single massless Dirac  sector. This coexistence naturally accounts for the observed value
\begin{equation}
	c_{\rm eff}=\frac{1}{2}+1=\frac{3}{2}.
\end{equation}
Interestingly, this conformal decomposition coincides with the field content of the second $\mathcal{N}=2$ superconformal minimal model, whose central charge is also $c_{\rm eff}=3/2$. Nevertheless, the present analysis establishes only the decomposition of the low-energy conformal degrees of freedom and the corresponding effective central charge. Demonstrating an emergent $\mathcal{N}=2$ superconformal symmetry would require additional evidence beyond the present mode-counting analysis, such as the organization of scaling operators into supermultiplets or the realization of the full superconformal algebra.

Therefore, our continuum analysis provides a microscopic explanation for the observed $c_{\rm eff}=3/2$ criticality in terms of coexisting Ising and Luttinger-liquid critical sectors. The agreement between this low-energy mode counting and the independently extracted effective central charge strongly supports this conformal decomposition, while leaving open the interesting possibility that a richer superconformal structure may emerge in the low-energy limit.

%\bibliography{ref}

\end{document}